\documentclass[12pt]{article}
\usepackage{floatrow}
\usepackage[top=1in, bottom=1in, left=1in, right=1in]{geometry}
\usepackage{setspace}
   
\fontsize{12}{0}
\usepackage{amsmath}
\usepackage{amssymb}
\usepackage{bm}
\usepackage{amsthm}
\usepackage{exscale}
\usepackage[mathscr]{eucal}
\usepackage{bm}
\usepackage{eqlist} 
\usepackage[final]{graphicx}
\usepackage[dvipsnames]{color}
\usepackage{verbatim}
\usepackage[square,sort,comma]{natbib}
\usepackage{caption}
\usepackage{enumerate}
\usepackage{subcaption}
\usepackage{algpseudocode}
\usepackage{rotating}
\usepackage{etoolbox}
\usepackage{hyperref}
\usepackage{color}
\usepackage[linesnumbered,ruled]{algorithm2e}

\makeatletter
\patchcmd{\env@cases}{1.2}{0.6}{}{}
\makeatother

\DeclareGraphicsExtensions{.pdf, .jpg}

\theoremstyle{definition}
\newtheorem{definition}{Definition}

\newtheorem{condition}{Condition}

\newtheorem{proposition.a}{Proposition A\ignorespaces}
\newtheorem{proposition.s}{Proposition S\ignorespaces}
\newtheorem{remark}{Remark}
\newtheorem{remark.s}{Remark S\ignorespaces}
\newtheorem{thm}{Theorem}
\newtheorem{thm.s}{Theorem S\ignorespaces}
\newtheorem{cor}{Corollary}

\newcommand{\x}{\bm{x}}
\newcommand{\w}{\bm{w}}
\newcommand{\y}{\bm{y}}
\newcommand{\X}{\bm{X}}
\newcommand{\Z}{\bm{Z}}
\newcommand{\W}{\bm{W}}
\newcommand{\bmS}{\bm{S}}
\newcommand{\bma}{\bm{a}}
\newcommand{\bmb}{\bm{b}}
\newcommand{\bmeps}{\bm{\epsilon}}
\newcommand{\bmbeta}{\bm{\beta}}
\newcommand{\bmalpha}{\bm{\alpha}}
\newcommand{\bmeta}{\bm{\eta}}
\newcommand{\1}{\bm{1_n}}
\newcommand{\tE}{\mbox{E}}
\newcommand{\var}{\mbox{Var}}
\newcommand{\tr}{\mbox{tr}}
\newcommand{\bbR}{\mathbb{R}}
\newcommand{\bmI}{\bm{I}}
\newcommand{\bigO}{O}

\usepackage{newfloat}
\DeclareFloatingEnvironment[name={Supplementary Fig.},fileext=lof]{suppfigure}
\DeclareFloatingEnvironment[name={Supplementary Table}]{supptable}
\usepackage[labelfont=bf]{caption}
\usepackage{tabularx}
\usepackage{lipsum}
\title{\LARGE On genetic correlation estimation with summary statistics from genome-wide association studies} 
\author{ Bingxin Zhao and Hongtu Zhu\\~\\
University of North Carolina at Chapel Hill}
\begin{document}
\maketitle
\date{}
\abstract{
Genome-wide association studies (GWAS) have been widely used to examine the association between single nucleotide polymorphisms (SNPs) and complex traits, where both the sample size $n$ and the number of SNPs $p$ can be very large. 
Recently, cross-trait polygenic risk score (PRS) method has gained extremely popular for assessing genetic correlation of complex traits based on GWAS summary statistics (e.g., SNP effect size).  
However, empirical evidence has shown a common bias phenomenon that even highly significant cross-trait PRS can only account for a very small amount of genetic variance ($R^2$ often $<1\%$). 
The aim of this paper is to develop a novel and powerful method to address the bias phenomenon of cross-trait PRS.    
We theoretically show that the estimated genetic correlation is asymptotically biased towards zero when complex traits are highly polygenic/omnigenic. 
When all $p$ SNPs are used to construct PRS, we show that the asymptotic bias of PRS estimator is independent of the unknown number of causal SNPs $m$.  
We propose a consistent PRS estimator to correct such asymptotic bias.  
We also develop a novel estimator of genetic correlation which is solely based on two sets of GWAS summary statistics. 
In addition, we investigate whether or not SNP screening by GWAS $p$-values can lead to improved estimation and show the effect of overlapping samples among GWAS.  
Our results may help demystify and tackle the puzzling ``missing genetic overlap'' phenomenon of cross-trait PRS for dissecting the genetic similarity of closely related heritable traits.
We illustrate the finite sample performance of our bias-corrected PRS estimator by using both numerical experiments   and  the UK Biobank data, in which we assess the genetic correlation between brain white matter tracts and neuropsychiatric disorders. 
\bigskip

}

\section{Introduction}\label{sec1}
The major aim of many
genome-wide association studies (GWAS) \citep{visscher201710} is to examine the  genetic influences on complex human traits given that most traits have a polygenic architecture \citep{fisher1919xv,gottesman1967polygenic,penrose1953genetical,orr1992genetics,wray2018common,hill2010understanding}. 
That is, a large number of single nucleotide polymorphisms (SNPs) have small but nonzero contributions to the phenotypic variation.
Many statistical methods have been developed on the use of individual-level GWAS SNP data to infer the heritability and cross-trait genetic correlation in general populations \citep{lee2012estimation,loh2015contrasting,chen2014estimating,golan2014measuring,guo2017optimal,lee2016mtg2,jiang2016high,yang2011gcta,yang2010common}. 
For instance, heritability $h^2$ can be estimated by aggregating the small contributions of a large number of SNP markers, resulting in the SNP heritability estimator \citep{yang2017concepts}. 
For two highly polygenic traits, cross-trait genetic correlation can be calculated as the correlation of the genetic effects of numerous SNPs on the two traits \citep{shi2017local,lu2017powerful,guo2017optimal,pasaniuc2017dissecting}.  
A growing number of empirical evidence  \citep{shi2016contrasting,chatterjee2016developing,ge2017phenome,dudbridge2016polygenic,yang2010common} supports the polygenicity of many complex human traits and verify that the common (minor allele frequency [MAF] $\ge0.05$) SNP can account for a large amount of heritability of many complex traits. 
The term ``omnigenic'' has been introduced to acknowledge the widespread causal genetic variants contributing to various complex human traits \citep{boyle2017expanded}.

Accessing individual-level SNP data is often inconvenient due to policy restrictions, and 
a recent standard practice in the genetic community is to share the summary association statistics, including the estimated effect size, standard error, $p$-value, and sample size $n$, of all genotyped SNPs after GWAS are published \citep{macarthur2016new,zheng2017ld}. 
Therefore, 
joint analysis of summary-level data of different GWAS provides new opportunities for further analyses and novel genetic discoveries, such as the shared genetic basis of complex traits.
It has became an active research area to examine the heritability and cross-trait genetic correlation based on GWAS summary statistics
\citep{bulik2015atlas,bulik2015ld,zhou2017unified,palla2015fast,weissbrod2018estimating,lu2017powerful,shi2017local,dudbridge2013power,lee2013genetic}.  
Among them, the cross-trait polygenic risk score (PRS) \citep{purcell2009common,power2015polygenic} has became a popular routine to measure genetic similarity of polygenic traits with widespread applications \citep{hagenaars2016shared,pouget2018cross,nivard2017genetic,clarke2016common,mistry2018use,socrates2017polygenic,bogdan2018polygenic}. 
Compared with other popular methods such as cross-trait LD score regression \citep{bulik2015atlas}, Bivariate GCTA
 \citep{lee2012estimation}, and BOLT-REML \citep{loh2015contrasting},  
cross-trait PRS offers at least two unique strengths as follows.   
First, cross-trait PRS only requires the GWAS summary statistics of one trait obtained from a large discovery GWAS, while it allows those of the other trait obtained from a much smaller GWAS dataset. 
In contrast, most other methods require large GWAS data for both traits on either summary or individual-level. 
Second, cross-trait PRS can provide genetic propensity for each sample in the testing dataset, enabling further prediction and treatment. 
However, given these strengths of cross-trait PRS, empirical evidence  has shown a common bias phenomenon that even highly significant cross-trait PRS can only account for a very small amount of variance ($R^2$ often $<1\%$) when dissecting the shared genetic basis among highly related heritable traits \citep{clarke2016common,mistry2018use,socrates2017polygenic,bogdan2018polygenic}.  
Except for some introductory studies \citep{daetwyler2008accuracy,dudbridge2013power,visscher2014statistical}, few attempts have ever been made to rigorously study cross-trait PRS and to explain such a counterintuitive phenomenon. 

This paper fills this significant gap with the following contributions. 
By comprehensively investigating the properties of cross-trait PRS for polygenic/omnigenic traits, 
our first contribution in Section~\ref{sec2} is to show that the estimated genetic correlation is asymptotically biased towards zero, uncovering that the underlying genetic overlap is seriously underestimated.
Furthermore, when all $p$ SNPs are used in cross-trait PRS, we show that the asymptotic bias is largely determined by the triple $(n,p,h^2)$ and is independent of the unknown number of causal SNPs of the two traits. 
Thus, our second contribution in Section~\ref{sec2} is to propose a consistent estimator by correcting such asymptotic bias in cross-trait PRS.
We also develop a novel estimator of genetic correlation which only requires two sets of summary statistics.

Next, in Section~\ref{sec3}, we show that 
when cross-trait PRS is constructed using $q$ top-ranked SNPs whose GWAS $p$-values pass a given threshold, in addition to $(n,p,h^2)$, the asymptotic bias will also be determined by the number of causal SNPs $m$, since the sparsity $m/p$ determines the quality of the $q$ selected SNPs. 
Particularly, for highly polygenic/omnigenic traits with dense SNP signals, such screening may fail, resulting in larger bias in genetic correlation estimation. 
In Section~\ref{sec4}, we generalize our results to quantify the influence of overlapping samples among GWAS. 
We show that our bias-corrected estimator for independent GWAS can be smoothly extended to GWAS with partially or even fully overlapping samples.

The remainder of this paper is structured as follows. Sections~\ref{sec2} and~\ref{sec3} study the cross-trait PRS with all SNPs and selected SNPs, respectively. 
Section~\ref{sec4} considers the effect of overlapping samples among different GWAS. 
Sections~\ref{sec5} and~\ref{sec6} summarize the numerical results on numerical experiments and real data analysis. The paper concludes with some discussions in Section~\ref{sec7}.

\section{Cross-trait PRS with all SNPs}\label{sec2}
Since cross-trait PRS is designed for polygenic traits based on their GWAS summary statistics, we first introduce the polygenic model and some properties of GWAS summary statistics. 
We note that the standard approach in GWAS is marginal screening. That is, the marginal association between the phenotype and single SNP is assessed each at a time, while adjusting for the same set of covariates including population stratification \citep{price2006principal}.  
Marginal screening procedures often work well to prioritize important variables given that the signals are sparse \citep{fan2008sure}, but they may have noisy outcomes when signals are dense \citep{fan2012variance}, which is often the case for GWAS of highly polygenic traits.

\subsection{Polygenic trait and GWAS summary statistics}\label{sec2.0}
Let $\X_{(1)}$ be an $n \times m$ matrix of the SNP data with nonzero effects, and 
$\X_{(2)}$ be an $n \times (p-m)$ matrix of the null SNPs, resulting in an $n\times p$ matrix of all SNPs, donated by $\X=[\X_{(1)},\X_{(2)}]=(\x_{1},\cdots,\x_{m},\x_{m+1}, \cdots,\x_{p})$, where $\x_{i}$ is an $n\times 1$ vector of the SNP $i$, $i=1,\cdots,p$.   
Columns of $\X$ are assumed to be independent after linkage disequilibrium (LD)-based pruning. 
Further, we assume column-wise normalization on $\X$ is performed such that each variable has sample mean zero and sample variance one. Therefore, we may introduce the following condition on SNP data:
\begin{condition}
\label{con1}
Entries of $\X=[\X_{(1)},\X_{(2)}]$ are real-value independent random variables with mean zero, variance one and a finite eighth order moment. 
\end{condition}
Let $\y$ be an $n \times 1$ vector of continuous polygenic phenotype. 
We assume a linear polygenic structure between $\y$ and $\X$ as follows:
\begin{flalign}
\y=\sum_{i=1}^{p}\x_i\beta_i+\bmeps=\sum_{i=1}^{m}\x_i\beta_i+\bmeps = \X_{(1)}\bmbeta_{(1)}+\bmeps,  
\label{equ1.1}
\end{flalign}
where $\bmbeta=(\beta_1,\cdots, \beta_{m}, \beta_{m+1},\cdots, \beta_p)^{T}=\big(\bmbeta_{(1)}^T,\bmbeta_{(2)}^T\big)$ is a vector of genetic effects such that  $\beta_i$ in $\bmbeta_{(1)}^T=(\beta_1,\cdots, \beta_{m})^T$ are random variables ($i=1,\cdots, m$),  
$\bmbeta_{(2)}^T=(\beta_{m+1},\cdots, \beta_p)^T$ are zeros, 
and $\bmeps$ represents the vector of independent non-genetic random errors. 
For simplicity, we assume that there are no other fixed effects in model~(\ref{equ1.1}), or equivalently, other covariates can be well observed and adjusted for. 

We allow flexible ratios among $(n,p,m)$. 
As $\mbox{min}(n,p) \to \infty$, we assume
\begin{flalign*}
\frac{m}{n}=\gamma \to \gamma_0 \quad \text{and} \quad \frac{m}{p}=\omega \to \omega_0 \quad \text{for} \quad 0 < \gamma_0 \le \infty \quad \text{and} \quad 0 \le  \omega_0 \le 1 ,
\end{flalign*}
which should satisfy most large-scale GWAS of polygenic traits. 
\begin{figure}
\includegraphics[page=1,width=0.75\linewidth]{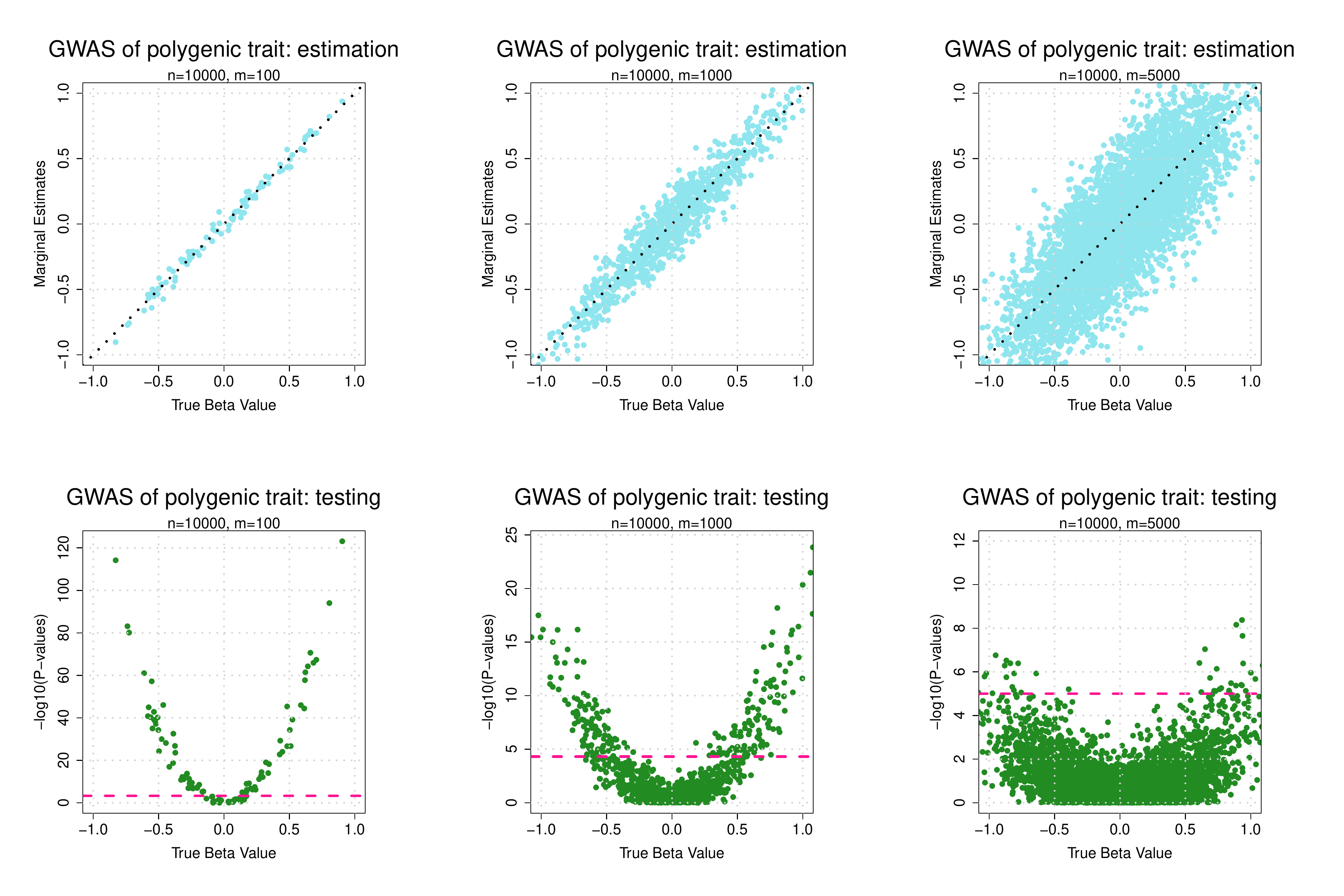}
  \caption{Estimation (upper panels) and testing (bottom panels) of marginal genetic effects in GWAS of polygenic traits. We set $n=10,000$ and $m=p=100$, $1000$ and $5000$.}
     \label{fig1}
\end{figure}
Most GWAS use ordinary least squares (OLS) to perform linear regression given by
\begin{flalign}
\y=\1\mu+\x_i\beta_i+\bmeps_i^*
\label{equ1.3}
\end{flalign}
for $i=1,\cdots,p$, where $\1$ is an $n\times 1$ vector of ones. Let $\widehat{\mu}$ and $\widehat{\beta}_i$ be the OLS estimates of $\mu$ and $\beta_i$, respectively, for $ i=1,\cdots,p$.
When $\y$ and $\x_i$ are normalized and both $n$ and $m\to \infty$, under {Condition~\ref{con1}} and model~(\ref{equ1.1}), 
it can be shown that 
\begin{flalign*}
\tE\big(\widehat{\mu}\big) =0, \qquad 
\tE\big(\widehat{\beta}_i\big) =\beta_i,
\qquad
\text{ and } 
\end{flalign*}
\begin{flalign}
\var\big(\widehat{\beta_i}\big) = \left\{ 
\begin{array}{ll}
n^{-1} \cdot \sum_{j \neq i}^{m} \beta_j^2=\bigO(m/n), & \mbox{for \quad$i \in [1,m]$}\mathbf{;}\\ 
n^{-1} \cdot \sum_{j =1}^{m} \beta_j^2= \bigO(m/n), & \mbox{for \quad $i \in [m+1,p]$} \mathbf{.} \\ 
\end{array} \right.
\label{equ1.5}
\end{flalign}
Equation~(\ref{equ1.5}) indicates that the variance or mean squared error (MSE) of $\widehat{\beta_i}$ calculated from model~(\ref{equ1.3}) moves up linearly as $m \to \infty$. 
Therefore, the $T$ scores for testing 
\begin{flalign*}
&H_{0i}: \beta_i=0 \qquad  \text{versus} \qquad H_{1i}:  \beta_i\neq0, \quad \text{for} \quad i=1,\cdots,p
\end{flalign*}
are given by
\begin{flalign*}
&T_i = \left\{ 
\begin{array}{ll}
\widehat{\beta_i}/(\sum_{j \neq i}^{m} \beta_j^2/n)^{1/2}
=\widehat{\beta_i} \cdot   \bigO(\sqrt{n/m}), & \mbox{for \quad$i \in [1,m]$}\mathbf{;}\\ 
\widehat{\beta_i}/(\sum_{j=1}^{m} \beta_j^2/n)^{1/2}
=\widehat{\beta_i} \cdot \bigO(\sqrt{n/m}), & \mbox{for \quad $i \in [m+1,p]$} \\
\end{array} \right.
\end{flalign*}
under $H_{0i}$, $i=1,\cdots,p$.  

\begin{remark}\label{rmk1}
The above simple derivations reveal important insights into the {challenge} of performing marginal screening for polygenic traits. 
For estimation, although $\widehat{\beta}_i$s are all unbiased given that $\X$ are independent, $\var(\widehat{\beta_i})$ is $\bigO(m/n)$ instead of $\bigO(1/n)$. 
Therefore,  
when $m/n$ is large, the variance (and MSE) of $\widehat{\beta}_i$s can be so overwhelming that $\widehat{\beta}_i$s might be dominated by their standard errors.
Note that all $\var(\widehat{\beta_i})$s are in the same scale regardless of whether their original $\beta_i$s are zeros or not. 
Thus, the $\widehat{\beta_i}$s from causal and null variants can be totally mixed up when $m/n$ is large. 
In addition, the test statistics $T_i$s may not well preserve the ranking of variables in $X$ when $m/n$ is large, resulting in potential low power and high false positive rate in detecting and prioritizing important SNPs. 
\end{remark}
Figure~\ref{fig1} demonstrates the estimation and testing of marginal genetic effects in GWAS with $n=10,000$ as $p=m$ increases from $100$, $1000$ to $5000$.  
Each entry of $\X$ is i.i.d generated from $N(0,1)$, 
elements of $\bmbeta_{(1)}$ are i.i.d generated from $N(0,0.4)$, and entries of $\bmeps$ are i.i.d from $N(0,1)$. Then, $\y$ is generated from model~(\ref{equ1.1}). 
The estimated genetic effects are unbiased in general, however, the uncertainty clearly moves up as $m$ increases. 
The relative contribution of each SNP decreases as $m$ increases, and thus the testing power drops as well. More simulations on GWAS summary statistics can be found in Section~\ref{sec5}.

As illustrated in later sections, these properties of GWAS summary statistics are closely related to the asymptotic bias of cross-trait PRS and the performance of SNP screening. 
Specifically, i) when cross-trait PRS is constructed with all $p$ SNPs, the $p$ $\var\big(\widehat{\beta_i}\big)$s are aggregated, resulting in inflated genetic variance and underestimated genetic correlation; and ii)
when cross-trait PRS is constructed with top-ranked SNPs that pass a pre-specified $p$-value threshold, it may have worse performance if GWAS marginal screening fails to prioritize the causal SNPs.

\subsection{General setup}\label{sec2.1}
In this subsection, we introduce the modelling framework to investigate the cross-trait PRS, including the genetic architecture of polygenic traits, distribution of genetic effects, and genetic correlation estimators.
\subsubsection{Polygenic traits}
Consider three independent GWAS that are conducted for three different traits as follows:
\begin{itemize}
\item Discovery GWAS-I: $(\X,\y_{\alpha})$, with $\X=[\X_{(1)},\X_{(2)}] \in \bbR^{n_{1}\times p}$, $\X_{(1)} \in \bbR^{n_{1} \times m_{\alpha}}$, and $\y_{\alpha} \in \bbR^{n_{1} \times 1}$. 
\item Discovery GWAS-II: $(\Z,\y_{\beta})$, with $\Z=[\Z_{(1)},\Z_{(2)}] \in \bbR^{n_{2}\times p}$, $\Z_{(1)} \in \bbR^{n_{2} \times m_{\beta}}$, and $\y_{\beta} \in \bbR^{n_{2} \times 1}$. 
\item Target testing GWAS: $(\W,\y_{\eta})$, with $\W=[\W_{(1)},\W_{(2)}] \in \bbR^{n_{3}\times p}$, $\W_{(1)} \in \bbR^{n_{3} \times m_{\eta}}$,  and $\y_{\eta} \in \bbR^{n_{3} \times 1}$. 
\end{itemize}
Here $\y_{\alpha}$, $\y_{\beta}$, and $\y_{\eta}$ are three different continuous phenotypes studied in three GWAS with sample sizes $n_1$, $n_2$, and $n_3$, respectively. Thus, $m_{\alpha}$, $m_{\beta}$, and $m_{\eta}$ are different numbers of causal SNPs in general.  
The $\X_{(1)}$, $\Z_{(1)}$, and $\W_{(1)}$ denote the causal SNPs of $\y_{\alpha}$, $\y_{\beta}$, and $\y_{\eta}$, respectively, 
and $\X_{(2)}$, $\Z_{(2)}$, and $\W_{(2)}$ donate the corresponding null SNPs. Thus, $\X$, $\Z$, and $\W$ are three matrices of $p$ SNPs.   
It is assumed that $\X$, $\Z$, and $\W$ have been normalized and  satisfy Condition~\ref{con1}. 
Similar to model~(\ref{equ1.1}), the linear polygenic model assumes
\begin{flalign}
\y_{\alpha}= \X\bmalpha+\bmeps_{\alpha},  \quad
\y_{\beta}= \Z\bmbeta+\bmeps_{\beta},   \quad \text{and}  \quad
\y_{\eta}= \W\bmeta+\bmeps_{\eta},  
\label{equ2.1}
\end{flalign}
where $\bmalpha$, $\bmbeta$, and $\bmeta$ are $p\times 1$ vectors of SNP effects, and $\bmeps_{\alpha}$, $\bmeps_{\beta}$, and $\bmeps_{\eta}$ represent independent random error vectors. 
The overall genetic heritability of $\y_{\alpha}$ is, therefore, given by  
\begin{flalign*}
h^2_{\alpha}=\frac{\var(\X\bmalpha)}{\var(\y_{\alpha})}=\frac{\var(\X_{(1)}\bmalpha_{(1)})}{\var(\X_{(1)}\bmalpha_{(1)})+\var(\bmeps_{\bmalpha})},
\end{flalign*}
which measures the proportion of variation in $\y_{\alpha}$ that can be explained by the genetic variation $\X\bmalpha$. 
The $\y_{\alpha}$ is fully heritable when $h^2_{\alpha}=1$.
Similarly, we can define the heritability $h^2_{\beta}$ of $\y_{\beta}$ and $h^2_{\eta}$ of $\y_{\eta}$, respectively.
We assume $h^2_{\alpha}$, $h^2_{\beta}$, and $h^2_{\eta} \in (0, 1]$. 
The genetic correlation in this paper is defined as the correlation of SNP effects on pairs of phenotypes \citep{lu2017powerful,pasaniuc2017dissecting,shi2017local,guo2017optimal}.
\begin{definition}[Genetic Correlation]\label{def1}
The genetic correlation between $\y_{\alpha}$ and $\y_{\eta}$ and that between $\y_{\alpha}$ and $\y_{\beta}$ are respectively given by
\begin{flalign*}
\varphi_{\alpha\eta}=\frac{\bmalpha^T\bmeta}{\Vert\bmalpha\Vert\cdot\Vert\bmeta\Vert}\cdot \bmI(\Vert\bmalpha\Vert\cdot\Vert\bmeta\Vert>0) \quad \text{and} \quad
\varphi_{\alpha\beta}=\frac{\bmalpha^T\bmbeta}{\Vert\bmalpha\Vert\cdot\Vert\bmbeta\Vert} \cdot \bmI(\Vert\bmalpha\Vert\cdot\Vert\bmbeta\Vert>0),
\end{flalign*}
where $\bmI(\cdot )$ is the indicator function, 
$\Vert\cdot\Vert$ is the $l_2$ norm of a vector, 
and  $\varphi_{\alpha\eta}$ and $\varphi_{\alpha\beta} \in [-1,1]$.
\end{definition}
\subsubsection{Genetic effects}
Since $m_{\alpha}$, $m_{\beta}$ and $m_{\eta}$ can be different and the causal SNPs of different phenotypes may partially overlap, we  
let $m_{\alpha\eta}$ be the number of overlapping causal SNPs of $\y_{\alpha}$ and $\y_{\eta}$, and $m_{\alpha\beta}$ be the number of overlapping causal SNPs of $\y_{\alpha}$ and $\y_{\beta}$. 
Let $F(0,V)$ represent a generic distribution with mean zero, (co)variance $V$,  and finite fourth order moments. Without loss of generality,
we introduce the following condition on genetic effects and random errors. 
\begin{condition}\label{con2}
$\alpha_i$, $\beta_j$, and $\eta_k$ are independent random variables satisfying
\begin{gather*}
\alpha_i \sim F(0,\sigma^2_{\alpha}), \quad i=1,...,m_{\alpha}; \qquad
\beta_j \sim F(0,\sigma^2_{\beta}), \quad j=1,...,m_{\beta};\\ 
 \eta_k \sim F(0,\sigma^2_{\eta}), \quad 
k=1,...,m_{\eta} \mathbf{.}
\end{gather*}
The $m_{\alpha\eta}$ overlapping nonzero effects $(\alpha_i,\eta_i)$s of ($\y_{\alpha}$,$\y_{\eta}$) 
and $m_{\alpha\beta}$ overlapping nonzero effects $(\alpha_j,\beta_j)$s of  
($\y_{\alpha}$,$\y_{\beta}$) 
satisfy
\begin{flalign*}
\begin{pmatrix} 
\alpha_i\\
\eta_i 
\end{pmatrix}
\sim F 
\left \lbrack
\begin{pmatrix} 
0\\
0 
\end{pmatrix},
\begin{pmatrix} 
\sigma^2_{\alpha} & \sigma_{\alpha\eta} \\
\sigma_{\alpha\eta} & \sigma^2_{\eta}
\end{pmatrix}
\right \rbrack
\quad \text{and} \quad
\begin{pmatrix} 
\alpha_j\\
\beta_j 
\end{pmatrix}
\sim F 
\left \lbrack
\begin{pmatrix} 
0\\
0 
\end{pmatrix},
\begin{pmatrix} 
\sigma^2_{\alpha} & \sigma_{\alpha\beta} \\
\sigma_{\alpha\beta} & \sigma^2_{\beta}
\end{pmatrix}
\right \rbrack \mathbf{,}
\end{flalign*} 
respectively.
And $\epsilon_{\alpha_i}$, $\epsilon_{\beta_j}$ and $\epsilon_{\eta_k}$ are independent random variables satisfying
\begin{gather*}
\epsilon_{\alpha i} \sim F(0,\sigma^2_{\epsilon_{\alpha}}), \quad i=1,...,n_{1} ; \qquad
\epsilon_{\beta j} \sim F(0,\sigma^2_{\epsilon_{\beta}}), \quad j=1,...,n_{2}; \\
\epsilon_{\eta k} \sim F(0,\sigma^2_{\epsilon_{\eta}}), \quad k=1,...,n_{3}; 
\end{gather*}
where 
$\sigma_{\alpha\eta}=\rho_{\alpha\eta} \cdot \sigma_{\alpha}\sigma_{\eta}$  and
$\sigma_{\alpha\beta}=\rho_{\alpha\beta}\cdot \sigma_{\alpha}\sigma_{\beta}$.
\end{condition}
Since the three GWAS have independent samples, we assume that their random errors are independent. Overlapping samples and the induced non-genetic correlation  {will be}  studied in Section~\ref{sec4}. 
Under {Condition~\ref{con2}}, when $n_1,n_3$, and $p \to \infty$, if $m_{\alpha\eta},m_{\alpha},$ and $m_{\eta} \to \infty$, and {$m_{\alpha\eta}/\sqrt{m_{\alpha}m_{\eta}}=\kappa_{\alpha\eta} \to \kappa_{0\alpha\eta} \in (0,1]$, }
then the genetic correlation between $\y_{\alpha}$ and $\y_{\eta}$ is asymptotically  given by 
\begin{flalign*}
\varphi_{\alpha\eta}=\frac{\bmalpha^T\bmeta}{\Vert\bmalpha\Vert\cdot\Vert\bmeta\Vert}
&=\frac{\sum_{i=1}^{m_{\alpha\eta}} \alpha_i\eta_i}{(\sum_{i=1}^{m_{\alpha}} \alpha_i^2)^{1/2}(\sum_{i=1}^{m_{\eta}}\eta_i^2)^{1/2}}\\
&=\frac{m_{\alpha\eta}}{~(m_{\alpha}m_{\eta})^{1/2}}\cdot\rho_{\alpha\eta}\cdot\{1+o(1)\}
=\kappa_{0\alpha\eta}\cdot\rho_{\alpha\eta}\cdot\{1+o(1)\}
\mathbf{.}
\end{flalign*}
Similarly,  when $n_1,n_2,n_3$, and $p \to \infty$, if $m_{\alpha\beta},m_{\alpha}$, and $m_{\beta} \to \infty$ and $m_{\alpha\beta}/\sqrt{m_{\alpha}m_{\beta}}=\kappa_{\alpha\beta}\to \kappa_{0\alpha\beta}\in (0,1]$,
then the genetic correlation between $\y_{\alpha}$ and $\y_{\beta}$ is asymptotically given by 
\begin{flalign*}
\varphi_{\alpha\beta}=\frac{\bmalpha^T\bmbeta}{\Vert\bmalpha\Vert\cdot\Vert\bmbeta\Vert}
=\frac{m_{\alpha\beta}}{~(m_{\alpha}m_{\beta})^{1/2}}\cdot \rho_{\alpha\beta}\cdot\{1+o(1)\}
=\kappa_{0\alpha\beta}\cdot \rho_{\alpha\beta}\cdot\{1+o(1)\} \mathbf{.}
\end{flalign*}
As in \cite{jiang2016high}, heritability $h^2_{\alpha}$, $h^2_{\beta}$, and $h^2_{\eta}$ can be asymptotically represented {as follows:}
\begin{flalign*}
h^2_{\alpha}=\frac{m_{\alpha}\sigma^2_{\alpha}}{m_{\alpha}\sigma^2_{\alpha}+\sigma^2_{\epsilon_{\alpha}}}, \quad
h^2_{\beta}=\frac{m_{\beta}\sigma^2_{\beta}}{m_{\beta}\sigma^2_{\beta}+\sigma^2_{\epsilon_{\beta}}}, 
\quad \text{and}  \quad
h^2_{\eta}=\frac{m_{\eta}\sigma^2_{\eta}}{m_{\eta}\sigma^2_{\eta}+\sigma^2_{\epsilon_{\eta}}} \mathbf{.}
\end{flalign*}
\subsubsection{Genetic correlation estimators}
Now we introduce the cross-trait PRS and genetic correlation estimators.
We need the following data.
As $n_1$, $n_2$, and $p \to \infty$, the summary association statistics for $\y_{\alpha}$ and $\y_{\beta}$ from Discovery GWAS-I \& II are given by
\begin{flalign*}
\widehat{\bmalpha}=\frac{1}{n_1}\X^T\big(\X_{(1)}\bmalpha_{(1)}+\bmeps_{\alpha}\big) \quad \text{and} \quad
\widehat{\bmbeta}=\frac{1}{n_2}\Z^T\big(\Z_{(1)}\bmbeta_{(1)}+\bmeps_{\beta}\big).
\end{flalign*}
We assume that the individual-level SNP $\W$ and phenotype $\y_{\eta}$ in the Target testing GWAS can be accessed. 
In addition, $h^2_{\alpha}$, $h^2_{\beta}$, and $h^2_{\eta}$ are assumed to be estimable, using either their corresponding individual-level data \citep{yang2011gcta,loh2015contrasting} or summary-level data \citep{bulik2015ld,weissbrod2018estimating,palla2015fast}, or can be found in the literature \citep{polderman2015meta}.
In summary, besides $(n_1,n_2,n_3,p)$, it is assumed that 
$\widehat{\bmalpha}$, $\widehat{\bmbeta}$, $\W$, $\y_{\eta}$, $\widehat{h}^2_{\alpha}$, $\widehat{h}^2_{\beta}$, and $\widehat{h}^2_{\eta}$
are available. 

We construct cross-trait PRSs as follows:
\begin{flalign*}
&\widehat{\bmS}_{\alpha}=\sum_{i=1}^{p}\w_i\widehat{a}_i=\W\widehat{\bma}
=\W_{(1,\alpha)}\widehat{\bma}_{(1)}+\W_{(2,\alpha)}\widehat{\bma}_{(2)}
\quad
\text{for $\y_{\alpha}$ and}\\
&\widehat{\bmS}_{\bmbeta}=\sum_{i=1}^{p}\w_i\widehat{b}_i=\W\widehat{\bmb}
=\W_{(1,\beta)}\widehat{\bmb}_{(1)}+\W_{(2,\beta)}\widehat{\bmb}_{(2)}
\quad
\text{for $\y_{\beta}$,}
\end{flalign*}
where $\widehat{\bma}=(\widehat{a}_1,\cdots,\widehat{a}_{m_{\alpha}},\widehat{a}_{m_{\alpha}+1},\cdots,\widehat{a}_p)^T=\big(\widehat{\bma}_{(1)}^{~T},\widehat{\bma}_{(2)}^{~T}\big)$, in which $\widehat{a}_i=\widehat{\alpha}_i\cdot \bmI(|\widehat{\alpha}_i|>c_{\alpha})$, 
$\widehat{\bmb}=(\widehat{b}_1,\cdots,\widehat{b}_{m_{\beta}},
\widehat{b}_{m_{\beta}+1},\cdots,\widehat{b}_p)^T
=\big(\widehat{\bmb}_{(1)}^{~T},\widehat{\bmb}_{(2)}^{~T}\big)$, 
in which $\widehat{b}_i=\widehat{\beta}_i\cdot \bmI(|\widehat{\beta}_i|>c_{\beta})$, and $c_{\alpha}$ and $c_{\beta}$ are given thresholds  used for SNP screening in order to calculate $\widehat{\bmS}_{\alpha}$ and $\widehat{\bmS}_{\beta}$. Moreover, we define
$\W_{(1,\alpha)}$ $=[\w_1,\cdots,\w_{m_{\alpha}}]$, 
$\W_{(2,\alpha)}$ $=[\w_{m_{\alpha}+1},\cdots,\w_{p}]$, 
$\W_{(1,\beta)}$ $=[\w_1,\cdots,\w_{m_{\beta}}]$, 
$\W_{(2,\beta)}$ $=[\w_{m_{\beta}+1},\cdots,\w_{p}]$, and  $\W$ $=[\W_{(1,\alpha)},\W_{(2,\alpha)}]$ $=[\W_{(1,\beta)},\W_{(2,\beta)}]$.

We estimate the genetic correlation between $\y_{\alpha}$ and $\y_{\eta}$ with $\big(\widehat{\bmS}_{\alpha}$,$\y_{\eta}\big)$ and that between $\y_{\alpha}$ and $\y_{\beta}$ with $\big(\widehat{\bmS}_{\alpha}$,$\widehat{\bmS}_{\beta}\big)$. 
They represent two common cases in real data applications.
For $\big(\widehat{\bmS}_{\alpha}$,$\y_{\eta}\big)$, individual-level data are available for one trait, but not for another one. It often occurs when the traits are studied in two different GWAS.  
For $\big(\widehat{\bmS}_{\alpha}$,$\widehat{\bmS}_{\beta}\big)$,
neither of the two traits has individual-level data.
This happens when we have GWAS summary statistics of two traits and estimate their genetic correction on an independent target dataset. 
The genetic correlation estimators are given by 
\begin{flalign*}
G_{\alpha\eta}=\frac{\y_{\eta}^T\widehat{\bmS}_{\alpha}}{\big\Vert\y_{\eta}\big\Vert\cdot\big\Vert\widehat{\bmS}_{\alpha}\big\Vert}
=
\frac{\big( \W_{(1)}\bmeta_{(1)}+\bmeps_{\eta}\big)^T\big(\W_{(1,\alpha)}\widehat{\bma}_{(1)} +\W_{(2,\alpha)}\widehat{\bma}_{(2)}\big)}
{\big\Vert\W_{(1)}\bmeta_{(1)}+\bmeps_{\eta}\big\Vert\cdot\big\Vert\W_{(1,\alpha)}\widehat{\bma}_{(1)} +\W_{(2,\alpha)}\widehat{\bma}_{(2)}\big\Vert} 
\end{flalign*}
for $\varphi_{\alpha\eta}$, and 
\begin{flalign*}
G_{\alpha\beta}=\frac{\widehat{\bmS}_{\beta}^T\widehat{\bmS}_{\alpha}}{\big\Vert\widehat{\bmS}_{\beta}\big\Vert \cdot \big\Vert\widehat{\bmS}_{\alpha}\big\Vert}
= \frac{\big(\W_{(1,\beta)}\widehat{\bmb}_{(1)}+\W_{(2,\beta)}\widehat{\bmb}_{(2)}\big)^T\big(\W_{(1,\alpha)}\widehat{\bma}_{(1)} +\W_{(2,\alpha)}\widehat{\bma}_{(2)}\big)}{ \big\Vert\W_{(1,\beta)}\widehat{\bmb}_{(1)}+\W_{(2,\beta)}\widehat{\bmb}_{(2)}\big\Vert\cdot
\big\Vert\W_{(1,\alpha)}\widehat{\bma}_{(1)} +\W_{(2,\alpha)}\widehat{\bma}_{(2)}\big\Vert} 
\end{flalign*}
for $\varphi_{\alpha\beta}$. 

\subsection{Asymptotic bias and correction}\label{sec2.2}
We first investigate $G_{\alpha\beta}$ and $G_{\alpha\eta}$ when all of the $p$ candidate SNPs are used, or when $c_{\alpha}=c_{\beta}=0$.  
Thus, $\widehat{\bma}_{(1)}=\widehat{\bmalpha}_{(1)}$,
$\widehat{\bma}_{(2)}=\widehat{\bmalpha}_{(2)}$, $\widehat{\bmb}_{(1)}=\widehat{\bmbeta}_{(1)}$, 
and $\widehat{\bmb}_{(2)}=\widehat{\bmbeta}_{(2)}$. 
Then, we have  
\begin{flalign*}
G_{\alpha\eta}
=\frac{\big(\W_{(1)}\bmeta_{(1)}+\bmeps_{\eta}\big)^T\W\X^T\big(\X_{(1)}\bmalpha_{(1)}+\bmeps_{\alpha}\big)}
{\big\Vert\W_{(1)}\bmeta_{(1)}+\bmeps_{\eta}\big\Vert \cdot
\big\Vert\big(\X_{(1)}\bmalpha_{(1)}+\bmeps_{\alpha}\big)^T\X\W^T \big\Vert}
\end{flalign*}
and 
\begin{flalign*}
&G_{\alpha\beta}
=\frac{\big(\Z_{(1)}\bmbeta_{(1)}+\bmeps_{\beta}\big)^T\Z\W^T\W\X^T\big(\X_{(1)}\bmalpha_{(1)}+\bmeps_{\alpha}\big)}{
\big\Vert\big(\Z_{(1)}\bmbeta_{(1)}+\bmeps_{\bmbeta}\big)^T\Z\W^T \big\Vert \cdot
\big\Vert\big(\X_{(1)}\bmalpha_{(1)}+\bmeps_{\alpha}\big)^T\X\W^T\big\Vert} \mathbf{.}
\end{flalign*}
We have the following results on the asymptotic properties of $G_{\alpha\eta}$, whose proof can be found in the Appendix~A.
\begin{thm}\label{thm1}
Under polygenic model~(\ref{equ2.1}) and Conditions~\ref{con1} and~\ref{con2}, suppose $m_{\alpha\eta},m_{\alpha}$, and
$m_{\eta}$ $\rightarrow \infty$ as $\mbox{min}(n_1, n_3,p)\rightarrow\infty$, and let $p=c \cdot (n_1n_3)^{a}$ for some constants $c>0$ and $ a  \in (0,\infty]$. 
If $a \in (0,1)$, then we have 
\begin{flalign}
G_{\alpha\eta}=\varphi_{\alpha\eta}+\bigg( \sqrt{\frac{n_1}{n_1+p/h^2_{\alpha}}}\cdot h_{\eta}-1\bigg) \cdot  \varphi_{\alpha\eta} \cdot\{1+o_p(1)\}
\mathbf{.}
\end{flalign}
If $ a  \in [1,\infty]$, then we have 
\begin{flalign}
G_{\alpha\eta} \cdot n_3=\bigO_p(1)\mathbf{.}
\end{flalign}
\end{thm}

\begin{remark}\label{rmk3}
For $a \in (0,1)$, $G_{\alpha\eta}$ is  a biased estimator of $\varphi_{\alpha\eta}$ since 
$\sqrt{n_1/(n_1+p/h^2_{\alpha})}\cdot h_{\eta}$
is smaller than $1$.
Interestingly, the asymptotic bias is independent of the unknown numbers $m_{\alpha}, m_{\eta}$, and $m_{\alpha\eta}$, and is only determined by $n_1$, $p$, $h^2_{\alpha}$ and $h^2_{\eta}$. 
When $n_1$ and $p$ are comparable, a consistent estimator of $\varphi_{\alpha\eta}$ is given by 
\begin{flalign*}
G_{\alpha\eta}^A=G_{\alpha\eta}\cdot \sqrt{\frac{n_1+p/h^2_{\alpha}}{n_1\cdot h_{\eta}^2}}=\varphi_{\alpha\eta} \cdot\{1+o_p(1)\} \mathbf{.}
\end{flalign*}
In addition, the testing sample size $n_3$ vanishes in $G_{\alpha\eta}$ for $a \in (0,1)$, which verifies that given the sample size $n_1$ of discovery GWAS is large, we can apply the summary statistics onto a much smaller set of target samples.

If $ a  \in [1,\infty]$, i.e., $p/(n_1n_3)$ is too large, then $G_{\alpha\eta}$ will have a zero asymptotic limit. 
In practice, this occurs when the sample size of discovery GWAS is too small to obtain reliable GWAS summary statistics. 
When these summary statistics are applied on an independent target dataset, the mean of genetic covariance $\y_{\eta}^T\widehat{\bmS}_{\alpha}$ cannot dominate its standard error. The genetic variance $\widehat{\bmS}_{\alpha}^T\widehat{\bmS}_{\alpha}$ is so overwhelming that $G_{\alpha\eta}$ goes to zero. Details can be found in Appendix~A. 
\end{remark}
\begin{figure}
\includegraphics[page=2,width=0.5\linewidth]{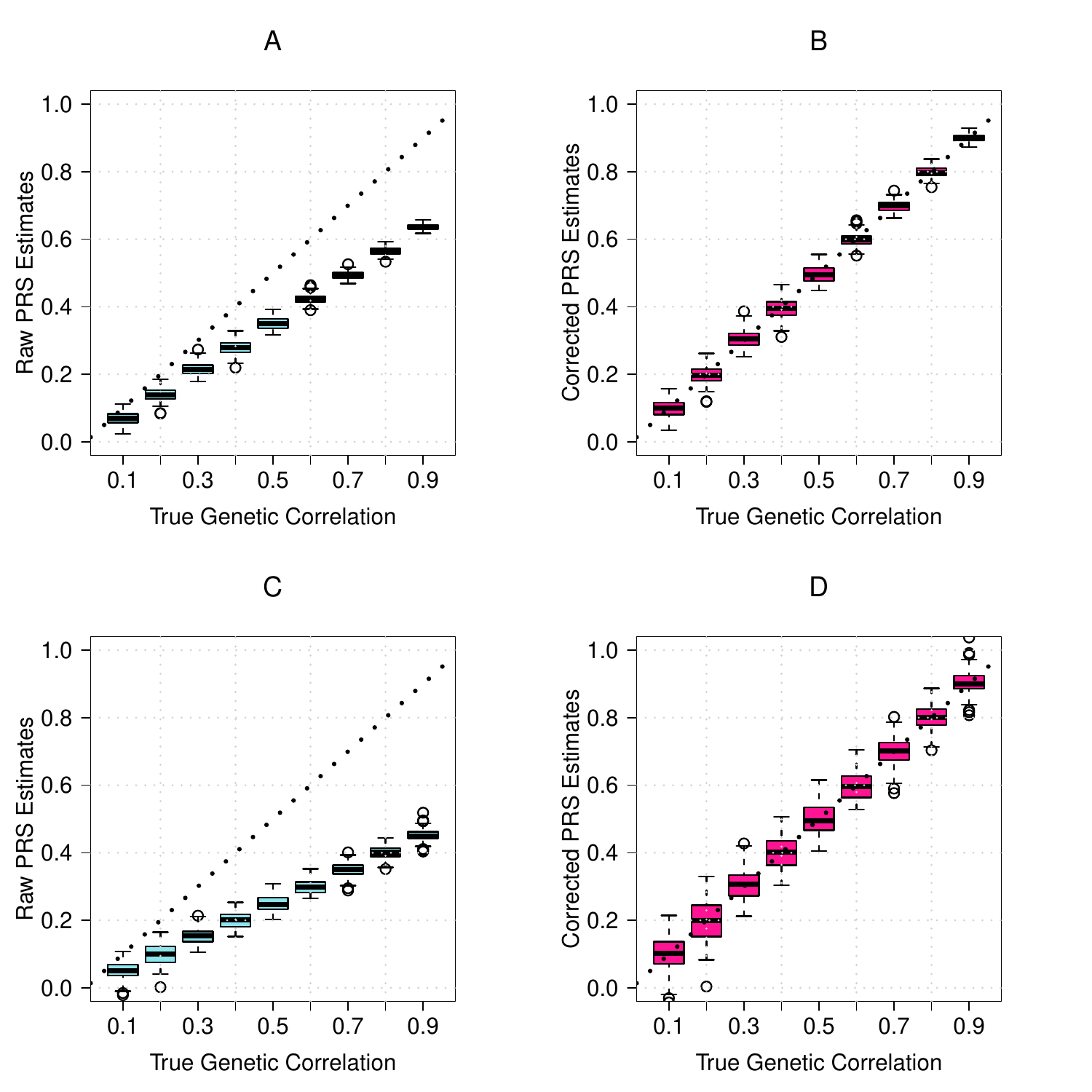}
  \caption{Raw genetic correlations estimated by cross-trait PRS with all SNPs (left panels, A: $G_{\alpha\eta}$, C: $G_{\alpha\beta}$) and the bias-corrected genetic correlation estimates (right panels, B: $G^A_{\alpha\eta}$, D: $G^A_{\alpha\beta}$). 
  We set $h^2_{\alpha}=h^2_{\beta}=h^2_{\eta}=1$, $n_1=n_2=n_3=p=10,000$, and $m=2000$.
}
\label{fig2}
\end{figure}
The asymptotic properties of 
$G_{\alpha\beta}$ are given as follows. 
\begin{thm}\label{thm2}
Under polygenic model~(\ref{equ2.1}) and Conditions~\ref{con1} and~\ref{con2}, 
suppose $m_{\alpha\beta},m_{\alpha}$, and 
$m_{\beta} \rightarrow \infty$  as $\mbox{min}(n_1,n_2, n_3,p)\rightarrow\infty$, and let $p^2=c \cdot (n_1n_2n_3)^{a}$ for some constants $c>0$ and $ a \in (0,\infty]$.
If $a  \in (0,1)$, then we have 
\begin{flalign*}
G_{\alpha\beta}=\varphi_{\alpha\beta}+ \bigg( \sqrt{\frac{n_1}{n_1+p/h^2_{\alpha}}\cdot \frac{n_2}{n_2+p/h^2_{\beta}}}-1\bigg)
\cdot \varphi_{\alpha\beta} \cdot\{1+o_p(1)\}\mathbf{.}
\end{flalign*}
If $ a  \in [1,\infty]$, then we have 
\begin{flalign*}
G_{\alpha\beta}\cdot \frac{n_3(n_1+p)(n_2+p)}{p^2} =\bigO_p(1)\mathbf{.}
\end{flalign*}
\end{thm}
\begin{remark}\label{rmk4}
For $a \in (0,1)$, $G_{\alpha\beta}$ is a biased estimator of $\varphi_{\alpha\beta}$ since $\sqrt{n_1/(n_1+p/h^2_{\alpha})}$ and $\sqrt{n_2/(n_2+p/h^2_{\beta})}$ are smaller than $1$.  
The asymptotic bias is independent of  $m_{\alpha}, m_{\beta}$, and $m_{\alpha\beta}$, and is determined by $n_1$, $n_2$, $p$, $h^2_{\alpha}$ and $h^2_{\beta}$. 
Giving that $n_1,n_2$, and $p$ are comparable, a consistent estimator of $\varphi_{\alpha\beta}$ is given by 
\begin{flalign*}
G_{\alpha\beta}^A=
G_{\alpha\beta}\cdot\sqrt{\frac{(n_1+p/h^2_{\alpha})\cdot(n_2+p/h^2_{\beta})}{n_1n_2}}=\varphi_{\alpha\beta} \cdot\{1+o_p(1)\}\mathbf{.}
\end{flalign*}	
\end{remark}
Now we propose a novel estimator of $\varphi_{\alpha\beta}$ that can be directly constructed by using two sets of summary statistics $\widehat{\bmalpha}$ and $\widehat{\bmbeta}$. Let 
\begin{flalign*}
\widehat{\varphi}_{\alpha\beta}&=\frac{\widehat{\bmalpha}^T\widehat{\bmbeta}}{\big\Vert \widehat{\bmalpha}\big\Vert \cdot\big\Vert \widehat{\bmbeta}\big\Vert }=\frac{\big(\X_{(1)}\bmalpha_{(1)}+\bmeps_{\alpha}\big)^T\X\Z^T\big(\Z_{(1)}\bmbeta_{(1)}+\bmeps_{\beta}\big)}{
\big\Vert 
\big(\X_{(1)}\bmalpha_{(1)}+\bmeps_{\alpha}\big)^T\X\big\Vert \cdot
\big\Vert 
\big(\Z_{(1)}\bmbeta_{(1)}+\bmeps_{\beta}\big)^T\Z
\big\Vert 
} ,
\end{flalign*}
we have the following asymptotic properties.
\begin{thm}\label{thm3}
Under polygenic model~(\ref{equ2.1}) and Conditions~\ref{con1} and~\ref{con2}, 
suppose $m_{\alpha\beta},m_{\alpha}$, and 
$m_{\beta} \rightarrow \infty$ as $\mbox{min}(n_1,n_2,p)\rightarrow\infty$, and let $p=c \cdot (n_1n_2)^{a}$ for some constants $c>0$ and $ a  \in (0,\infty]$. 
If $a  \in (0,1)$, then we have 
\begin{flalign*}
\widehat{\varphi}_{\alpha\beta}= \varphi_{\alpha\beta}+
\bigg( \sqrt{\frac{n_1}{n_1+p/h^2_{\alpha}}\cdot \frac{n_2}{n_2+p/h^2_{\beta}}}-1\bigg)
\cdot \varphi_{\alpha\beta} \cdot\{1+o_p(1)\}\mathbf{.}
\end{flalign*}
If $a  \in [1,\infty]$, then we have 
\begin{flalign*}
\widehat{\varphi}_{\alpha\beta}\cdot \frac{(n_1+p)(n_2+p)}{p} =\bigO_p(1)\mathbf{.}
\end{flalign*}
\end{thm}
It follows from Theorem~\ref{thm3} that a consistent estimator of $\varphi_{\alpha\beta}$ is given by 
\begin{flalign*}
\widehat{\varphi}_{\alpha\beta}^A=
\widehat{\varphi}_{\alpha\beta}\cdot\sqrt{\frac{(n_1+p/h^2_{\alpha})\cdot(n_2+p/h^2_{\beta})}{n_1n_2}}=\varphi_{\alpha\beta} \cdot\{1+o_p(1)\}
\mathbf{.}
\end{flalign*}	
Since $\widehat{\varphi}_{\alpha\beta}$ and $G_{\alpha\beta}$ have similar asymptotic properties, 
in what follows we will focus on $G_{\alpha\beta}$ and the general conclusions of $G_{\alpha\beta}$ remain the same for $\widehat{\varphi}_{\alpha\beta}$.

\section{SNP screening}\label{sec3}
As shown in Theorems~\ref{thm1} and~\ref{thm2}, in addition to heritability, the asymptotic bias of $G_{\alpha\eta}$ or $G_{\alpha\beta}$ is largely affected by $n/p$. 
These results intuitively suggest to select a subset of 
$p$ SNPs to construct cross-trait PRS. 
The common approach in practice is to screen the SNPs according to their GWAS $p$-values. We investigate this strategy in this section.  

For a given threshold $c_{\alpha}>0$, 
let  $q_{\alpha}=p\cdot \pi_{\alpha}$ $=q_{\alpha1} + q_{\alpha2} ~ (\pi_{\alpha} \in (0,1])$ be  the number of top-ranked SNPs selected for $\y_{\alpha}$, among  which there are $q_{\alpha1}$ true causal SNPs and the remaining $q_{\alpha2}$ are null SNPs, 
and we let $q_{\alpha\eta}$ be the number of overlapping causal SNPs of $\y_{\alpha}$ and $\y_{\eta}$. 
Similarly, given a threshold $c_{\beta}>0$, let $q_{\beta}=p\cdot \pi_{\beta}$ $=q_{\beta1} + q_{\beta2}~ (\pi_{\beta} \in (0,1])$ be the number of top-ranked SNPs selected for $\y_{\beta}$, among  which there are $q_{\beta1}$ true causal SNPs and the remaining $q_{\beta2}$ are null SNPs, 
and we let $q_{\alpha\beta}$ be the number of overlapping causal SNPs of $\y_{\alpha}$ and $\y_{\beta}$. 
Thus, $q_{\alpha1}\ge q_{\alpha\eta}$ and $\mbox{min}(q_{\beta1}, q_{\alpha1})\ge q_{\alpha\beta}$. 

The SNP data are defined accordingly. 
We write 
$\X_{(1)}=[\X_{(11)},\X_{(12)}]$, $\X_{(2)}=[\X_{(21)},\X_{(22)}]$, 
$\Z_{(1)}=[\Z_{(11)},\Z_{(12)}]$, $\Z_{(2)}=[\Z_{(21)},\Z_{(22)}]$,
$\W_{(1,\alpha)}=[\W_{(11,\alpha)},\W_{(12,\alpha)}]$, $\W_{(2,\alpha)}=[\W_{(21,\alpha)},\W_{(22,\alpha)}]$,
$\W_{(1,\beta)}=[\W_{(11,\beta)},\W_{(12,\beta)}]$, and  $\W_{(2,\beta)}=[\W_{(21,\beta)},\W_{(22,\beta)}]$. 
Here $\X_{(11)}$ and $\W_{(11,\alpha)}$ are the selected  $q_{\alpha1}$ causal SNPs of $\y_{\alpha}$, and $\Z_{(11)}$ and $\W_{(11,\beta)}$ are the  selected $q_{\beta1}$ causal SNPs of $\y_{\beta}$. 
Similarly,  
$\X_{(21)}$ and $\W_{(21,\alpha)}$ are the selected $q_{\alpha2}$ null SNPs of $\y_{\alpha}$, and 
$\Z_{(21)}$ and $\W_{(21,\beta)}$ are the selected $q_{\beta2}$ null SNPs of $\y_{\beta}$. 
In addition, we let 
$\widehat{\bmalpha}_{(1)}=[\widehat{\bmalpha}_{(11)},\widehat{\bmalpha}_{(12)}]$,
$\widehat{\bmalpha}_{(2)}=[\widehat{\bmalpha}_{(21)},\widehat{\bmalpha}_{(22)}]$,
$\widehat{\bmbeta}_{(1)}=[\widehat{\bmbeta}_{(11)},\widehat{\bmbeta}_{(12)}]$, and $\widehat{\bmbeta}_{(2)}=[\widehat{\bmbeta}_{(21)},\widehat{\bmbeta}_{(22)}]$,
where $\widehat{\bmalpha}_{(11)}$ and $\widehat{\bmbeta}_{(11)}$ 
correspond to the selected causal SNPs of $\y_{\alpha}$ and $\y_{\beta}$, respectively, 
and $\widehat{\bmalpha}_{(21)}$ and $\widehat{\bmbeta}_{(21)}$ 
correspond to the selected null ones. 
Then we have 
\begin{flalign*}
G_{T\alpha\eta}
&=\frac{\big(\W_{(1)}\bmeta_{(1)}+\bmeps_{\eta}\big)^T\big(\W_{(11,\alpha)}\widehat{\bmalpha}_{(11)}+\W_{(21,\alpha)}\widehat{\bmalpha}_{(21)}\big)}
{\big\Vert \W_{(1)}\bmeta_{(1)}+\bmeps_{\eta} \big\Vert\cdot\big\Vert \W_{(11,\alpha)}\widehat{\bmalpha}_{(11)}+\W_{(21,\alpha)}\widehat{\bmalpha}_{(21)}\big\Vert}
=\frac{C_{T\alpha\eta}}
{V_{\eta} \cdot
V_{T\alpha}}
\end{flalign*}
where $~V_{\eta}=\big\Vert \W_{(1)}\bmeta_{(1)}+\bmeps_{\eta}\big\Vert$, 
\begin{flalign*}
&V_{T\alpha}=\big\Vert \W_{(11,\alpha)}\X_{(11)}^T\big(\X_{(1)}\bmalpha_{(1)}+\bmeps_{\alpha}\big)+\W_{(21,\alpha)}\X_{(21)}^T\big(\X_{(1)}\bmalpha_{(1)}+\bmeps_{\alpha}\big) \big\Vert,  \mbox{ and} \\
&C_{T\alpha\eta}=\big(\W_{(1)}\bmeta_{(1)}+\bmeps_{\eta}\big)^T\W_{(11,\alpha)}\X_{(11)}^T\big(\X_{(1)}\bmalpha_{(1)}+\bmeps_{\alpha}\big)+\\
&
\qquad \big(\W_{(1)}\bmeta_{(1)}+\bmeps_{\eta}\big)^T\W_{(21,\alpha)}\X_{(21)}^T\big(\X_{(1)}\bmalpha_{(1)}+\bmeps_{\alpha}\big).
\end{flalign*}
\begin{cor}\label{cor1}
Under polygenic model~(\ref{equ2.1}) and Conditions~\ref{con1} and~\ref{con2}, suppose that $\mbox{min}(m_{\alpha\eta}$, $m_{\alpha}$,
$m_{\eta})\rightarrow \infty$ and $\mbox{min}(q_{\alpha\eta},q_{\alpha 1},q_{\alpha 2})\rightarrow \infty$ as $\mbox{min}(n_1, n_3,p)\rightarrow\infty$, 
further if $\big\{m_{\alpha\eta}^2(q_{\alpha 1}+q_{\alpha 2})\big\}/(q_{\alpha\eta}^2n_1n_3)$ 
$\to 0$, then we have 
\begin{flalign*}
G_{T\alpha\eta}=\varphi_{\alpha\eta} +
\bigg(
\sqrt{\frac{n_1m_{\alpha}}{n_1q_{\alpha1}+m_{\alpha}q_{\alpha}/h^2_{\alpha}}}\cdot \frac{q_{\alpha\eta}}{m_{\alpha\eta}}  \cdot h_{\eta}-1\bigg) \cdot \varphi_{\alpha\eta} \cdot\{1+o_p(1)\} \mathbf{.}
\end{flalign*}
\end{cor}
\begin{figure}
\includegraphics[page=1,width=0.95\linewidth]{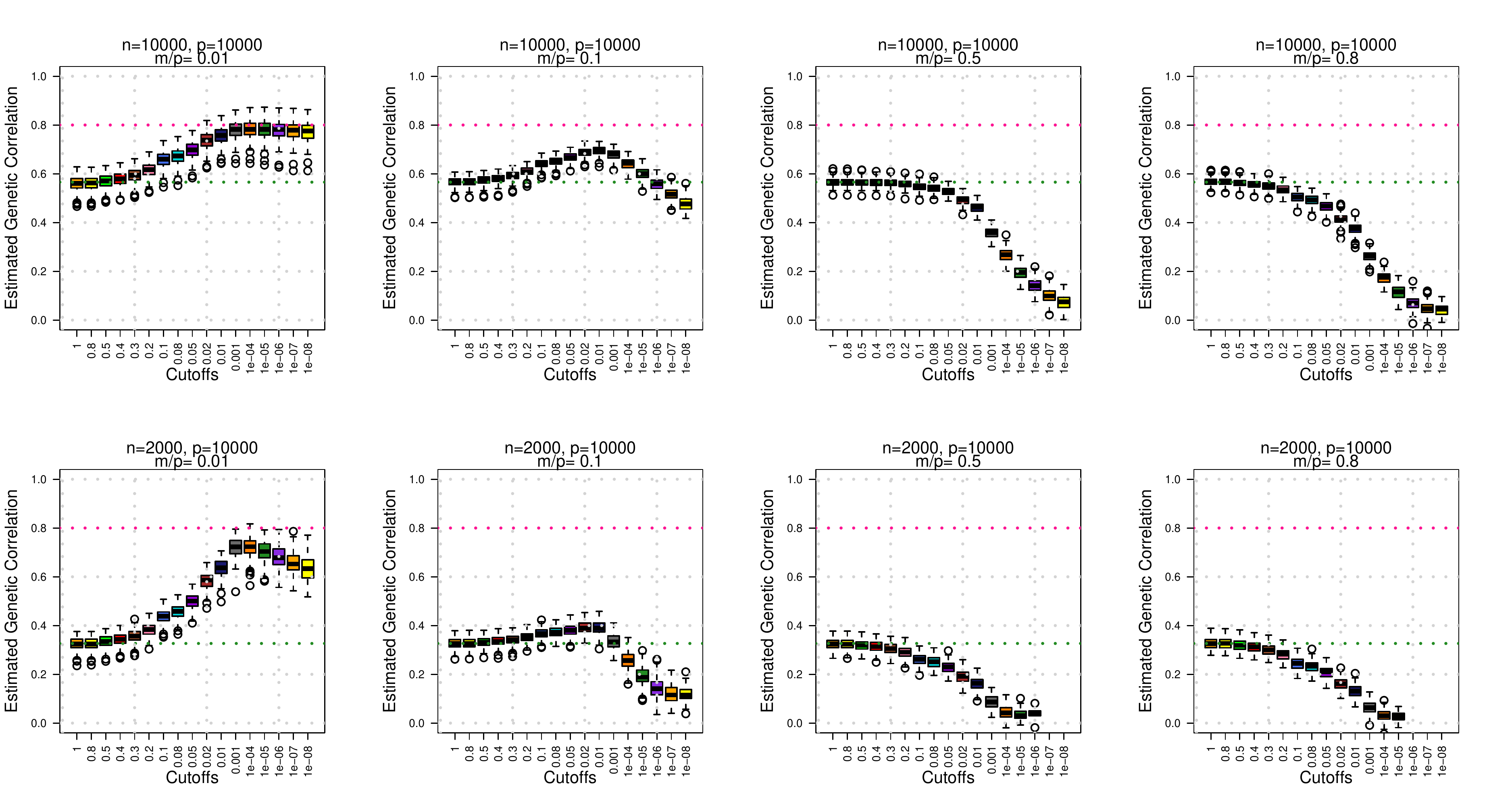}
  \caption{Raw genetic correlation $G_{T\alpha\eta}$ estimated by cross-trait PRS with selected SNPs under different sparsity $m/p$ and sample size $n$. We set  $h^2_{\alpha}=h^2_{\eta}=1$, $\varphi_{\alpha\eta}=0.8$, $p=10,000$, and $n=10,000$ (upper panels) or $2000$ (lower panels).
 }
\label{fig3}
\end{figure}
Corollary~\ref{cor1} shows the trade-off of SNP screening. 
Given $n_1$, $m_{\alpha}$, $m_{\alpha\eta}$, $h_{\alpha}$, and  $h_{\eta}$,
the bias of $G_{T\alpha\eta}$ is also affected by $q_{\alpha}$, $q_{\alpha1}$ and $q_{\alpha\eta}$.
As more SNPs are selected, the numerator of  $\sqrt{(n_1m_{\alpha})/(n_1q_{\alpha1}+m_{\alpha}q_{\alpha}/h^2_{\alpha})}\cdot (q_{\alpha\eta}/m_{\alpha\eta})$ increases with $q_{\alpha\eta}$, while the denominator increases with $\sqrt{q_{\alpha}}$ (and $\sqrt{q_{\alpha1}}$). 
Therefore,
whether or not SNP screening can improve the estimation is largely affected by the quality of the selected SNPs, which is highly related to the properties of the GWAS summary statistics. 
In the optimistic case where $q_{\alpha\eta}=m_{\alpha\eta}$ and $q_{\alpha}=q_{\alpha1}=m_{\alpha}$, $G_{T\alpha\eta}$ becomes 
\begin{flalign*}
\sqrt{\frac{n_1}{n_1+m_{\alpha}/h^2_{\alpha}}} \cdot h_{\eta} \cdot \varphi_{\alpha\eta}, 
\end{flalign*}
which is the theoretical upper limit. We note that this optimistic upper limit is still biased towards zero.
Another interesting case is that the GWAS summary statistics of causal and null SNPs are totally mixed up, which may occur when $n_1=o(m_{\alpha})$ (i.e., sample size is small or trait is highly polygenic/omnigenic) according to~(\ref{equ1.5}).
Therefore, we have $q_{\alpha1}/q_{\alpha} \approx m_{\alpha}/p $.  Suppose also $q_{\alpha\eta}/q_{\alpha1} \approx m_{\alpha\eta}/m_{\alpha}$,
we have 
\begin{flalign*}
G_{T\alpha\eta} \approx \sqrt{\frac{n_1}{n_1p+p^2/h^2_{\alpha}} \cdot q_{\alpha}} \cdot h_{\eta} \cdot \varphi_{\alpha\eta},  
\end{flalign*}
which increases with $q_{\alpha}$. 

As $q_{\alpha}=p$,  $G_{T\alpha\eta}$ reaches its upper bound 
\begin{flalign*}
\sqrt{\frac{n_1}{n_1+p/h^2_{\alpha}}} \cdot h_{\eta} \cdot \varphi_{\alpha\eta}.
\end{flalign*}
That is, $G_{T\alpha\eta}$ achieves the best performance when the cross-trait PRS is constructed \textit{without SNP screening}. 
For example, in the left two panels of Figure~\ref{fig3}, we set $m/p=0.01$ to reflect the sparse signal case, in which causal and null SNPs can be easily separated by SNP screening. 
Thus, SNP screening can reduce the bias of $G_{\alpha\eta}$ when signals are sparse.  
However, as the number of causal SNPs increase (from left to right in Figure~\ref{fig3}), it becomes much hard to separate causal and null SNPs by their GWAS $p$-values. 
Therefore, SNP screening will enlarge the bias. 

Similarly, we have 
\begin{flalign*}
G_{T\alpha\beta}
=\frac{\big(\W_{(11,\beta)}\widehat{\bmbeta}_{(11)}+\W_{(21,\beta)}\widehat{\bmbeta}_{(21)}\big)^T
\big(\W_{(11,\alpha)}\widehat{\bmalpha}_{(11)}+\W_{(21,\alpha)}\widehat{\bmalpha}_{(21)}\big)}{
\Vert \W_{(11,\beta)}\widehat{\bmbeta}_{(11)}+\W_{(21,\beta)}\widehat{\bmbeta}_{(21)} \Vert\cdot
\Vert \W_{(11,\alpha)}\widehat{\bmalpha}_{(11)}+\W_{(21,\alpha)}\widehat{\bmalpha}_{(21)}\Vert
}
=\frac{C_{T\alpha\beta}}
{V_{T\alpha} \cdot
V_{T\beta}}, 
\end{flalign*}
where 
\begin{flalign*}
C_{T\alpha\beta}&=
\big\{\W_{(11,\beta)}\Z_{(11)}^T\big(\Z_{(1)}\bmbeta_{(1)}+\bmeps_{\beta}\big)+\W_{(21,\beta)}\Z_{(21)}^T\big(\Z_{(1)}\bmbeta_{(1)}+\bmeps_{\beta}\big)\big\}^T\\
&
\big\{\W_{(11,\alpha)}\X_{(11)}^{T}\big(\X_{(1)}\bmalpha_{(1)}+\bmeps_{\alpha}\big)+
\W_{(21,\alpha)}\X_{(21)}^{T}\big(\X_{(1)}\bmalpha_{(1)}+\bmeps_{\alpha}\big)\big\}
\end{flalign*}
and $V_{T\beta}=\big\Vert \W_{(11,\beta)}\Z_{(11)}^T\big(\Z_{(1)}\bmbeta_{(1)}+\bmeps_{\beta}\big)+\W_{(21,\beta)}\Z_{(21)}^T\big(\Z_{(1)}\bmbeta_{(1)}+\bmeps_{\beta}\big)\big\Vert$.

\begin{cor}\label{cor2}
Under polygenic model~(\ref{equ2.1}) and Conditions~\ref{con1} and~\ref{con2}, 
suppose that $\mbox{min}(m_{\alpha\beta}$, $m_{\alpha}$,
$m_{\beta}) \rightarrow \infty$ and 
$\mbox{min}(q_{\alpha\beta},q_{\alpha 1},q_{\alpha 2},q_{\beta 1},q_{\beta 2})\rightarrow \infty$ as $\mbox{min}(n_1,n_2, n_3,p)\rightarrow\infty$. Further if $\big\{m^2_{\alpha\beta}(q_{\alpha1}+q_{\alpha2})(q_{\beta1}+q_{\beta2})\big\}/(q_{\alpha\beta}^2n_1n_2n_3)\to 0$, then we have 
\begin{flalign*}
G_{\alpha\beta}=\varphi_{\alpha\beta}+
\bigg(
\sqrt{\frac{n_1m_{\alpha}}{n_1q_{\alpha1}+m_{\alpha}q_{\alpha}/h^2_{\alpha}}
\cdot \frac{n_2m_{\beta}}{n_2q_{\beta1}+m_{\beta}q_{\beta}/h^2_{\beta}}}
\cdot \frac{q_{\alpha\beta}}{m_{\alpha\beta}}-1\bigg) \cdot \varphi_{\alpha\beta} \cdot\{1 + o_p(1)\}\mathbf{.}
\end{flalign*}
\end{cor}
Corollary~\ref{cor2} shows the trade-off of SNP screening for $G_{\alpha\beta}$. 
Given $n_1$, $n_2$, $m_{\alpha}$, $m_{\beta}$, $m_{\alpha\beta}$, $h_{\alpha}$, and $h_{\beta}$,
the bias of $G_{T\alpha\eta}$ is also affected by $q_{\alpha}$, $q_{\alpha1}$, $q_{\beta}$, $q_{\beta1}$ and $q_{\alpha\beta}$.
As more SNPs are selected, the numerator of
$q_{\alpha\beta}/m_{\alpha\beta}$ increases with $q_{\alpha\beta}$, while the denominator of $\sqrt{(n_1m_{\alpha})/(n_1q_{\alpha1}+m_{\alpha}q_{\alpha}/h^2_{\alpha})\cdot(n_2m_{\beta})/(n_2q_{\beta1}+m_{\beta}q_{\beta}/h^2_{\beta})}$ increases with $\sqrt{q_{\alpha}}$ and $\sqrt{q_{\beta}}$ (also $\sqrt{q_{\alpha1}}$ and $\sqrt{q_{\beta1}}$). 
In the optimistic case where $q_{\alpha\beta}=m_{\alpha\beta}$, $q_{\alpha}=q_{\alpha1}=m_{\alpha}$ and $q_{\beta}=q_{\beta1}=m_{\beta}$, $G_{T\alpha\beta}$ reduces to
\begin{flalign*}
\sqrt{\frac{n_1}{n_1+m_{\alpha}/h^2_{\alpha}}\cdot \frac{n_2}{n_2+m_{\beta}/h^2_{\beta}}} \cdot \varphi_{\alpha\beta}, 
\end{flalign*}
which is the theoretical upper limit. 
On the other hand,
suppose $q_{\alpha\beta}/q_{\alpha1} \approx  m_{\alpha\beta}/m_{\alpha}$ and $q_{\alpha\beta}/q_{\beta1} \approx m_{\alpha\beta}/m_{\beta}$, when $n_1=o(m_{\alpha})$, $n_2=o(m_{\beta})$, i.e., the causal SNPs and null SNPs are totally mixed, we have $q_{\alpha1}/q_{\alpha} \approx m_{\alpha}/p $, $q_{\beta1}/q_{\beta} \approx  m_{\beta}/p $, and 
\begin{flalign*}
G_{T\alpha\beta}\approx \sqrt{\frac{n_1}{n_1p+p^2/h^2_{\alpha}}\cdot \frac{n_2}{n_2p+p^2/h^2_{\alpha}} \cdot q_{\alpha}q_{\beta}} \cdot \varphi_{\alpha\beta} , 
\end{flalign*}
which increases with $q_{\alpha}$ and $q_{\beta}$. 
Therefore, as $q_{\alpha}=q_{\beta}=p$, $G_{T\alpha\beta}$ reaches its upper bound 
\begin{flalign*}
\sqrt{\frac{n_1}{n_1+p/h^2_{\alpha}}\cdot\frac{n_2}{n_2+p/h^2_{\beta}}} \cdot \varphi_{\alpha\beta}.
\end{flalign*}
In conclusion, when causal SNP and null SNP can be easily separated by GWAS, the top-ranked SNPs are more likely to be causal ones, that is, SNP screening helps. 
However, for highly polygenic complex traits whose $m/n$ is large, SNP screening may result in larger bias and should be used with caution.

\section{Overlapping samples}\label{sec4}
In real data applications, different GWAS may share a subset of participants. 
It is often inconvenient to recalculate the GWAS summary statistics after removing the overlapping samples.  
In this section, we examine the effect of overlapping samples on the bias of cross-trait PRS, which provides more insights into the bias phenomenon of cross-trait PRS. 
Particularly, we focus on two distinct cases which are both common in practice:
i) $n_s$ overlapping samples between discovery GWAS and Target testing data for $\varphi_{\alpha\eta}$ estimation; and 
ii) $n_s$ overlapping samples between two discovery GWAS for $\varphi_{\alpha\beta}$ estimation. 

\subsection*{Case i)}
We add $n_s$ overlapping samples into Discovery GWAS-I and Target testing GWAS, resulting in the following two new datasets: 
\begin{itemize}
\item Dataset~IV: $(\X,\bmS,\y_{\alpha})$, with $\X\in \bbR^{n_{1}\times p}$, $\bmS\in \bbR^{n_{s}\times p}$, and $\y_{\alpha}^T=(\y_{\alpha_X}^T,\y_{\alpha_S}^T) \in \bbR^{(n_{1}+n_{s}) \times 1}$.  
\item Dataset~V: $(\W,\bmS,\y_{\eta})$, with $\W\in \bbR^{n_{3}\times p}$, $\bmS\in \bbR^{n_{s}\times p}$, and $\y_{\eta}^T=(\y_{\eta_W}^T,\y_{\eta_S}^T) \in \bbR^{(n_{3}+n_{s}) \times 1}$.  
\end{itemize}
Mimicking $h^2$, we define $h_{\alpha\eta} \in (0,1]$ as  the proportion of phenotypic correlation that can be explained by the correlation of their genetic components
\begin{flalign*}
h_{\alpha\eta}=\frac{m_{\alpha \eta}\sigma_{\alpha\eta}}{m_{\alpha\eta}\sigma_{\alpha\eta}+\sigma_{\epsilon_{\alpha}\epsilon_{\eta}}} \mathbf{.}
\end{flalign*}
On the overlapping samples, we allow nonzero correlation between random errors to capture the non-genetic contribution to phenotypic correlation. 
We introduce an additional condition on random errors. 
\begin{condition}\label{con3}
On $n_{s}$ overlapping samples, 
$\epsilon_{\alpha_j}$ and $\epsilon_{\eta_j}$ are independent random variables satisfying
\begin{flalign*}
\begin{pmatrix} 
\epsilon_{\alpha_j}\\
\epsilon_{\eta_j} 
\end{pmatrix}
\sim F 
\left \lbrack
\begin{pmatrix} 
0\\
0 
\end{pmatrix},
\begin{pmatrix} 
\sigma^2_{\epsilon_{\alpha}} & \sigma_{\epsilon_{\alpha}\epsilon_{\eta}} \\
\sigma_{\epsilon_{\alpha}\epsilon_{\eta}} & \sigma^2_{\epsilon_{\eta}}
\end{pmatrix}
\right \rbrack
\end{flalign*} 
for $j=1,...,n_{s}$, where 
$\sigma_{\epsilon_{\alpha}\epsilon_{\eta}}=\rho_{\epsilon_{\alpha}\epsilon_{\eta}} \cdot \sigma_{\epsilon_{\alpha}}\sigma_{\epsilon_{\eta}}$. 
\end{condition}

\begin{thm}\label{thm6}
Under polygenic model~(\ref{equ2.1}) and Conditions~\ref{con1}~-~\ref{con3}, suppose $\mbox{min}(m_{\alpha\eta}$, $m_{\alpha}$,
$m_{\eta})\rightarrow \infty$ as $\mbox{min}\{(n_1+n_s), (n_3+n_s),p\}\rightarrow\infty$, 
and let $p=c \cdot \{(n_1+n_s)(n_3+n_s)\}^{a}$ for some constants $c>0$ and $a\in (0,\infty]$. 
If $a\in (0,1)$, then $G_{S\alpha\eta}$ can be written as
\begin{flalign*}
&\frac{ \big[ 1+n_sp/\{(n_1+n_s)(n_3+n_s)\cdot h_{\alpha\eta}\}\big]\cdot \big[ h_{\eta} \cdot \varphi_{\alpha\eta} \cdot\{1 + o_p(1)\} \big]
}{ \big[1+p/\{(n_1+n_s)\cdot h^2_{\alpha}\}+2n_sp/\{(n_1+n_s)(n_3+n_s)\}+
n_sp^2/\{(n_1+n_s)^2(n_3+n_s)\cdot h^2_{\alpha}\}\big]^{1/2} }
 \mathbf{.}
\end{flalign*}
If $ a  \in [1,\infty]$, then we have $G_{S\alpha\eta} =o_p(1)$.
\end{thm}
\begin{remark}\label{rmk8}
Theorem~\ref{thm6} shows the effect  of $n_s$ overlapping samples on the estimation of $\varphi_{\alpha\eta}$. 
Both sample sizes $(n_1+n_s)$ and $(n_3+n_s)$ are involved in the bias. 
A consistent estimator $G^A_{S\alpha\eta}$ can be derived given that  $h_{\alpha\eta}$ is estimable. 
An interesting special case is when the two GWAS are fully overlapped, then we have 
\begin{flalign*}
G_{S\alpha\eta}&=
\frac{n_s+p/h_{\alpha\eta}}{\big\{n_s^2+2n_sp+p(p+n_s)/h^2_{\alpha}\big\}^{1/2}} 
\cdot h_{\eta} \cdot \varphi_{\alpha\eta} \cdot\{1 + o_p(1)\} \mathbf{.}
\end{flalign*}
In the optimal situation where $h^2_{\alpha}=h^2_{\eta}=h_{\alpha\eta}=1$, we have
\begin{flalign*}
G_{S\alpha\eta}&=
\bigg(1+\frac{1}{p/n_s+n_s/p+2} \bigg)^{-1/2} \cdot \varphi_{\alpha\eta} \cdot\{1 + o_p(1)\} \mathbf{.}
\end{flalign*}
Therefore, $G_{S\alpha\eta}$ is asymptoticly biased unless either $p=o(n_s)$ or $n_s=o(p)$ holds, neither of which is the case in modern GWAS. 
As $n_s$ and $p$ are more comparable, the asymptotic bias in $G_{S\alpha\eta}$ increases and the largest bias occurs as $p=n_s \to \infty$. 
\begin{figure}
\includegraphics[page=2,width=0.5\linewidth]{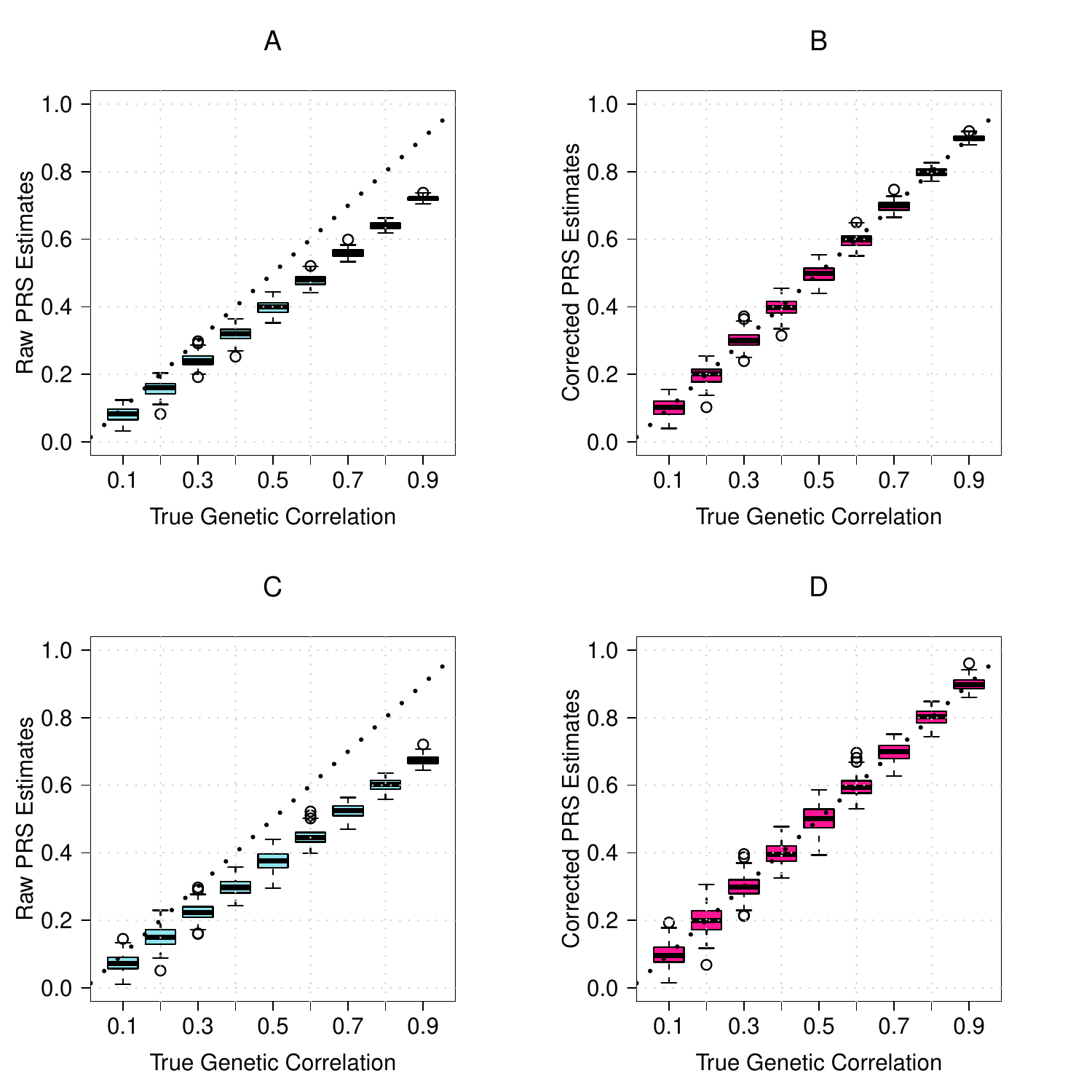}
  \caption{Raw genetic correlations estimated by cross-trait PRS with all SNPs (left panels, A: $G_{S\alpha\eta}$, C: $G_{S\alpha\beta}$) and bias-corrected genetic correlation estimates (right panels, B: $G^A_{S\alpha\eta}$, D: $G^A_{S\alpha\beta}$). 
  We set $h^2_{\alpha}=h^2_{\beta}=h^2_{\eta}=1$, $n_1=n_s=n_2=n_3=5000$ (half samples overlap), $p=10,000$, and $m=2000$.
 }
\label{fig4}
\end{figure}
Note that it is not recommended to estimate the genetic correlation between two traits with (fully) overlapping samples due to concerns such as confounding and overfitting  \citep{pasaniuc2017dissecting,dudbridge2013power}. In our analysis, such concern is quantified by the value of $h_{\alpha\eta}$. That is, when non-genetic correlation exists in error terms,
we have $h_{\alpha\eta}<1$, and the estimation of genetic correlation is inflated.
However, on the other hand, our results show that even in an optimal overlapping setting with $h^2_{\alpha}=h^2_{\eta}=h_{\alpha\eta}=1$, the cross-trait PRS estimator based on GWAS summary statistics is biased towards zero. 
\end{remark}
\subsection*{Case ii)}
In this case, we add $n_s$ overlapping samples into Discovery GWAS-I and II, resulting in the following two new datasets: 
\begin{itemize}
\item Dataset~IV: $(\X,\bmS,\y_{\alpha})$, with $\X\in \bbR^{n_{1}\times p}$, $\bmS\in \bbR^{n_{s}\times p}$, and  $\y_{\alpha}^T=(\y_{\alpha_X}^T,\y_{\alpha_S}^T) \in \bbR^{(n_{1}+n_{s}) \times 1}$.
\item Dataset~VI: $(\Z,\bmS,\y_{\beta})$, with $\Z\in \bbR^{n_{2}\times p}$, $\bmS\in R^{n_{s}\times p}$, and $\y_{\beta}^T=(\y_{\beta_Z}^T,\y_{\beta_S}^T) \in \bbR^{(n_{2}+n_{s}) \times 1}$.  
\end{itemize}
Then we define $h_{\alpha\beta} \in (0,1]$ as 
\begin{flalign*}
h_{\alpha\beta}=\frac{m_{\alpha \beta}\sigma_{\alpha\beta}}{m_{\alpha\beta}\sigma_{\alpha\beta}+\sigma_{\epsilon_{\alpha}\epsilon_{\beta}}} , 
\end{flalign*}
which quantifies the contribution of genetic correlation to the phenotypic correlation. 
We introduce the following additional condition on random errors. 
\begin{condition}\label{con4}
On $n_{s}$ overlapping samples, 
$\epsilon_{\alpha_j}$ and $\epsilon_{\beta_j}$ are independent random variables satisfying 
\begin{flalign*}
\begin{pmatrix} 
\epsilon_{\alpha_j}\\
\epsilon_{\beta_j} 
\end{pmatrix}
\sim F 
\left \lbrack
\begin{pmatrix} 
0\\
0 
\end{pmatrix},
\begin{pmatrix} 
\sigma^2_{\epsilon_{\alpha}} & \sigma_{\epsilon_{\alpha}\epsilon_{\beta}} \\
\sigma_{\epsilon_{\alpha}\epsilon_{\beta}} & \sigma^2_{\epsilon_{\beta}}
\end{pmatrix}
\right \rbrack
\end{flalign*} 
for $j=1,...,n_{s}$, where 
$\sigma_{\epsilon_{\alpha}\epsilon_{\beta}}=\rho_{\epsilon_{\alpha}\epsilon_{\beta}} \cdot \sigma_{\epsilon_{\alpha}}\sigma_{\epsilon_{\beta}}$. 
\end{condition}

\begin{thm}\label{thm7}
Under polygenic model~(\ref{equ2.1}) and Conditions~\ref{con1},~\ref{con2} and~\ref{con4}, 
suppose $\mbox{min}(m_{\alpha\beta}$, $m_{\alpha}$, 
$m_{\beta}) \rightarrow \infty$ as $\mbox{min}\{(n_1+n_s),(n_2+n_s), n_3,p\}\rightarrow\infty$, and let $p=c \cdot \{(n_1+n_s)(n_2+n_s)n_3\}^{a}$ for some constants $c>0$ and $ a  \in (0,\infty]$. 
If $a  \in (0,1)$, then $G_{S\alpha\beta}$ is given by 
\begin{flalign*}
&\frac{(n_1+n_s)^{1/2}(n_2+n_s)^{1/2}+n_sp/\{(n_1+n_s)^{1/2}(n_2+n_s)^{1/2}\cdot h_{\alpha\beta}\}}
{\big\{ (n_1+n_s+p/h^2_{\alpha})\cdot(n_2+n_s+p/h^2_{\beta})\big\}^{1/2} } \cdot \varphi_{\alpha\beta} \cdot\{1 + o_p(1)\} \mathbf{.}
\end{flalign*}
If $a  \in [1,\infty]$, then we have $ G_{S\alpha\beta}=o_p(1)$.
\end{thm}
\begin{remark}\label{rmk9}
Theorem~\ref{thm7} shows the effect of $n_s$ overlapping samples on the estimation of $\varphi_{\alpha\beta}$. 
Since $n_3$ vanishes in the bias, when $(n_1+n_s)$ and $(n_2+n_s)$ are large, a consistent estimator $G^A_{S\alpha\beta}$ can be derived given that $h_{\alpha\beta}$ is estimable. 
When the two discovery GWAS are fully overlapped, i.e., the two set of summary statistics are generated from the same GWAS, then we have
\begin{flalign*}
&G_{S\alpha\beta}=\frac{n_s +p/h_{\alpha\beta}}
{\big\{ (n_s+p/h^2_{\alpha})\cdot(n_s+p/h^2_{\beta})\big\} ^{1/2} } \cdot \varphi_{\alpha\beta} \cdot\{1 + o_p(1)\} \mathbf{.}
\end{flalign*}
In the optimal situation with $h^2_{\alpha}=h^2_{\beta}=h_{\alpha\beta}=1$, we have
$G_{S\alpha\beta}= \varphi_{\alpha\beta} \cdot\{1 + o_p(1)\}$.
Thus, $G_{S\alpha\beta}$ is a consistent estimator and we may have an unbiased estimator of genetic correlation. 

In summary, above analyses reveal that the bias in cross-trait PRS estimator may
result from the following facts: 
i) summary statistics are generated from independent GWAS, where the induced bias is largely determined by the $n/p$ ratio; 
ii) phenotypes are not fully heritable, i.e., heritability is less than one;
and iii) non-genetic correlation exists in the random errors of overlapping samples. This may happen, for example,  when confounding effects are not fully adjusted. 
The first two facts may bias the genetic correlation estimator towards zero, while the last fact may inflate the estimated genetic correlation.
In the supplementary file, we further investigate several other specific overlapping cases,  which can be useful for quantifying potential bias and perform correction in real data. 
\end{remark}
\section{Numerical experiments}\label{sec5}
\subsection{GWAS of polygenic traits}
We first numerically evaluate the marginal effect size estimates in GWAS with $p=100,000$ and $n=10,000$ or $1000$.  
Each entry of $\X$ is independently generated from $N(0,1)$. 
We vary the ratio $m/p$ from $0.001$ to $0.8$ to reflect a wide range of sparsity. 
The nonzero SNP effects in $\bmbeta_{(1)}$ are independently generated from $N(0,1)$. 
Entries of $\bmeps$ are independently generated from 
$N(0,1)$. 
A continuous phenotype $\y$ is then generated from model~(\ref{equ1.1}) 
and we apply model~(\ref{equ1.3}) to estimate the marginal effects. 
A total of $200$ replicates was conducted. 
We calculated the sum of the MSE of regression coefficients $\widehat{\bmbeta}$, the area under curve (AUC) and power of test statistics $T_i$ ($i=1,\cdots,p$), and enrichment, which is the proportion of true causal SNPs among the top $(10\% \times p)$-ranked SNPs. 
As expected, when sparsity $m/p$ increasing, the MSE of $\widehat{\bmbeta}$ is inflated, and both AUC and power of $T_i$s decrease dramatically (Supplementary Figure~\ref{fig.s1}). When $m/p$ is larger than $0.5$, AUC is close to $0.5$ and power is near zero. 
Enrichment is high when $m/p$ is small, but it drops dramatically as $m/p$ increases. Finally, enrichment becomes similar to $m/p$, reflecting that marginal screening can well preserve the rank of variants only when signals are very sparse. These results indicate that causal and null SNPs may be highly mixed in the ranking list of SNP for polygenic traits. 
\subsection{Cross-trait PRS with all SNPs}
To illustrate the finite sample performance of our theoretical results, we simulate $10,000$ uncorrelated SNPs. 
The MAF of each SNP, $f$, is independently generated from Uniform $[0.05, 0.45]$ based on which the SNP genotypes are independently sampled from $\{0,1,2\}$ with probabilities $\{ (1-f)^2,2f(1-f),f^2\}$, respectively. 
We set the same $2000$ causal SNPs on each trait 
and the nonzero genetic effects are generated from Normal distribution according to {Condition~\ref{con2}} with $\sigma_{\alpha}=\sigma_{\eta}=\sigma_{\beta}=1$.
We set all heritability to one and vary $\varphi_{\alpha\eta}$ and $\varphi_{\alpha\beta}$ from $0.1$ to $0.9$. 
Model~(\ref{equ2.1}) is used to generate continuous phenotypes.
We generated 
$10,000$ samples in each dataset and a total of $200$ replicates was conducted. 
Cross-trait PRS was built with all SNPs.  
We calculated the raw estimators $G_{\alpha\eta}$ and $G_{\alpha\beta}$ studied in Theorems~\ref{thm1}~-~\ref{thm2}, and the corresponding bias-corrected estimators $G^A_{\alpha\eta}$ and $G^A_{\alpha\beta}$.
The performance of $G_{\alpha\eta}$ and $G_{\alpha\beta}$ is displayed in the left panels of Figure~\ref{fig2}. It is clear that these raw estimates are biased towards zero. 
For example, when $\varphi_{\alpha\eta}=\varphi_{\alpha\beta}=0.9$, $G_{\alpha\eta}$ is around $0.6$ while $G_{\alpha\beta}$ is less than $0.45$.
The performance of $G^A_{\alpha\eta}$ and $G^A_{\alpha\beta}$ is displayed in the right panels of Figure~\ref{fig2}, which indicates that the two bias-corrected estimators perform well and are close to the true value of  $\varphi_{\alpha\eta}$ and $\varphi_{\alpha\beta}$, respectively.

To verify that our results are independent of the signal sparsity, we set $m_{\alpha}=m_{\beta}=m_{\eta}=p\cdot a_\alpha$ and vary the sparsity $a_{\alpha}=0.01,0.02,0.05,0.1,0.2,0.5,0.6,0.7$ and $0.8$ to generate sparse and dense signals. 
Next, we fix $a_\alpha=0.2$ and set $m_{\beta}=m_{\eta}=k \cdot m_{\alpha}$ to allow phenotypes to have different number of causal SNPs, where $k=0.3,0.4,0.5,0.8,1,1.25,2,2.5$ and $3.3$. 
We set all heritability to one and let  $\varphi_{\alpha\eta}=\varphi_{\alpha\beta}=0.5$.
Sample size is set to either $2000$ or $10,000$. 
The performance of $G_{\alpha\eta}$ is displayed in the upper panels of Supplementary Figure~\ref{fig.s2}. The bias of $G_{\alpha\eta}$ is independent of the sparsity $a_{\alpha}$ of a trait or the ratio of sparsity $k$ between two traits, which verifies our results of Theorem~\ref{thm1}. 
The bottom panels of Supplementary Figure~\ref{fig.s2} display the performance of $G^A_{\alpha\eta}$. It is clear that $G^A_{\alpha\eta}$ is unbiased regardless of $a_{\alpha}$ and $k$. 
The Supplementary Figure~\ref{fig.s3} shows a similar pattern in $G^A_{\alpha\eta}$ as heritability $h^2_{\alpha}=h^2_{\eta}=0.5$.
The performance of $G_{\alpha\beta}$ and $G^A_{\alpha\beta}$ is displayed in Supplementary Figure~\ref{fig.s4} and supports our results in Theorem~\ref{thm2}.
Finally, we illustrate the  performance of  $\widehat{\varphi}_{\alpha\beta}$ and $\widehat{\varphi}^A_{\alpha\beta}$ in Supplementary Figure~\ref{fig.s5}, verifying our results in Theorem~\ref{thm3} and the unbiasedness of $\widehat{\varphi}^A_{\alpha\beta}$.
\subsection{SNP screening and overlapping samples}
Instead of using all the $10,000$ SNPs, we construct cross-trait PRS with the top-ranked SNPs whose GWAS $p$-values pass a pre-specified threshold. 
We consider a series of thresholds $\{1,0.8,$ $0.5,$ $0.4,$ $0.3,$ $0.2,$ $0.1,$ $0.08,$ $0.05,$ $0.02,$ $0.01,$ $10^{-3},$ $10^{-4},$ $10^{-5},$ $10^{-6},$ $10^{-7},$ $10^{-8}\}$ and generate a series of $G_{T\alpha\eta}$ accordingly. 
We set heritability to one and $\varphi_{\alpha\eta}=0.8$. 
Four levels of sparsity $m_{\alpha}/p=m_{\eta}/p=0.01, 0.1, 0.5$ and $0.8$ are examined. 
Figure~\ref{fig3} displays the performance of $G_{T\alpha\eta}$ across a series of thresholds.
As expected, the pattern of $G_{T\alpha\eta}$ varies dramatically with the sparsity.
When signals are sparse, SNP screening helps and $G_{T\alpha\eta}$ performs better than $G_{\alpha\eta}$. 
However, when signals are dense, the performance of $G_{T\alpha\eta}$ drops 
as the threshold decreases. $G_{T\alpha\eta}$ has the best performance as all SNPs are selected, i.e., the same as $G_{\alpha\eta}$, which confirms our results of $G_{T\alpha\eta}$ in Corollary~\ref{cor1}. 
In addition, we examine our analyses of overlapping samples. 
For $G_{S\alpha\eta}$ and $G_{S\alpha\beta}$, half of the $10,000$ samples are set to be overlapping.
Other settings remain the same as those of Figure~\ref{fig2}.
The performance of $G_{S\alpha\eta}$,  $G_{S\alpha\beta}$, $G^A_{S\alpha\eta}$ and $G^A_{S\alpha\beta}$
is displayed in Figure~\ref{fig4}, which fully support the results in Theorems~\ref{thm6}~-~\ref{thm7}.
\section{UK Biobank data analysis}\label{sec6}
We apply our bias-corrected estimator on the United Kingdom (UK)
Biobank data \citep{sudlow2015uk} to assess the genetic correlation between brain white matter (WM) tracts and several neuropsychiatric disorders. 
The structural changes of WM tracts are measured and quantified in diffusion tensor imaging (dMRI). 
We run the TBSS-ENIGMA pipeline \citep{thompson2014enigma} to generate tract-based diffusion tensor imaging (DTI) parameters from dMRI of UK Biobank samples.
Seven DTI parameters, FA, MD, MO, RD, L1, L2, and L3 (Supplementary Table~\ref{tab.s1}) are derived in each of the $18$ WM tracts (Supplementary Table~\ref{tab.s2}, Supplementary Figure~\ref{fig.s6}), thus there are $7\times 18=126$ DTI parameters in total.
We use the unimputed UK Biobank SNP data released in July 2017.
Detailed genetic data collection/processing procedures and quality control prior to the release of data are documented at \url{ http://www.ukbiobank.ac.uk/scientists-3/genetic-data/}.
We take all autosomal SNPs and apply the standard quality control procedures using the Plink tool set \citep{purcell2007plink}: excluding subjects with more than $10\%$ missing genotypes, only including SNPs with MAF $>0.01$, genotyping rate $> 90\%$, and passing Hardy-Weinberg test ($P > 1\times 10^{-7}$). The number of SNPs are $461,488$ after these steps.  
We further removed non-European subjects if any. To avoid including closely related relatives, we excluded one of any pair of individuals with estimated genetic relationship larger than $0.025$.  
We then select subjects that have DTI data as well, which yields a final dataset consisting of $7979$ UK Biobank samples with age range $[47,80]$ (mean=$64.26$ years, sd=$7.44$), and the proportion of female is $0.526$.  

Cross-trait PRSs of three psychiatric disorders are constructed on these UK Biobank samples by using their published GWAS summary statistics, including attention-deficit /hyperactivity disorder (ADHD, sample size $55,374$), bipolar disorder (BD, $41,653$), and Schizophrenia (SCZ, $65,967$).
The original GWAS \citep{demontis2017discovery,ruderfer2018genomic} have no overlapping samples with the UK Biobank data used in this study. 
The GWAS summary data of these disorders are downloaded from the Psychiatric Genomics Consortium \citep{sullivan2017psychiatric}. 
To obtain independent SNPs, we perform LD pruning
with $R^2=0.2$ and window size $50$.  
There are $230,072$ SNPs remain after LD pruning and they are used in later steps as candidates for constructing PRS. 
We generate one PRS separately for each disorder by summarizing across all the pruned candidates SNPs, weighed by their GWAS effect sizes (log odds ratios). 
The number of overlapping SNPs is $204,367$ for SCZ, $215,655$ for BD,  and $129,052$ for ADHD. 
Plink tool set \citep{purcell2007plink} is used to generate these scores. 
The association between each pair of PRS and DTI parameter is estimated and tested in linear regression, adjusting for age, sex and ten genetic principal components of the UK Biobank. 
There are $7\times 18\times 3=378$ tests and we correct for multiple testing using the false discovery rate (FDR) method \citep{storey2002direct} at $0.05$ level.  
\begin{figure}
\includegraphics[page=1,width=0.95\linewidth]{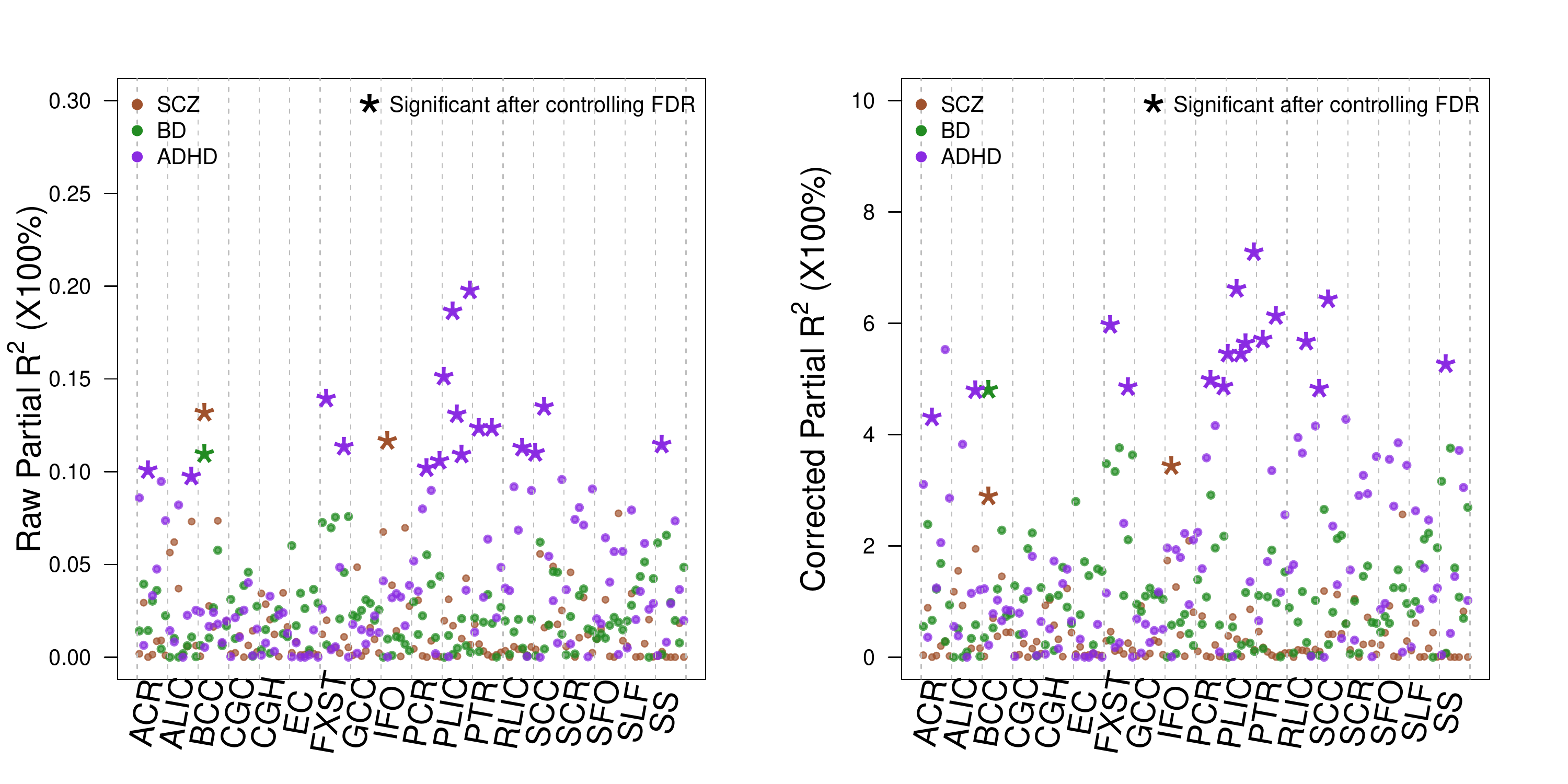}
  \caption{Raw partial $R^2$ of fitting psychiatric disorder PRS on $18$ brain WM tracts (listed in $x$ axis) in the UK Biobank data (left panel) and corrected ones based on our formulas (right panel). Each tract has seven DTI parameters (FA, MD, MO, RD, L1, L2, L3). 
Partial $R^2$ measures the variance in DTI parameter that can be explained by the PRS, adjusting for age, sex and ten genetic principal components of the UK Biobank.}
\label{fig5}
\end{figure}

We focus on the $20$ significant associations after controlling for FDR: $17$ for ADHD, $1$ for BD, and $2$ for SCZ (Supplementary Figure~\ref{fig.s7}, Supplementary Table~\ref{tab.s3}). 
On these significant DTI-Disorder pairs, 
the proportions of variation in DTI parameter that can be explained by PRS of disorder (partial $R^2$) are all less than $0.2\%$ (mean=$0.125\%$, max=$0.197\%$, left panel of Figure~\ref{fig5}). 
Partial $R^2$ is the square of the estimated genetic correlation between PRS and WM tract DTI parameters after adjusting for other covariates and is often interpreted as the genetic overlap or shared genetic etiology between the two traits. 
Such small $R^2$s are widely reported in similar studies for highly heritable psychiatric disorders \citep{clarke2016common,guo2017polygenic,mistry2018use,bogdan2018polygenic,power2015polygenic}. 

Next, we correct these estimates with our formula in Theorem~\ref{thm1}. 
We applied the heritability estimates of psychiatric disorders reported in a recent large-scale study \citep{anttila2018analysis}: $0.256$ for SCZ, $0.205$ for BD, and $0.100$ for ADHD. 
We estimate heritability of the $126$ DTI parameters with the individual-level UK Biobank data using the GCTA tool set \citep{yang2011gcta}. 
These heritability estimates range from $0.224$ to $0.733$ with mean=$0.532$ and sd=$0.087$, and are reported in  \cite{zhao2018large}. 
Plugging in these heritability estimates, sample sizes and number of  SNPs, 
the updated partial $R^2$s are much larger than previous ones (mean=$5.260\%$, max=$ 7.270\%$, right panel of Figure~\ref{fig5}). 
These corrected partial $R^2$s are within $4\%$ to $7.5\%$ for ADHD,
$2.5\%$ to $3.5\%$ for SCZ, and is $4.8\%$ for BD. 
In conclusion, we detect the significant association between genetic risk scores of psychiatric disorders and brain WM microstructure changes in UK Biobank participants sampled from the general population. 
Compared to the originally estimated partial $R^2$s, the corrected partial $R^2$s may better reflect the degree of genetic similarity between the two set of traits and suggest the potential prediction power of brain imaging markers on these disorders. 
\section{Discussion}\label{sec7}
Understanding the genetic similarity among human complex traits is essential to model biological mechanisms, improve genetic risk prediction, and design personalized prevention/treatment. 
Cross-trait PRS \citep{purcell2009common,power2015polygenic} is one of the most popular methods for genetic correlation estimation with thousands of publications. 
This paper empirically and theoretically studies the asymptotic properties of cross-trait PRS. 
Our analyses demystify the commonly observed small $R^2$ in real data applications, and help avoid over- or under-interpreting of research findings.
More importantly, the asymptotic bias is largely independent of the unknown genetic architecture if we use all SNPs in cross-trait PRS, which enables bias correction.
As the sample size of discovery GWAS becomes much larger in the last few years \citep{lee2018gene,evangelou2018genetic} and may keep on increasing in the future, our bias-corrected estimators can be used to recover the underlying genetic correlation of many complex traits.
We also discuss the popular SNP screening strategy and illustrate that this procedure may enlarge the bias for highly polygenic traits, and thus should be used with caution. 
Influence of overlapping samples is also quantified in several practical cases. 

The training-testing design employed by cross-trait PRS may help avoid the inflation caused by non-genetic correlation, but results in systematic bias due to the restricted prediction power of GWAS summary statistics in testing data.
The behavior of cross-trait PRS studied in this paper is closely related to the properties of GWAS summary statistics, which have received little scrutiny in statistical genetics. Our research should bring attentions to the potential unexpected results when analyzing summary-level data of different GWAS for polygenic traits, and call to thoroughly (re)study the statistical properties of other popular GWAS summary statistics-based methods. 

\section*{Acknowledgement}
We would like to thank Haiyan Deng from North Carolina State University for helpful discussion. 
This research was partially supported by U.S. NIH grants MH086633 and MH116527, and a grant from the Cancer Prevention Research Institute of Texas. 
We thank the individuals represented in the UK Biobank (\url{http://www.ukbiobank.ac.uk/}) for their participation and the research teams for their work in collecting, processing and disseminating these datasets for analysis. 
This research has been conducted using the UK Biobank resource (application number $22783$), subject to a data transfer agreement. We thank the Psychiatric Genomics Consortium (PGC, \url{http://www.med.unc.edu/pgc/}) for providing GWAS summary-level data used in the real data analysis.
The authors acknowledge the Texas Advanced Computing Center (TACC, \url{http://www.tacc.utexas.edu/}) at The University of Texas at Austin for providing HPC and storage resources that have contributed to the research results reported within this paper. 

\section*{Appendix~A: Proofs}
In this appendix, we highlight the key steps and results to prove our main theorems. More proofs and technical details can be found in the supplementary file.  
\begin{proposition.a}\label{pop.a1}
Under polygenic model~(\ref{equ2.1}) and Conditions~\ref{con1} and~\ref{con2}, if $m_{\alpha\eta},m_{\alpha}$, and 
$m_{\eta}$ $\rightarrow \infty$ as $\mbox{min}(n_1, n_3,p)\rightarrow\infty$, then we have 
\begin{flalign*}
&\frac{\big(\W_{(1)}\bmeta_{(1)}+\bmeps_{\eta}\big)^T\big(\W_{(1)}\bmeta_{(1)}+\bmeps_{\eta}\big)}{n_3m_{\eta}\cdot{\sigma^2_{\eta}}+n_3\cdot{\sigma^2_{\epsilon_{\eta}}}} =1 + o_p(1),  \\
&\frac{\big(\X_{(1)}\bmalpha_{(1)}+\bmeps_{\alpha}\big)^T\X\W^T\W\X^T\big(\X_{(1)}\bmalpha_{(1)}+\bmeps_{\alpha}\big)}{\{n_1n_3m_{\alpha}(p-m_{\alpha})+n_1n_3m_{\alpha}(m_{\alpha}+n_1)\}\cdot{\sigma^2_{\alpha}}+n_1n_3p\cdot{\sigma^2_{\epsilon_{\alpha}}}} = 1 + o_p(1) \mathbf{.}
\end{flalign*}
Further if $p/(n_1n_3) \to 0$, then we have 
\begin{flalign*}
\frac{\big(\W_{(1)}\bmeta_{(1)}+\bmeps_{\eta}\big)^T\W\X^T\big(\X_{(1)}\bmalpha_{(1)}+\bmeps_{\alpha}\big)}{n_1n_3m_{\alpha\eta}\cdot{\sigma_{\alpha\eta}}} =1 + o_p(1)\mathbf{.}
\end{flalign*}
\end{proposition.a}
By continuous mapping theorem, we have
\begin{flalign*}
G_{\alpha\eta}=\sqrt{\frac{n_1}{n_1+p/h^2_{\alpha}}}\cdot h_{\eta} \cdot  \varphi_{\alpha\eta} \cdot\{1 + o_p(1)\} \mathbf{.}
\end{flalign*}
Then Theorem~\ref{thm1} holds for $a \in(0,1)$. 
When $a \in[1,\infty]$, i.e., $p/(n_1n_3) \not\to 0$,  we note 
\begin{flalign*}
\big(\W_{(1)}\bmeta_{(1)}+\bmeps_{\eta}\big)^T\W\X^T\big(\X_{(1)}\bmalpha_{(1)}+\bmeps_{\alpha}\big)=
\bigO_p\big\{(n_1^{1/2}n_3^{1/2}m_{\alpha\eta}p^{1/2}+n_1n_3m_{\alpha\eta})\cdot\sigma_{\alpha\eta}\big\}\mathbf{.}
\end{flalign*}
It follows that 
\begin{flalign*}
G_{\alpha\eta}^2&=\frac{\bigO_p\big\{(n_1n_3m_{\alpha\eta}^2p+n_1^2n_3^2m_{\alpha\eta}^2)\cdot\sigma^2_{\alpha\eta}\big\}}
{\big[n_3m_{\eta} \cdot{\sigma^2_{\eta}}\cdot\{1+o(1)\} \big]\cdot \big[n_1n_3m_{\alpha}(n_1+p) \cdot{\sigma^2_{\alpha}} \cdot\{1+o(1)\}\big]}\\
&=\bigO_p \bigg\{ \frac{n_1n_3p+n_1^2n_3^2}{n_3^2n_1(n_1+p)}\cdot \varphi_{\alpha\eta}^2\bigg\}=\bigO_p(\frac{1}{n_3})\mathbf{.}
\end{flalign*}
Thus, Theorem~\ref{thm1} is proved. 

\begin{proposition.a}\label{pop.a2}
Under polygenic model~(\ref{equ2.1}) and Conditions~\ref{con1} and~\ref{con2}, 
if $m_{\alpha\beta},m_{\alpha}$, and 
$m_{\beta} \rightarrow \infty$  as $\mbox{min}(n_1,n_2, n_3,p)\rightarrow\infty$, then we have 
\begin{flalign*}
&\frac{\big(\Z_{(1)}\bmbeta_{(1)}+\bmeps_{\beta}\big)^T\Z\W^T\W\Z^T\big(\Z_{(1)}\bmbeta_{(1)}+\bmeps_{\beta}\big)}{\{n_2n_3m_{\beta}(p-m_{\beta})+n_2n_3m_{\beta}(m_{\beta}+n_2)\}\cdot{\sigma^2_{\beta}}+n_2n_3p\cdot{\sigma^2_{\epsilon_{\beta}}}} = 1 + o_p(1) \mathbf{.}
\end{flalign*}
Further if $p^2/(n_1n_2n_3) \to 0$, then we have 
\begin{flalign*}
\frac{\big(\Z_{(1)}\bmbeta_{(1)}+\bmeps_{\beta}\big)^T\Z\W^T\W\X^T\big(\X_{(1)}\bmalpha_{(1)}+\bmeps_{\alpha}\big)}{n_1n_2n_3m_{\alpha\beta}\cdot{\sigma_{\alpha\beta}}} =1 + o_p(1)\mathbf{.}
\end{flalign*}
\end{proposition.a}
It follows that Theorem~\ref{thm2} holds for $a \in(0,1)$.
When $a \in[1,\infty]$, we have 
\begin{flalign*}
G_{\alpha\beta}^2&=\frac{\bigO_p\big\{(n_1n_2n_3m_{\alpha\beta}^2p^2+n_1^2n_2^2n_3^2m_{\alpha\beta}^2)\cdot\sigma_{\alpha\beta}\big\}}
{\big[n_2n_3m_{\beta}(n_2+p) \cdot{\sigma^2_{\beta}} \cdot\{1+o(1)\} \big]\cdot \big[n_1n_3m_{\alpha}(n_1+p) \cdot{\sigma^2_{\alpha}} \cdot\{1+o(1)\}\big]}\\
&=\bigO_p\bigg\{ \frac{n_1n_2n_3p^2+n_1^2n_2^2n_3^2}{n_3^2n_1n_2(n_1+p)(n_2+p)}\bigg\}=\bigO_p\Big\{\frac{p^2}{n_3(n_1+p)(n_2+p)}\Big\}\mathbf{.}
\end{flalign*}
Thus, Theorem~\ref{thm2} is proved. 

\begin{proposition.a}\label{pop.a3}
Under polygenic model~(\ref{equ2.1}) and Conditions~\ref{con1} and~\ref{con2}, 
if $m_{\alpha\beta},m_{\alpha}$, and  
$m_{\beta} \rightarrow \infty$ as $\mbox{min}(n_1,n_2,p)\rightarrow\infty$, then we have 
\begin{flalign*}
&\frac{\big(\X_{(1)}\bmalpha_{(1)}+\bmeps_{\alpha}\big)^T\X\X^T\big(\X_{(1)}\bmalpha_{(1)}+\bmeps_{\alpha}\big)}
{\{n_1m_{\alpha}(n_1+m_{\alpha})+n_1m_{\alpha}(p-m_{\alpha})\}\cdot{\sigma^2_{\alpha}}+n_1p\cdot \sigma^2_{\epsilon_{\alpha}}} =1 + o_p(1),  \\
&\frac{\big(\Z_{(1)}\bmbeta_{(1)}+\bmeps_{\beta}\big)^T\Z\Z^T\big(\Z_{(1)}\bmbeta_{(1)}+\bmeps_{\beta}\big)}
{\big\{n_2m_{\beta}(n_2+m_{\beta})+n_2m_{\beta}(p-m_{\beta})\big\}\cdot{\sigma^2_{\beta}}+n_2p\cdot \sigma^2_{\epsilon_{\beta}}
} =1 + o_p(1)\mathbf{.}
\end{flalign*}
Further if $p/(n_1n_2) \to 0$, then  we have 
\begin{flalign*}
\frac{\big(\X_{(1)}\bmalpha_{(1)}+\bmeps_{\alpha}\big)^T\X\Z^T\big(\Z_{(1)}\bmbeta_{(1)}+\bmeps_{\beta}\big)}{n_1n_2m_{\alpha\beta}\cdot{\sigma_{\alpha\beta}}} =1 + o_p(1) \mathbf{.}
\end{flalign*}
\end{proposition.a}
Therefore Theorem~\ref{thm3} holds for $a\in(0,1)$.
When $a \in[1,\infty]$, we have 
\begin{flalign*}
\widehat{\varphi}_{\alpha\beta}^2&=\frac{\bigO_p\big\{(n_1n_2m_{\alpha\beta}^2p+n_1^2n_2^2m_{\alpha\beta}^2)\cdot \sigma_{\alpha\beta}\big\}}
{\big[n_2m_{\beta}(n_2+p) \cdot{\sigma^2_{\beta}} \cdot\{1+o(1)\}\big]\cdot \big[n_1m_{\alpha}(n_1+p) \cdot{\sigma^2_{\alpha}} \cdot\{1+o(1)\}\big]}\\
&=\bigO_p\bigg\{ \frac{n_1n_2p+n_1^2n_2^2}{n_1n_2(n_1+p)(n_2+p)}\bigg\} =\bigO_p\Big\{\frac{p}{(n_1+p)(n_2+p)}\Big\} \mathbf{.}
\end{flalign*}
Thus, Theorem~\ref{thm3} is proved. 
Corollaries~\ref{cor1}~and~\ref{cor2} follow from the two  propositions below. The proofs of overlapping samples can be found in the supplementary file.
\begin{proposition.a}\label{pop.a4}
Under polygenic model~(\ref{equ2.1}) and Conditions~\ref{con1} and~\ref{con2}, if $\mbox{min}(m_{\alpha\eta},m_{\alpha}$, 
$m_{\eta})\rightarrow \infty$, $\mbox{min}(q_{\alpha\eta},q_{\alpha 1},q_{\alpha 2})\rightarrow \infty$ when $\mbox{min}(n_1, n_3,p)\rightarrow\infty$, then we have 
\begin{flalign*}
&\frac{V_{T\alpha}}
{\{n_1n_3m_{\alpha}q_{\alpha2}+n_1n_3q_{\alpha1}(m_{\alpha}+n_1)\}\cdot{\sigma^2_{\alpha}}+
n_1n_3q_{\alpha}\cdot \sigma^2_{\epsilon_{\alpha}}} = 1 + o_p(1) \mathbf{.}
\end{flalign*}
Further if $\{m_{\alpha\eta}^2(q_{\alpha 1}+q_{\alpha 2})\}/(q_{\alpha\eta}^2n_1n_3) \to 0$, then we have 
\begin{flalign*}
\frac{C_{T\alpha\eta}}{n_1n_3q_{\alpha\eta}\cdot{\sigma_{\alpha\eta}}} =1 + o_p(1)\mathbf{.}
\end{flalign*}
\end{proposition.a}

\begin{proposition.a}\label{pop.a5}
Under polygenic model~(\ref{equ2.1}) and Conditions~\ref{con1} and~\ref{con2}, 
if $\mbox{min}(m_{\alpha\beta},m_{\alpha}$,
$m_{\beta}) \rightarrow \infty$, 
$\mbox{min}(q_{\alpha\beta},q_{\alpha 1},q_{\alpha 2},q_{\beta 1},q_{\beta 2})\rightarrow \infty$ when $\mbox{min}(n_1,n_2, n_3,p)\rightarrow\infty$, then we have 
\begin{flalign*}
&\frac{V_{T\beta}}
{\{n_2n_3m_{\beta}q_{\beta2}+n_2n_3q_{\beta1}(m_{\beta}+n_2)\}\cdot{\sigma^2_{\beta}}+
n_1n_3q_{\beta}\cdot \sigma^2_{\epsilon_{\beta}}} = 1 + o_p(1)\mathbf{.}
\end{flalign*}
Further if $\{m^2_{\alpha\beta}(q_{\alpha1}+q_{\alpha2})(q_{\beta1}+q_{\beta2})\}/(q_{\alpha\beta}^2n_1n_2n_3)\to 0$, then we have 
\begin{flalign*}
\frac{C_{T\alpha\beta}}{n_1n_2n_3q_{\alpha\beta}\cdot{\sigma_{\alpha\beta}}} =1 + o_p(1)\mathbf{.}
\end{flalign*}
\end{proposition.a}
\bibliographystyle{rss}
\bibliography{sample.bib}
\clearpage
\section*{\LARGE Supplementary Material}\label{sec8}
\section{More proofs}
\begin{proposition.s}\label{pop.o1}
Under polygenic model~(\ref{equ2.1}) and Conditions~\ref{con1}~-~\ref{con3}, suppose $m_{\alpha\eta},m_{\alpha}$, and 
$m_{\eta}\rightarrow \infty$ as $(n_1+n_s), (n_3+n_s),p\rightarrow\infty$,  then we have 
\begin{flalign*}
\frac{\y_{\eta}^T\y_{\eta}}{(n_3+n_s)m_{\eta}\cdot{\sigma^2_{\eta}}+(n_3+n_s)\cdot{\sigma^2_{\epsilon_{\eta}}}} =1 + o_p(1)  
\qquad \text{and} \qquad
\frac{\widehat{\bmS}_{S\alpha}^T\widehat{\bmS}_{S\alpha}}{v_{S\alpha}} = 1 + o_p(1), 
\end{flalign*}
where 
\begin{flalign*}
\y_{\eta}^T\y_{\eta}=
\big(\W_{(1,\eta)}\bmeta_{(1)}+\bmeps_{\eta w}\big)^T\big(\W_{(1,\eta)}\bmeta_{(1)}+\bmeps_{\eta w}\big)+
\big(\bmS_{(1,\eta)}\bmeta_{(1)}+\bmeps_{\eta s}\big)^T\big(\bmS_{(1,\eta)}\bmeta_{(1)}+\bmeps_{\eta s}\big),
\end{flalign*}
\begin{flalign*}
\widehat{\bmS}_{S\alpha}^T\widehat{\bmS}_{S\alpha}=&
\big(\X_{(1,\alpha)}\bmalpha_{(1)}+\bmeps_{\alpha x}\big)^T\X\W^T\W\X^T\big(\X_{(1,\alpha)}\bmalpha_{(1)}+\bmeps_{\alpha x}\big)+&\\
&2\big(\X_{(1,\alpha)}\bmalpha_{(1)}+\bmeps_{\alpha x}\big)^T\X\W^T\W\bmS^T\big(\bmS_{(1,\alpha)}\bmalpha_{(1)}+\bmeps_{\alpha s}\big)+&\\
&\big(\bmS_{(1,\alpha)}\bmalpha_{(1)}+\bmeps_{\alpha s}\big)^T\bmS\W^T\W\bmS^T\big(\bmS_{(1,\alpha)}\bmalpha_{(1)}+\bmeps_{\alpha s}\big)+&\\
&\big(\X_{(1,\alpha)}\bmalpha_{(1)}+\bmeps_{\alpha x}\big)^T\X\bmS^T\bmS\X^T\big(\X_{(1,\alpha)}\bmalpha_{(1)}+\bmeps_{\alpha x}\big)+&\\
&2\big(\X_{(1,\alpha)}\bmalpha_{(1)}+\bmeps_{\alpha x}\big)^T\X\bmS^T\bmS\bmS^T\big(\bmS_{(1,\alpha)}\bmalpha_{(1)}+\bmeps_{\alpha s}\big)+&\\
&\big(\bmS_{(1,\alpha)}\bmalpha_{(1)}+\bmeps_{\alpha s}\big)^T\bmS\bmS^T\bmS\bmS^T\big(\bmS_{(1,\alpha)}\bmalpha_{(1)}+\bmeps_{\alpha s}\big),
\end{flalign*}
and 
\begin{flalign*}
v_{S\alpha}=&\{n_1n_3m_{\alpha}(p+n_1)\cdot \sigma^2_{\alpha}+n_1n_3p\cdot\sigma^2_{\epsilon_{\alpha} }\}+2\{n_1n_3n_sm_{\alpha}\cdot\sigma^2_{\alpha}\}+\\
&\{n_sn_3m_{\alpha}(p+n_s)\cdot\sigma^2_{\alpha}+n_sn_3p\cdot\sigma^2_{\epsilon_{\alpha} }\}+\{n_1n_sm_{\alpha}(p+n_1)\cdot \sigma^2_{\alpha}+n_1n_sp\cdot\sigma^2_{\epsilon_{\alpha} }\}\\
&+2\{n_1n_sm_{\alpha}(n_s+p)\cdot\sigma^2_{\alpha}\}+\{n_sm_{\alpha}(n^2_s+p^2+3n_sp)\cdot \sigma^2_{\alpha}+n_sp(n_s+p)\cdot\sigma^2_{\epsilon_{\alpha}}\} \mathbf{.}
\end{flalign*}
Further if $p/\{(n_1+n_S)(n_3+n_s)\} \to 0$, then we have 
\begin{flalign*}
\frac{\y_{\eta}^T\widehat{\bmS}_{S\alpha}}{(n_1+n_s)n_3m_{\alpha\eta}\cdot{\sigma_{\alpha\eta}}+
\{n_sm_{\alpha\eta}(p+n_s)\cdot\sigma_{\alpha\eta}+n_sp\cdot\sigma_{\epsilon_{\alpha}\epsilon_{\eta}}\}+
n_sn_1m_{\alpha\eta}\cdot\sigma_{\alpha\eta}
} =1 + o_p(1),
\end{flalign*}
where $\y_{\eta}^T\widehat{\bmS}_{S\alpha}$ is given by 
\begin{flalign*}
&\big(\bmeps_{\eta w}^T+\bmeta_{(1)}^T\W_{(1,\eta)}^T\big)\W\X^T\big(\X_{(1,\alpha)}\bmalpha_{(1)}+\bmeps_{\alpha x}\big)+ 
\big(\bmeps_{\eta w}^T+\bmeta_{(1)}^T\W_{(1,\eta)}^T\big)\W\bmS^T\big(\bmS_{(1,\alpha)}\bmalpha_{(1)}+\bmeps_{\alpha s}\big)\\
&+\big(\bmeps_{\eta s}^T+\bmeta_{(1)}^T\bmS_{(1,\eta)}^T\big)\bmS\bmS^T\big(\bmS_{(1,\alpha)}\bmalpha_{(1)}+\bmeps_{\alpha s}\big)+
\big(\bmeps_{\eta s}^T+\bmeta_{(1)}^T\bmS_{(1,\eta)}^T\big)\bmS\X^T\big(\X_{(1,\alpha)}\bmalpha_{(1)}+\bmeps_{\alpha x}\big)\mathbf{.}
\end{flalign*}
\end{proposition.s}

\begin{proposition.s}\label{pop.o2}
Under polygenic model~(\ref{equ2.1}) and Conditions~\ref{con1},~\ref{con2} and~\ref{con4}, 
suppose $m_{\alpha\beta},m_{\alpha}$, and 
$m_{\beta} \rightarrow \infty$ as $(n_1+n_s),(n_2+n_s), n_3,p\rightarrow\infty$,  then we have 
\begin{flalign*}
&\frac{\widehat{\bmS}_{S\alpha}^T\widehat{\bmS}_{S\alpha}}{v_{S\alpha}} = 1 + o_p(1)
\qquad \text{and} \qquad
\frac{\widehat{\bmS}_{S\beta}^T\widehat{\bmS}_{S\beta}}{v_{S\beta}} = 1 + o_p(1), 
\end{flalign*}
where 
\begin{flalign*}
\widehat{\bmS}_{S\alpha}^T\widehat{\bmS}_{S\alpha}=&\big(\X_{(1,\alpha)}\bmalpha_{(1)}+\bmeps_{\alpha x}\big)^T\X\W^T\W\X^T\big(\X_{(1,\alpha)}\bmalpha_{(1)}+\bmeps_{\alpha x}\big)+\\
&2\big(\X_{(1,\alpha)}\bmalpha_{(1)}+\bmeps_{\alpha x}\big)^T\X\W^T\W\bmS^T\big(\bmS_{(1,\alpha)}\bmalpha_{(1)}+\bmeps_{\alpha s}\big)+\\
&\big(\bmS_{(1,\alpha)}\bmalpha_{(1)}+\bmeps_{\alpha s}\big)^T\bmS\W^T\W\bmS^T\big(\bmS_{(1,\alpha)}\bmalpha_{(1)}+\bmeps_{\alpha s}\big),
\end{flalign*}
\begin{flalign*}
\widehat{\bmS}_{S\beta}^T\widehat{\bmS}_{S\beta}=&\big(\Z_{(1,\beta)}\bmbeta_{(1)}+\bmeps_{\beta z}\big)^T\Z\W^T\W\Z^T\big(\Z_{(1,\beta)}\bmbeta_{(1)}+\bmeps_{\beta z}\big)+\\
&2\big(\Z_{(1,\beta)}\bmbeta_{(1)}+\bmeps_{\beta z}\big)^T\Z\W^T\W\bmS^T\big(\bmS_{(1,\beta)}\bmbeta_{(1)}+\bmeps_{\beta s}\big)+\\
&\big(\bmS_{(1,\beta)}\bmbeta_{(1)}+\bmeps_{\beta s}\big)^T\bmS\W^T\W\bmS^T\big(\bmS_{(1,\beta)}\bmbeta_{(1)}+\bmeps_{\beta s}\big),
\end{flalign*}
\begin{flalign*}
v_{S\alpha}=&n_1n_3m_{\alpha}(p+n_1)\cdot\sigma^2_{\alpha}+n_1n_3p\cdot \sigma^2_{\epsilon_{\alpha}}+
2n_1n_3n_sm_{\alpha}\cdot \sigma^2_{\alpha}+ \\
&n_sn_3m_{\alpha}(p+n_s)\cdot\sigma^2_{\alpha}+n_sn_3p\cdot \sigma^2_{\epsilon_{\alpha}},
\end{flalign*}
and 
\begin{flalign*}
v_{S\beta}=&n_2n_3m_{\beta}(p+n_2)\cdot\sigma^2_{\beta}+n_2n_3p\cdot \sigma^2_{\epsilon_{\beta}}+
2n_2n_3n_sm_{\beta}\cdot \sigma^2_{\beta}+ \\
&n_sn_3m_{\beta}(p+n_s)\cdot\sigma^2_{\beta}+n_sn_3p\cdot \sigma^2_{\epsilon_{\beta}}\mathbf{.}
\end{flalign*}
Further if $p^2/\{(n_1+n_s)(n_2+n_s)n_3\} \to 0$, then we have 
\begin{flalign*}
\frac{\widehat{\bmS}_{S\alpha}^T\widehat{\bmS}_{S\beta}}{v_{S\alpha\beta}
} =1 + o_p(1),
\end{flalign*}
where 
\begin{flalign*}
\widehat{\bmS}_{S\alpha}^T\widehat{\bmS}_{S\beta}&=\big(\X_{(1,\alpha)}\bmalpha_{(1)}+\bmeps_{\alpha x}\big)^T\X\W^T\W\Z^T\big(\Z_{(1,\beta)}\bmbeta_{(1)}+\bmeps_{\beta z}\big)+\\
&\big(\X_{(1,\alpha)}\bmalpha_{(1)}+\bmeps_{\alpha x}\big)^T\X\W^T\W\bmS^T\big(\bmS_{(1,\beta)}\bmbeta_{(1)}+\bmeps_{\beta s}\big)+\\
&\big(\bmS_{(1,\alpha)}\bmalpha_{(1)}+\bmeps_{\alpha s}\big)^T\bmS\W^T\W\Z^T\big(\Z_{(1,\beta)}\bmbeta_{(1)}+\bmeps_{\beta z}\big)+\\
&\big(\bmS_{(1,\alpha)}\bmalpha_{(1)}+\bmeps_{\alpha s}\big)^T\bmS\W^T\W\bmS^T\big(\bmS_{(1,\beta)}\bmbeta_{(1)}+\bmeps_{\beta s}\big),
\end{flalign*}
and 
\begin{flalign*}
v_{S\alpha\beta}=&n_1n_2n_3m_{\alpha\beta}\cdot{\sigma_{\alpha\beta}}+n_1n_sn_3m_{\alpha\beta}\cdot{\sigma_{\alpha\beta}}+
n_sn_2n_3m_{\alpha\beta}\cdot\sigma_{\alpha\beta}+\\
&\{n_s(p+n_s)n_3m_{\alpha\beta}\cdot\sigma_{\alpha\beta}+n_sn_3p\cdot\sigma_{\epsilon_{\alpha}\epsilon_{\beta}}\}\mathbf{.}
\end{flalign*}
\end{proposition.s}
Then Theorems~\ref{thm6}~and~\ref{thm7} follow from continuous mapping theorem and similar arguments in the Appendix~A. 
\section{More overlapping cases}
This section provides more analyses on the overlapping samples. We consider several additional cases that might occur in real data applications.
\subsection*{Case iii)}
When the two GWAS are fully overlapped, i.e., the two set of summary statistics $\widehat{\bmalpha}$  and $\widehat{\bmbeta}$ are generated from the same GWAS data 
\begin{itemize}
\item Dataset~VII: $(\X,\y_{\alpha},\y_{\beta})$, with 
$\X=[\X_{(1,\alpha)},\X_{(2,\alpha)}]=[\X_{(1,\beta)},\X_{(2,\beta)}]
\in \bbR^{n_{1}\times p}$,
$\X_{(1,\alpha)} \in \bbR^{n_{1} \times m_{\alpha}}$, 
$\X_{(1,\beta)} \in \bbR^{n_{1} \times m_{\beta}}$, 
$\y_{\alpha} \in \bbR^{n_{1} \times 1}$, and $\y_{\beta} \in \bbR^{n_{1} \times 1}$. 
\end{itemize}
We assume that $\y_{\alpha}$ and $\y_{\beta}$ have polygenic architectures. 
We estimate 
$\varphi_{\alpha\beta}$ directly by estimating the correlation of $\widehat{\bmalpha}$ and $\widehat{\bmbeta}$ 
\begin{flalign*}
&\widehat{\varphi}_{X\alpha\beta}=\frac{\widehat{\bmalpha}^T\widehat{\bmbeta}}{\big\Vert\widehat{\bmalpha}\big\Vert \cdot \big\Vert\widehat{\bmbeta}\big\Vert}\\
&=\frac{\big(\X_{(1,\alpha)}\bmalpha_{(1)}+\bmeps_{\alpha}\big)^T\X\X^T\big(\X_{(1,\beta)}\bmbeta_{(1)}+\bmeps_{\beta}\big)}{
\big\{
\big(\X_{(1,\alpha)}\bmalpha_{(1)}+\bmeps_{\alpha}\big)^T\X\X^T\big(\X_{(1,\alpha)}\bmalpha_{(1)}+\bmeps_{\alpha}\big)\big\}^{1/2}
\big\{
\big(\X_{(1,\beta)}\bmbeta_{(1)}+\bmeps_{\beta}\big)^T\X\X^T\big(\X_{(1,\beta)}\beta_{(1)}+\bmeps_{\beta}\big)
\big\}^{1/2}} \mathbf{.}
\end{flalign*}
\begin{proposition.s}\label{pops1}
Under polygenic models and Conditions~\ref{con1},~\ref{con2} and~\ref{con4}, 
if $m_{\alpha\beta},m_{\alpha}$, and $m_{\beta}\rightarrow \infty$, as $n_1,p\rightarrow\infty$, then we have 
\begin{flalign*}
&\frac{\big(\X_{(1,\alpha)}\bmalpha_{(1)}+\bmeps_{\alpha}\big)^T\X\X^T\big(\X_{(1,\alpha)}\bmalpha_{(1)}+\bmeps_{\alpha}\big)}
{\{n_1m_{\alpha}(n_1+m_{\alpha})+n_1m_{\alpha}(p-m_{\alpha})\}\cdot{\sigma^2_{\alpha}}+n_1p\cdot \sigma^2_{\epsilon_{\alpha}}} =1 + o_p(1),  \\
&\frac{\big(\X_{(1,\beta)}\bmbeta_{(1)}+\bmeps_{\beta}\big)^T\X\X^T\big(\X_{(1,\beta)}\bmbeta_{(1)}+\bmeps_{\beta}\big)}
{\{n_1m_{\beta}(n_1+m_{\beta})+n_1m_{\beta}(p-m_{\beta})\}\cdot{\sigma^2_{\beta}}+n_1p\cdot \sigma^2_{\epsilon_{\beta}}
} =1 + o_p(1),
\end{flalign*}
and
\begin{flalign*}
\frac{\big(\X_{(1,\alpha)}\bmalpha_{(1)}+\bmeps_{\alpha}\big)^T\X\X^T\big(\X_{(1,\beta)}\bmbeta_{(1)}+\bmeps_{\beta}\big)}{n_1m_{\alpha\beta}(p+n_1)\cdot{\sigma_{\alpha\beta}}+n_1p\cdot \sigma_{\epsilon_{\alpha\beta}}} =1 + o_p(1) \mathbf{.}
\end{flalign*}
Thus, we have 
\begin{flalign*}
\widehat{\varphi}_{X\alpha\beta}=\frac{n_1+p/h_{\alpha\beta}}{(n_1+p/h_{\alpha}^2)^{1/2}(n_1+p/h_{\beta}^2)^{1/2}} \cdot \varphi_{\alpha\beta} \cdot\{1 + o_p(1)\}\mathbf{.}
\end{flalign*}
It follows that $\widehat{\varphi}_{X\alpha\beta}$ is asymptotically unbiased as
$h^2_{\alpha}=h^2_{\beta}=h_{\alpha\beta}=1$. Otherwise, $\widehat{\varphi}_{X\alpha\beta}$ may be biased towards zero.
\end{proposition.s}
\subsection*{Case iv)}
Again, the two set of summary statistics $\widehat{\bmalpha}$ and $\widehat{\bmbeta}$ are generated from the same GWAS dataset 
\begin{itemize}
\item Dataset~VII: $(\X,\y_{\alpha},\y_{\beta})$, with
$\X=[\X_{(1,\alpha)},\X_{(2,\alpha)}]=[\X_{(1,\beta)},\X_{(2,\beta)}]
\in \bbR^{n_{1}\times p}$,
$\X_{(1,\alpha)} \in \bbR^{n_{1} \times m_{\alpha}}$, 
$\X_{(1,\beta)} \in \bbR^{n_{1} \times m_{\beta}}$, 
$\y_{\alpha} \in \bbR^{n_{1} \times 1}$, and $\y_{\beta} \in \bbR^{n_{1} \times 1}$.
\end{itemize}
And we construct two PRSs $\widehat{\bmS}_{X\alpha}$ and $\widehat{\bmS}_{X\beta}$ on $\X$.

\begin{proposition.s}\label{pop.s2}
Under polygenic models and Conditions~\ref{con1},~\ref{con2} and~\ref{con4}, 
if $m_{\alpha\beta},m_{\alpha}$, and $m_{\beta}\rightarrow \infty$, as $n_1,p\rightarrow\infty$, then we have 
\begin{flalign*}
&\frac{\widehat{\bmS}_{X\alpha}^T\widehat{\bmS}_{X\alpha}}{v_{X\alpha}} = 1 + o_p(1)
\qquad  \text{and}  \qquad
\frac{\widehat{\bmS}_{X\beta}^T\widehat{\bmS}_{X\beta}}{v_{X\beta}} = 1 + o_p(1), 
\end{flalign*}
where 
\begin{flalign*}
&\widehat{\bmS}_{X\alpha}^T\widehat{\bmS}_{X\alpha}=(\X_{(1,\alpha)}\bmalpha_{(1)}+\bmeps_{\alpha})^T\X\X^T\X\X^T(\X_{(1,\alpha)}\bmalpha_{(1)}+\bmeps_{\alpha}),\\
&\widehat{\bmS}_{X\beta}^T\widehat{\bmS}_{X\beta}=(\X_{(1,\beta)}\bmbeta_{(1)}+\bmeps_{\beta})^T\X\X^T\X\X^T(\X_{(1,\beta)}\bmbeta_{(1)}+\bmeps_{\beta}),\\
&v_{X\alpha}=n_1m_{\alpha}\{(n_1+p)^2+n_1p\}\cdot \sigma^2_{\alpha}+n_1p(n_1+p) \cdot \sigma^2_{\epsilon_{\alpha}}, \quad \mbox{and} \\
&v_{X\beta}=n_1m_{\beta}\{(n_1+p)^2+n_1p\}\cdot \sigma^2_{\beta}+n_1p(n_1+p) \cdot \sigma^2_{\epsilon_{\beta}}. 
\end{flalign*}
Similarly, we have 
\begin{flalign*}
\frac{\widehat{\bmS}_{X\beta}^T\widehat{\bmS}_{X\alpha}}{v_{X\alpha\beta}}=1 + o_p(1),
\end{flalign*}
where $\quad v_{X\alpha\beta}=n_1m_{\alpha\beta}\{(n_1+p)^2+n_1p\}\cdot \sigma_{\alpha\beta}+n_1p(n_1+p) \cdot \sigma_{\epsilon_{\alpha\beta}} \quad$ and 
\begin{flalign*}
&\widehat{\bmS}_{X\beta}^T\widehat{\bmS}_{X\alpha}=(\X_{(1,\beta)}\bmbeta_{(1)}+\bmeps_{\beta})^T\X\X^T\X\X^T(\X_{(1,\alpha)}\bmalpha_{(1)}+\bmeps_{\alpha})
\mathbf{.}
\end{flalign*}
Thus, we have 
\begin{flalign*}
G_{X\alpha\beta}=\frac{n_1^2+2n_1p+p(n_1+p)/h_{\alpha\beta}}{\big\{n_1^2+2n_1p+p(n_1+p)/h_{\alpha}^2\big\}^{1/2}\big\{n_1^2+2n_1p+p(n_1+p)/h_{\beta}^2\big\}^{1/2}} \cdot \varphi_{\alpha\beta} \cdot\{1 + o_p(1)\} \mathbf{.}
\end{flalign*}
It follows that $G_{X\alpha\beta}$ is asymptotically unbiased if
$h^2_{\alpha}=h^2_{\beta}=h_{\alpha\beta}=1$.  Otherwise, $G_{X\alpha\beta}$ may be biased towards zero.
\end{proposition.s}

\subsection*{Case v)}
The two set of GWAS summary statistics $\widehat{\bmalpha}$ and $\widehat{\bmbeta}$ are generated from the following two independent datasets: 
\begin{itemize}
\item Dataset~VIII: $(\X,\y_{\alpha})$, with $\X=[\X_{(1)},\X_{(2)}] \in \bbR^{n_{1}\times p}$, $\X_{(1)} \in \bbR^{n_{1} \times m_{\alpha}}$, and $\y_{\alpha} \in \bbR^{n_{1} \times 1}$.
\end{itemize}
\begin{itemize}
\item Dataset~IX: $(\Z,\y_{\beta})$, with $\Z=[\Z_{(1)},\Z_{(2)}] \in \bbR^{n_{2}\times p}$, $\Z_{(1)} \in \bbR^{n_{2} \times m_{\beta}}$, and $\y_{\beta} \in \bbR^{n_{2} \times 1}$.
\end{itemize}
We construct two PRSs $\widehat{\bmS}_{X\alpha}$ and $\widehat{\bmS}_{X\beta}$ on $\X$. 

\begin{proposition.s}\label{pop.s3}
Under polygenic models and Conditions~\ref{con1},~\ref{con2} and~\ref{con4}, 
if $m_{\alpha\beta},m_{\alpha}$, and $m_{\beta}\rightarrow \infty$, as $n_1,n_2,p\rightarrow\infty$, then we have 
\begin{flalign*}
&\frac{\widehat{\bmS}_{X\alpha}^T\widehat{\bmS}_{X\alpha}}{v_{X\alpha}} = 1 + o_p(1)
\qquad  \text{and} \qquad
\frac{\widehat{\bmS}_{X\beta}^T\widehat{\bmS}_{X\beta}}{v_{X\beta}} = 1 + o_p(1) ,
\end{flalign*}
where 
\begin{flalign*}
&\widehat{\bmS}_{X\alpha}^T\widehat{\bmS}_{X\alpha}=\big(\X_{(1,\alpha)}\bmalpha_{(1)}+\bmeps_{\alpha}\big)^T\X\X^T\X\X^T\big(\X_{(1,\alpha)}\bmalpha_{(1)}+\bmeps_{\alpha}\big),\\
&\widehat{\bmS}_{X\beta}^T\widehat{\bmS}_{X\beta}=\big(\Z_{(1,\beta)}\bmbeta_{(1)}+\bmeps_{\beta}\big)^T\Z\X^T\X\Z^T\big(\Z_{(1,\beta)}\bmbeta_{(1)}+\bmeps_{\beta}\big),\\
&v_{X\alpha}=n_1m_{\alpha}\{(n_1+p)^2+n_1p\}\cdot \sigma^2_{\alpha}+n_1p(n_1+p) \cdot \sigma^2_{\epsilon_{\alpha}},\quad \mbox{and} \\
&v_{X\beta}=n_1n_2m_{\beta}(p+n_2)\cdot \sigma^2_{\beta}+n_1n_2p \cdot \sigma^2_{\epsilon_{\beta}}.
\end{flalign*}
Further if $p/(n_1n_2) \to 0$, then we have 
\begin{flalign*}
\frac{\widehat{\bmS}_{X\beta}^T\widehat{\bmS}_{X\alpha}}{v_{X\alpha\beta}}=1 + o_p(1), 
\end{flalign*}
where $\quad v_{X\alpha\beta}=n_1n_2m_{\alpha\beta}(n_1+p)\cdot \sigma_{\alpha\beta} \quad$ and 
\begin{flalign*}
\widehat{\bmS}_{X\beta}^T\widehat{\bmS}_{X\alpha}=\big(\X_{(1,\alpha)}\bmalpha_{(1)}+\bmeps_{\alpha}\big)^T\X\X^T\X\Z^T\big(\Z_{(1,\beta)}\bmbeta_{(1)}+\bmeps_{\beta}\big).
\end{flalign*}
Let $p=c \cdot (n_1n_2)^{a}$ for some constants $c>0$ and $a  \in (0,\infty]$. As $n_1$ and $n_2$ increase to $\infty$, if $a \in (0,1)$, then we have 
\begin{flalign*}
G_{X\alpha\beta}=\frac{(n_1+p)\cdot n_2^{1/2}}{\big\{n_1^2+2n_1p+p(n_1+p)/h_{\alpha}^2\big\}^{1/2}\big\{n_2+p/h^2_{\beta}\big\}^{1/2}} \cdot \varphi_{\alpha\beta} \cdot\{1 + o_p(1)\}\mathbf{.}
\end{flalign*}
\end{proposition.s}

\section{Intermediate results}
\subsubsection*{Cross-trait PRS with all SNPs}
\begin{proposition.s}\label{pop.i1}
Under polygenic model~(\ref{equ2.1}) and Conditions~\ref{con1} and~\ref{con2}, if $m_{\alpha\eta},m_{\alpha}$, and 
$m_{\eta}$ $\rightarrow \infty$ as  $n_1, n_3,p\rightarrow\infty$, then we have 
\begin{flalign*}
&\tE \Big\{\big(\W_{(1)}\bmeta_{(1)}+\bmeps_{\eta}\big)^T\W\X^T\big(\X_{(1)}\bmalpha_{(1)}+\bmeps_{\alpha}\big) \Big\} = n_1n_3m_{\alpha\eta} \cdot{\sigma_{\alpha\eta}},& \\
&\var\Big\{\big(\W_{(1)}\bmeta_{(1)}+\bmeps_{\eta}\big)^T\W\X^T\big(\X_{(1)}\bmalpha_{(1)}+\bmeps_{\alpha}\big)\Big\}
=\{(n_1n_3m_{\alpha\eta}^2p+&\\
&\qquad 2n_1^2n_3m_{\alpha\eta}^2+
2n_1n_3^2m_{\alpha\eta}^2)\cdot \sigma^2_{\alpha\eta}+n_1^2n_3^2m_{\alpha\eta}\cdot (a_{22}-\sigma^2_{\alpha\eta})\}\cdot\{1+o(1)\}, &\\~\\
&\tE\Big\{\big(\W_{(1)}\bmeta_{(1)}+\bmeps_{\eta}\big)^T\big(\W_{(1)}\bmeta_{(1)}+\bmeps_{\eta}\big)\Big\}=n_3m_{\eta} \cdot{\sigma^2_{\eta}}+ n_3\cdot{\sigma^2_{\epsilon_{\eta}}}, &\\
&\var\Big\{\big(\W_{(1)}\bmeta_{(1)}+\bmeps_{\eta}\big)^T\big(\W_{(1)}\bmeta_{(1)}+\bmeps_{\eta}\big)\Big\}=o(n_3^2m_{\eta}^2\cdot{\sigma^4_{\eta}}),&\\~\\
&\tE\Big\{\big(\X_{(1)}\bmalpha_{(1)}+\bmeps_{\alpha}\big)^T\X\W^T\W\X^T\big(\X_{(1)}\bmalpha_{(1)}+\bmeps_{\alpha}\big)\Big\}&\\
&=\big[n_1n_3m_{\alpha}(n_1+m_{\alpha}) \cdot\{1+o(1)\} + n_1n_3m_{\alpha}(p-m_{\alpha})\big]\cdot{\sigma^2_{\alpha}}+n_1n_3p \cdot \sigma^2_{\epsilon_{\alpha}},&\\
&\var\Big\{\big(\X_{(1)}\bmalpha_{(1)}+\bmeps_{\alpha}\big)^T\X\W^T\W\X^T\big(\X_{(1)}\bmalpha_{(1)}+\bmeps_{\alpha}\big)\Big\}
=o\{n_1^2n_3^2m_{\alpha}^2(n_1+p)^2\cdot{\sigma^4_{\alpha}}\}, &
\end{flalign*}
where $a_{22}=\tE(\alpha^2\eta^2)<\infty$. 
\end{proposition.s}
\begin{proposition.s}\label{pop.i2}
Under polygenic model~(\ref{equ2.1}) and Conditions~\ref{con1} and~\ref{con2}, 
if $m_{\alpha\beta},m_{\alpha}$, and 
$m_{\beta} \rightarrow \infty$  as $n_1,n_2, n_3,p\rightarrow\infty$, then we have 
\begin{flalign*}
&\tE\Big\{\big(\Z_{(1)}\bmbeta_{(1)}+\bmeps_{\beta}\big)^T\Z\W^T\W\X^T\big(\X_{(1)}\bmalpha_{(1)}+\bmeps_{\alpha}\big)\Big\} = n_1n_2n_3m_{\alpha\beta} \cdot{\sigma_{\alpha\beta}},& \\
&\var\Big\{\big(\Z_{(1)}\bmbeta_{(1)}+\bmeps_{\beta}\big)^T\Z\W^T\W\X^T\big(\X_{(1)}\bmalpha_{(1)}+\bmeps_{\alpha}\big)\Big\}=\bigO(n_1n_2n_3m_{\alpha\beta}^2p^2)+o(n_1^2n_2^2n_3^2m^2_{\alpha\beta}), &
\end{flalign*}
\begin{flalign*}
&\tE\Big\{\big(\Z_{(1)}\bmbeta_{(1)}+\bmeps_{\beta}\big)^T\Z\W^T\W\Z^T\big(\Z_{(1)}\bmbeta_{(1)}+\bmeps_{\beta}\big)\Big\}&\\
&=\big[n_2n_3m_{\beta}(n_2+m_{\beta}) \cdot\{1+o(1)\} + n_2n_3m_{\beta}(p-m_{\beta})\big]\cdot{\sigma^2_{\beta}}+n_2n_3p \cdot \sigma^2_{\epsilon_{\beta}},&\\
&\var\Big\{\big(\Z_{(1)}\bmbeta_{(1)}+\bmeps_{\beta}\big)^T\Z\W^T\W\Z^T\big(\Z_{(1)}\bmbeta_{(1)}+\bmeps_{\beta}\big)\Big\}
=o\{n_2^2n_3^2m_{\beta}^2(n_2+p)^2\cdot{\sigma^4_{\beta}}\} \mathbf{.} &
\end{flalign*}
\end{proposition.s}
\begin{proposition.s}\label{pop.i3}
Under polygenic model~(\ref{equ2.1}) and Conditions~\ref{con1} and~\ref{con2}, 
if $m_{\alpha\beta},m_{\alpha}$, and 
$m_{\beta} \rightarrow \infty$ as  $n_1,n_2,p\rightarrow\infty$, then we have 
\begin{flalign*}
&\tE\Big\{\big(\Z_{(1)}\bmbeta_{(1)}+\bmeps_{\beta}\big)^T\Z\X^T\big(\X_{(1)}\bmalpha_{(1)}+\bmeps_{\alpha}\big)\Big\} = n_1n_2m_{\alpha\beta} \cdot{\sigma_{\alpha\beta}},&\\
&\var\Big\{\big(\Z_{(1)}\bmbeta_{(1)}+\bmeps_{\beta}\big)^T\Z\X^T\big(\X_{(1)}\bmalpha_{(1)}+\bmeps_{\alpha}\big)\Big\}
=\bigO(n_1n_2m_{\alpha\beta}^2p)+o(n_1^2n_2^2m^2_{\alpha\beta}),&\\~\\
&\tE\Big\{\big(\X_{(1)}\bmalpha_{(1)}+\bmeps_{\alpha}\big)^T\X\X^T\big(\X_{(1)}\bmalpha_{(1)}+\bmeps_{\alpha}\big)\Big\}&\\
&=\big[n_1m_{\alpha}(n_1+m_{\alpha}) \cdot\{1+o(1)\} + n_1m_{\alpha}(p-m_{\alpha})\big]\cdot{\sigma^2_{\alpha}}+n_1p \cdot \sigma^2_{\epsilon_{\alpha}},&\\
&\var\Big\{\big(\X_{(1)}\bmalpha_{(1)}+\bmeps_{\alpha}\big)^T\X\X^T\big(\X_{(1)}\bmalpha_{(1)}+\bmeps_{\alpha}\big)\Big\}
=o\{n_1^2m_{\alpha}^2(n_1+p)^2\cdot{\sigma^4_{\alpha}}\}, &\\~\\
&\tE\Big\{\big(\Z_{(1)}\bmbeta_{(1)}+\bmeps_{\beta}\big)^T\Z\Z^T\big(\Z_{(1)}\bmbeta_{(1)}+\bmeps_{\beta}\big)\Big\}&\\
&=\big[n_2m_{\beta}(n_2+m_{\beta}) \cdot\{1+o(1)\} + n_2m_{\beta}(p-m_{\beta})\big]\cdot{\sigma^2_{\beta}}+n_2p \cdot \sigma^2_{\epsilon_{\beta}},&\\
&\var\Big\{\big(\Z_{(1)}\bmbeta_{(1)}+\bmeps_{\beta}\big)^T\Z\Z^T\big(\Z_{(1)}\bmbeta_{(1)}+\bmeps_{\beta}\big)\Big\}
=o\{n_2^2m_{\beta}^2(n_2+p)^2\cdot{\sigma^4_{\beta}}\} \mathbf{.} &
\end{flalign*}
Then Propositions A\ref{pop.a1}~-~A\ref{pop.a3} follow from Markov's inequality.  
\end{proposition.s}
\subsubsection*{Cross-trait PRS with selected SNPs}
\begin{proposition.s}\label{pop.i4}
Under polygenic model~(\ref{equ2.1}) and Conditions~\ref{con1} and~\ref{con2}, 
suppose $m_{\alpha\eta}$, $m_{\alpha\beta}$, $m_{\alpha},m_{\eta}$, and 
$m_{\beta} \rightarrow \infty$, 
$q_{\alpha\beta},q_{\alpha 1},q_{\alpha 2},q_{\beta 1},q_{\beta 2}$, and $q_{\alpha\eta}\rightarrow \infty$ as $n_1,n_2, n_3,p\rightarrow\infty$, then we have 
\begin{flalign*}
&\tE(C_{\alpha\eta}) = n_1n_3q_{\alpha\eta} \cdot{\sigma_{\alpha\eta}},&\\
&\var(C_{\alpha\eta})
=\bigO(m_{\alpha\eta}^2n_1n_3q_{\alpha})+o(n_1^2n_3^2q_{\alpha\eta}^2), &\\~\\
&\tE(V_{\alpha})=\{n_1n_3m_{\alpha}q_{\alpha2}+n_1n_3q_{\alpha1}(m_{\alpha}+n_1)\}\cdot{\sigma^2_{\alpha}}\cdot\{1+o(1)\}+
n_1n_3q_{\alpha}\cdot \sigma^2_{\epsilon_{\alpha}},&\\
&\var(V_{\alpha})
=o\big[\{n_1n_3m_{\alpha}q_{\alpha2}+n_1n_3q_{\alpha1}(m_{\alpha}+n_1)\}^2\big], &\\~\\
&\tE(C_{\alpha\beta}) = n_1n_2n_3q_{\alpha\beta} \cdot{\sigma_{\alpha\beta}},&\\
&\var(C_{\alpha\eta})
=\bigO(m_{\alpha\beta}^2n_1n_2n_3q_{\alpha}q_{\beta})+o(n_1^2n_2^2n_3^2q_{\alpha\beta}^2), &\\~\\
&\tE(V_{\beta})=\{n_2n_3m_{\beta}q_{\beta2}+n_2n_3q_{\beta1}(m_{\beta}+n_2)\}\cdot{\sigma^2_{\beta}}\cdot\{1+o(1)\}+
n_2n_3q_{\alpha}\cdot \sigma^2_{\epsilon_{\alpha}},&\\
&\var(V_{\alpha})
=o\big[\{n_2n_3m_{\beta}q_{\beta2}+n_2n_3q_{\beta1}(m_{\beta}+n_2)\}^2\big] \mathbf{.}&
\end{flalign*}
Then Propositions A\ref{pop.a4}~and~A\ref{pop.a5} follow from Markov's inequality.  
\end{proposition.s}
\subsubsection*{Overlapping samples}
\begin{proposition.s}\label{pop.i5}
Under polygenic model~(\ref{equ2.1}) and Conditions~\ref{con1}~-~\ref{con3}, suppose $m_{\alpha\eta},m_{\alpha}$, and 
$m_{\eta}\rightarrow \infty$ as $(n_1+n_s), (n_3+n_s),p\rightarrow\infty$,  then we have 
\begin{flalign*}
&\tE\big(\y_{\eta}^T\widehat{\bmS}_{S\alpha}\big) =
(n_3+n_s)(n_1+n_s)m_{\alpha\eta}\cdot\sigma_{\alpha\eta}+
n_sm_{\alpha\eta}p\cdot\sigma_{\alpha\eta}+n_sp\cdot\sigma_{\epsilon_{\alpha}\epsilon_{\eta}},& \\
&\var\big(\y_{\eta}^T\widehat{\bmS}_{S\alpha}\big)
=\bigO\big\{(n_3+n_s)(n_1+n_s)m_{\alpha\eta}^2p\big\}+o\big\{\tE^2(\y_{\eta}^T\widehat{\bmS}_{S\alpha})\big\}, &
\end{flalign*}
\begin{flalign*}
&\tE\big(\y^T_{\eta}\y_{\eta}\big)=(n_3+n_s)m_{\eta} \cdot{\sigma^2_{\eta}}+ (n_3+n_s)\cdot{\sigma^2_{\epsilon_{\eta}}}, &\\
&\var\big(\y^T_{\eta}\y_{\eta}\big)=o\big\{(n_3+n_s)^2m_{\eta}^2\cdot{\sigma^4_{\eta}}\big\},&\\~\\
&\tE\big(\widehat{\bmS}_{S\alpha}^T\widehat{\bmS}_{S\alpha}\big)
=\big\{n_1n_3m_{\alpha}(p+n_1)\cdot \sigma^2_{\alpha}+n_1n_3p\cdot\sigma^2_{\epsilon_{\alpha} }\big\}+2\big\{n_1n_3n_sm_{\alpha}\cdot\sigma^2_{\alpha}\big\}+\\
&\qquad \big\{n_sn_3m_{\alpha}(p+n_s)\cdot\sigma^2_{\alpha}+n_sn_3p\cdot\sigma^2_{\epsilon_{\alpha} }\big\}+\big\{n_1n_sm_{\alpha}(p+n_1)\cdot \sigma^2_{\alpha}+n_1n_sp\cdot\sigma^2_{\epsilon_{\alpha} }\big\}\\
&\qquad +2\big\{n_1n_sm_{\alpha}(n_s+p)\cdot\sigma^2_{\alpha}\big\}+\big\{n_sm_{\alpha}(n^2_s+p^2+3n_sp)\cdot \sigma^2_{\alpha}+n_sp(n_s+p)\cdot\sigma^2_{\epsilon_{\alpha}}\big\},&\\
&\var\big(\widehat{\bmS}_{S\alpha}^T\widehat{\bmS}_{S\alpha}\big)
=o\big\{\tE^2(\widehat{\bmS}_{S\alpha}^T\widehat{\bmS}_{S\alpha})\big\} \mathbf{.}&
\end{flalign*}
\end{proposition.s}
\begin{proposition.s}\label{pop.i6}
Under polygenic model~(\ref{equ2.1}) and Conditions~\ref{con1},~\ref{con2} and~\ref{con4}, 
suppose $m_{\alpha\beta},m_{\alpha}$, and 
$m_{\beta} \rightarrow \infty$ as  $(n_1+n_s),(n_2+n_s), n_3,p\rightarrow\infty$,  then we have 
\begin{flalign*}
&\tE\big(\widehat{\bmS}_{S\alpha}^T\widehat{\bmS}_{S\beta}\big)
=(n_1+n_s)(n_2+n_s)n_3m_{\alpha\beta}\cdot{\sigma_{\alpha\beta}}+n_sn_3m_{\alpha\beta}p\cdot\sigma_{\alpha\beta}+n_sn_3p\cdot\sigma_{\epsilon_{\alpha}\epsilon_{\beta}},&\\
&\var\big(\widehat{\bmS}_{S\alpha}^T\widehat{\bmS}_{S\beta}\big)
=\bigO\big\{(n_1+n_s)(n_2+n_s)n_3m_{\alpha\eta}^2p^2\big\}+o\big\{\tE^2(\widehat{\bmS}_{S\alpha}^T\widehat{\bmS}_{S\beta})\big\},&
\end{flalign*}
\begin{flalign*}
&\tE(\widehat{\bmS}_{S\alpha}^T\widehat{\bmS}_{S\alpha})=n_1n_3m_{\alpha}(p+n_1)\cdot\sigma^2_{\alpha}+n_1n_3p\cdot \sigma^2_{\epsilon_{\alpha}}+
2n_1n_3n_sm_{\alpha}\cdot \sigma^2_{\alpha}+ \\
&\qquad n_sn_3m_{\alpha}(p+n_s)\cdot\sigma^2_{\alpha}+n_sn_3p\cdot \sigma^2_{\epsilon_{\alpha}}, &\\
&\var(\widehat{\bmS}_{S\alpha}^T\widehat{\bmS}_{S\alpha})=o\big\{\tE^2(\widehat{\bmS}_{S\alpha}^T\widehat{\bmS}_{S\alpha})\big\},&\\~\\
&\tE(\widehat{\bmS}_{S\beta}^T\widehat{\bmS}_{S\beta})=n_2n_3m_{\beta}(p+n_2)\cdot\sigma^2_{\beta}+n_2n_3p\cdot \sigma^2_{\epsilon_{\beta}}+
2n_2n_3n_sm_{\beta}\cdot \sigma^2_{\beta}+ \\
&\qquad n_sn_3m_{\beta}(p+n_s)\cdot\sigma^2_{\beta}+n_sn_3p\cdot \sigma^2_{\epsilon_{\beta}},&\\
&\var(\widehat{\bmS}_{S\beta}^T\widehat{\bmS}_{S\beta})
=o\big\{\tE^2(\widehat{\bmS}_{S\beta}^T\widehat{\bmS}_{S\beta})\big\}\mathbf{.} &
\end{flalign*}
Then Propositions~S\ref{pop.o1}~and~S\ref{pop.o2} follow from Markov's inequality.  
\end{proposition.s}
\section{Additional technical details}
The following technical details are useful in proving our theoretical results. 
Most of them involve in calculating the asymptotic expectation of the trace of the product of multiple large random matrices. 
\subsubsection*{First moment of covariance term}
\begin{flalign*}
&\tE\Big\{\big(\W_{(1)}\bmeta_{(1)}+\bmeps_{\eta}\big)^T\W\X^T\big(\X_{(1)}\bmalpha_{(1)}+\bmeps_{\alpha}\big)\Big\}&\\
&=\tE\Big\{\tr(\bmeta_{(1)}^T\W_{(1)}^T\W\X^T\X_{(1)}\bmalpha_{(1)})\Big\}
= n_1n_3m_{\alpha\eta} \cdot{\sigma_{\alpha\eta}}.&
\end{flalign*}
\subsubsection*{First moment of variance terms}
\begin{flalign*}
&\tE\Big\{\big(\W_{(1)}\bmeta_{(1)}+\bmeps_{\eta}\big)^T\big(\W_{(1)}\bmeta_{(1)}+\bmeps_{\eta}\big)\Big\}=\tE\big\{\tr(\W_{(1)}^T\W_{(1)})\big\} \cdot{\sigma^2_{\eta}}+\bmeps_{\eta}^T\bmeps_{\eta}
=n_3m_{\eta} \cdot{\sigma^2_{\eta}}+ n_3\cdot{\sigma^2_{\epsilon_{\eta}}}. &
\end{flalign*}
\begin{flalign*}
&\tE\Big\{\big(\X_{(1)}\bmalpha_{(1)}+\bmeps_{\alpha}\big)^T\X\W^T\W\X^T\big(\X_{(1)}\bmalpha_{(1)}+\bmeps_{\alpha}\big)\Big\}&\\
&=\tE\Big(\bmalpha_{(1)}^T\X_{(1)}^T\X_{(1)}\W_{(1)}^T\W_{(1)}\X_{(1)}^T\X_{(1)}\bmalpha_{(1)}\Big) +\tE\Big(\bmalpha_{(1)}^T\X_{(1)}^T\X_{(2)}\W_{(2)}^T\W_{(2)}\X_{(2)}^T\X_{(1)}\bmalpha_{(1)}\Big)&\\
&\qquad +2\tE\Big(\bmalpha_{(1)}^T\X_{(1)}^T\X_{(1)}\W_{(1)}^T\W_{(2)}\X_{(2)}^T\X_{(1)}\bmalpha_{(1)}\Big)
+\tE\Big(\bmeps_{\alpha}^T\X\W^T\W\X^T\bmeps_{\alpha}\Big)
&\\
&=\big[n_1n_3m_{\alpha}(n_1+m_{\alpha}) \cdot\{1+o(1)\} + n_1n_3m_{\alpha}(p-m_{\alpha})\big]\cdot{\sigma^2_{\alpha}}+n_1n_3p \cdot \sigma^2_{\epsilon_{\alpha}}.&
\end{flalign*}
\subsubsection*{Second moment of covariance term}
\begin{flalign*}
&\tE\Big(\bmeta_{(1)}^T\W_{(1)}^T\W\X^T\X_{(1)}\bmalpha_{(1)}\bmeta_{(1)}^T\W_{(1)}^T\W\X^T\X_{(1)}\bmalpha_{(1)}\Big)&\\
&=\tE\Big(\tr(\bmeta_{(1)}^T\W_{(1)}^T\W_{(1)}\X_{(1)}^T\X_{(1)}\bmalpha_{(1)}\bmeta_{(1)}^T\W_{(1)}^T\W_{(1)}\X_{(1)}^T\X_{(1)}\bmalpha_{(1)})\Big)&\\
& +\tE\Big(\tr(\bmeta_{(1)}^T\W_{(1)}^T\W_{(2)}\X_{(2)}^T\X_{(1)}\bmalpha_{(1)}\bmeta_{(1)}^T\W_{(1)}^T\W_{(2)}\X_{(2)}^T\X_{(1)}\bmalpha_{(1)})\Big)&\\
& +2\tE\Big(\tr(\bmeta_{(1)}^T\W_{(1)}^T\W_{(1)}\X_{(1)}^T\X_{(1)}\bmalpha_{(1)}\bmeta_{(1)}^T\W_{(1)}^T\W_{(2)}\X_{(2)}^T\X_{(1)}\bmalpha_{(1)})\Big)&\\
&=(n_1^2n_3^2m_{\alpha\eta}^2+n_1n_3m_{\alpha\eta}^2p+
2n_1^2n_3m_{\alpha\eta}^2+
2n_1n_3^2m_{\alpha\eta}^2)\cdot \sigma^2_{\alpha\eta}+n_1^2n_3^2m_{\alpha\eta}\cdot (a_{22}-\sigma^2_{\alpha\eta}). 
\end{flalign*}
It follows that 
\begin{flalign*}
& \var\Big(\bmeta_{(1)}^T\W_{(1)}^T\W\X^T\X_{(1)}\bmalpha_{(1)}\Big)=n_1n_3m_{\alpha\eta}^2p\cdot \sigma^2_{\alpha\eta}\cdot\{1+o(1)\} +o(n_1^2n_3^2m_{\alpha\eta}^2\cdot \sigma^2_{\alpha\eta}).& 
\end{flalign*}
\subsubsection*{Second moment of variance terms}
\begin{flalign*}
&\tE\Big(\bmeta_{(1)}^T\W_{(1)}^T\W_{(1)}\bmeta_{(1)}\bmeta_{(1)}^T\W_{(1)}^T\W_{(1)}\bmeta_{(1)}\Big)&\\
&=n_3^2m_{\eta}^2 \cdot{\sigma^4_{\eta}}+ n_3m_{\eta}\{2m_{\eta}{\sigma^4_{\eta}}+n_3(b_4-{\sigma^4_{\eta}})+c_4b_4-2{\sigma^4_{\eta}}-b_4\} =n_3^2m_{\eta}^2 \cdot{\sigma^4_{\eta}}\cdot\{1+o(1)\}.&
\end{flalign*}
and 
\begin{flalign*}
&\tE\Big(\bmalpha_{(1)}^T\X_{(1)}^T\X\W^T\W\X^T\X_{(1)}\bmalpha_{(1)}
\bmalpha_{(1)}^T\X_{(1)}^T\X\W^T\W\X^T\X_{(1)}\bmalpha_{(1)}\Big)&\\
&=\tE\Big(\bmalpha_{(1)}^T\X_{(1)}^T\X_{(1)}\W_{(1)}^T\W_{(1)}\X_{(1)}^T\X_{(1)}\bmalpha_{(1)}
\bmalpha_{(1)}^T\X_{(1)}^T\X_{(1)}\W_{(1)}^T\W_{(1)}\X_{(1)}^T\X_{(1)}\bmalpha_{(1)}\Big) &\\
&+\tE\Big(\bmalpha_{(1)}^T\X_{(1)}^T\X_{(2)}\W_{(2)}^T\W_{(2)}\X_{(2)}^T\X_{(1)}\bmalpha_{(1)}
\bmalpha_{(1)}^T\X_{(1)}^T\X_{(2)}\W_{(2)}^T\W_{(2)}\X_{(2)}^T\X_{(1)}\bmalpha_{(1)}\Big) &\\
&+4\tE\Big(\bmalpha_{(1)}^T\X_{(1)}^T\X_{(1)}\W_{(2)}^T\W_{(2)}\X_{(1)}^T\X_{(1)}\bmalpha_{(1)}
\bmalpha_{(1)}^T\X_{(1)}^T\X_{(1)}\W_{(1)}^T\W_{(2)}\X_{(2)}^T\X_{(1)}\bmalpha_{(1)}\Big) &\\
&+2\tE\Big(\bmalpha_{(1)}^T\X_{(1)}^T\X_{(1)}\W_{(1)}^T\W_{(1)}\X_{(1)}^T\X_{(1)}\bmalpha_{(1)}
\bmalpha_{(1)}^T\X_{(1)}^T\X_{(2)}\W_{(2)}^T\W_{(2)}\X_{(2)}^T\X_{(1)}\bmalpha_{(1)}\Big) &\\
&=n_1^2n_3^2m_{\alpha}^2(n_1+p)^2\cdot{\sigma^4_{\alpha}}\cdot\{1+o(1)\}.
\end{flalign*}
Similarly, we have 
\begin{flalign*}
&\tE\Big(\bmbeta_{(1)}^T\Z_{(1)}^T\Z\W^T\W\X^T\Z_{(1)}\bmbeta_{(1)}\bmbeta_{(1)}^T\Z_{(1)}^T\Z\W^T\W\Z^T\Z_{(1)}\bmbeta_{(1)}\Big)&\\
&=n_2^2n_3^2m_{\beta}^2(n_2+p)^2\cdot{\sigma^4_{\beta}}\cdot\{1+o(1)\}. 
\end{flalign*}
Thus, we have  
\begin{flalign*}
&\var\Big(\bmeta_{(1)}^T\W_{(1)}^T\W_{(1)}\bmeta_{(1)}\Big)=o(n_3^2m_{\eta}^2 \cdot{\sigma^4_{\eta}}), &\\
& \var\Big(\bmalpha_{(1)}^T\X_{(1)}^T\X\W^T\W\X^T\X_{(1)}\bmalpha_{(1)}\Big)=o\big\{n_1^2n_3^2m_{\alpha}^2(n_1+p)^2\cdot{\sigma^4_{\alpha}}\big\}, & \\
& \var\Big(\bmbeta_{(1)}^T\Z_{(1)}^T\Z\W^T\W\X^T\Z_{(1)}\bmbeta_{(1)}\Big)=o\big\{n_2^2n_3^2m_{\beta}^2(n_2+p)^2\cdot{\sigma^4_{\beta}}\big\}. & 
\end{flalign*}
\subsubsection*{First moment of covariance term}
\begin{flalign*}
&\tE\Big\{\big(\Z_{(1)}\bmbeta_{(1)}+\bmeps_{\beta}\big)^T\Z\W^T\W\X^T\big(\X_{(1)}\bmalpha_{(1)}+\bmeps_{\alpha}\big)\Big\}
=\tE\Big(\bmbeta_{(1)}^T\Z_{(1)}^T\Z\W^T\W\X^T\X_{(1)}\bmalpha_{(1)}\Big)&\\
&=\tE\Big(\tr(\bmbeta_{(1)}^T\Z_{(1)}^T\Z_{(1)}\W_{(1)}^T\W_{(1)}\X_{(1)}^T\X_{(1)}\bmalpha_{(1)})\Big)
=n_1n_2n_3m_{\alpha\beta} \cdot{\sigma_{\alpha\beta}}.&
\end{flalign*}
\subsubsection*{Second moment of covariance term}
\begin{flalign*}
&\tE\Big(\bmbeta_{(1)}^T\Z_{(1)}^T\Z\W^T\W\X^T\X_{(1)}\bmalpha_{(1)}\bmbeta_{(1)}^T\Z_{(1)}^T\Z\W^T\W\X^T\X_{(1)}\bmalpha_{(1)}\Big)&\\
&=\tE\Big(\bmbeta_{(1)}^T\Z_{(1)}^T\Z_{(1)}\W_{(1)}^T\W_{(1)}\X_{(1)}^T\X_{(1)}\bmalpha_{(1)}\bmbeta_{(1)}^T\Z_{(1)}^T\Z_{(1)}\W_{(1)}^T\W_{(1)}\X_{(1)}^T\X_{(1)}\bmalpha_{(1)}\Big)&\\
&+\tE\Big(\bmbeta_{(1)}^T\Z_{(1)}^T\Z_{(1)}\W_{(1)}^T\W_{(2)}\X_{(2)}^T\X_{(1)}\bmalpha_{(1)}\bmbeta_{(1)}^T\Z_{(1)}^T\Z_{(1)}\W_{(1)}^T\W_{(2)}\X_{(2)}^T\X_{(1)}\bmalpha_{(1)}\Big)&\\
&+\tE(\bmbeta_{(1)}^T\Z_{(1)}^T\Z_{(2)}\W_{(2)}^T\W_{(1)}\X_{(1)}^T\X_{(1)}\bmalpha_{(1)}\bmbeta_{(1)}^T\Z_{(1)}^T\Z_{(2)}\W_{(2)}^T\W_{(1)}\X_{(1)}^T\X_{(1)}\bmalpha_{(1)}\Big)&\\
&+\tE\Big(\bmbeta_{(1)}^T\Z_{(1)}^T\Z_{(2)}\W_{(2)}^T\W_{(2)}\X_{(2)}^T\X_{(1)}\bmalpha_{(1)}\bmbeta_{(1)}^T\Z_{(1)}^T\Z_{(2)}\W_{(2)}^T\W_{(2)}\X_{(2)}^T\X_{(1)}\bmalpha_{(1)}\Big)&\\
&=\bigO\Big[n_1n_2n_3m^2_{\alpha\beta}\{(p-m_{\alpha})(p-m_{\beta})+
(p-m_{\alpha})m_{\beta}+m_{\alpha}(p-m_{\beta})+m_{\alpha}m_{\alpha}\}\Big]&\\
&\qquad +n_1^2n_2^2n_3^2m^2_{\alpha\beta}\cdot{\sigma^2_{\alpha\beta}}\cdot\{1+o(1)\}&\\
&=\bigO(n_1n_2n_3m^2_{\alpha\beta}p^2)+n_1^2n_2^2n_3^2m^2_{\alpha\beta}\cdot{\sigma^2_{\alpha\beta}}\cdot\{1+o(1)\}.&
\end{flalign*}
It follows that 
\begin{flalign*}
& \var\Big(\bmbeta_{(1)}^T\Z_{(1)}^T\Z\W^T\W\X^T\X_{(1)}\bmalpha_{(1)}\Big)
=\bigO(n_1n_2n_3m^2_{\alpha\beta}p^2)+o(n_1^2n_2^2n_3^2m^2_{\alpha\beta}\cdot{\sigma^2_{\alpha\beta}}) .&
\end{flalign*}
\subsubsection*{First moment of covariance term}
\begin{flalign*}
&\tE\Big\{\big(\Z_{(1)}\bmbeta_{(1)}+\bmeps_{\beta}\big)^T\Z\X^T\big(\X_{(1)}\bmalpha_{(1)}+\bmeps_{\alpha}\big)\Big\}&\\
&=\tE\Big(\tr(\bmbeta_{(1)}^T\Z_{(1)}^T\Z\X^T\X_{(1)}\bmalpha_{(1)})\Big)
=n_1n_2m_{\alpha\beta} \cdot{\sigma_{\alpha\beta}}.&
\end{flalign*}
\subsubsection*{First moment of variance terms}
\begin{flalign*}
&\tE\Big\{\big(\X_{(1)}\bmalpha_{(1)}+\bmeps_{\alpha}\big)^T\X\X^T\big(\X_{(1)}\bmalpha_{(1)}+\bmeps_{\alpha}\big)\Big\}&\\
&=\tE\Big(\bmalpha_{(1)}^T\X_{(1)}^T\X_{(1)}\X_{(1)}^T\X_{(1)}\bmalpha_{(1)}\Big)+
\tE\Big(\bmalpha_{(1)}^T\X_{(1)}^T\X_{(2)}\X_{(2)}^T\X_{(1)}\bmalpha_{(1)}\Big)+
\tE\Big(\bmeps_{\alpha}^T\X\X^T\bmeps_{\alpha}\Big)&\\
&=\big[n_1m_{\alpha}(n_1+m_{\alpha})\cdot\{1+o(1)\}+ n_1m_{\alpha}(p-m_{\alpha})\big]\cdot{\sigma_{\alpha}^2}+n_1p \cdot \sigma^2_{\epsilon_{\alpha}}.&
\end{flalign*}
Similarly, we have 
\begin{flalign*}
&\tE\Big\{\big(\Z_{(1)}\bmbeta_{(1)}+\bmeps_{\beta}\big)^T\Z\Z^T\big(\Z_{(1)}\bmbeta_{(1)}+\bmeps_{\beta}\big)\Big\}&\\
&=\big[n_2m_{\beta}(n_2+m_{\beta}) \cdot\{1+o(1)\} + n_2m_{\beta}(p-m_{\beta})\big]\cdot{\sigma^2_{\beta}}+n_2p \cdot \sigma^2_{\epsilon_{\beta}}.&
\end{flalign*}
\subsubsection*{Second moment of covariance term}
\begin{flalign*}
&\tE\Big(\bmbeta_{(1)}^T\Z_{(1)}^T\Z\X^T\X_{(1)}\bmalpha_{(1)}\bmbeta_{(1)}^T\Z_{(1)}^T\Z\X^T\X_{(1)}\bmalpha_{(1)}\Big)&\\
&=\tE\Big(\bmbeta_{(1)}^T\Z_{(1)}^T\Z_{(1)}\X_{(1)}^T\X_{(1)}\bmalpha_{(1)}\bmbeta_{(1)}^T\Z_{(1)}^T\Z_{(1)}\X_{(1)}^T\X_{(1)}\bmalpha_{(1)}\Big)&\\
&+\tE\Big(\bmbeta_{(1)}^T\Z_{(1)}^T\Z_{(1)}\X_{(1)}^T\X_{(1)}\bmalpha_{(1)}\bmbeta_{(1)}^T\Z_{(1)}^T\Z_{(2)}\X_{(2)}^T\X_{(1)}\bmalpha_{(1)}\Big)&\\
&+\tE\Big(\bmbeta_{(1)}^T\Z_{(1)}^T\Z_{(2)}\X_{(2)}^T\X_{(1)}\bmalpha_{(1)}\bmbeta_{(1)}^T\Z_{(1)}^T\Z_{(1)}\X_{(1)}^T\X_{(1)}\bmalpha_{(1)}\Big)&\\
&+\tE\Big(\bmbeta_{(1)}^T\Z_{(1)}^T\Z_{(2)}\X_{(2)}^T\X_{(1)}\bmalpha_{(1)}\bmbeta_{(1)}^T\Z_{(1)}^T\Z_{(2)}\X_{(2)}^T\X_{(1)}\bmalpha_{(1)}\Big)&\\
&=\bigO(n_1n_2m^2_{\alpha\beta}p)+n_1^2n_2^2m^2_{\alpha\beta}\cdot{\sigma^2_{\alpha\beta}}\cdot\{1+o(1)\}.&
\end{flalign*}
It follows that 
\begin{flalign*}
&\var\Big(\bmbeta_{(1)}^T\Z_{(1)}^T\Z\X^T\X_{(1)}\bmalpha_{(1)}\Big)
=\bigO(n_1n_2m^2_{\alpha\beta}p)+o(n_1^2n_2^2m^2_{\alpha\beta}\cdot{\sigma^2_{\alpha\beta}}).&
\end{flalign*}
\subsubsection*{Second moment of variance terms}
\begin{flalign*}
&\tE\Big(\bmalpha_{(1)}^T\X_{(1)}^T\X\X^T\X_{(1)}\bmalpha_{(1)}\bmalpha_{(1)}^T\X_{(1)}^T\X\X^T\X_{(1)}\bmalpha_{(1)}\Big)&\\
&=\tE\Big(\bmalpha_{(1)}^T\X_{(1)}^T\X_{(1)}\X_{(1)}^T\X_{(1)}\bmalpha_{(1)}\bmalpha_{(1)}^T\X_{(1)}^T\X_{(1)}\X_{(1)}^T\X_{(1)}\bmalpha_{(1)}\Big)&\\
&+\tE\Big(\bmalpha_{(1)}^T\X_{(1)}^T\X_{(2)}\X_{(2)}^T\X_{(1)}\bmalpha_{(1)}\bmalpha_{(1)}^T\X_{(1)}^T\X_{(1)}\X_{(1)}^T\X_{(1)}\bmalpha_{(1)}\Big)&\\
&+\tE\Big(\bmalpha_{(1)}^T\X_{(1)}^T\X_{(1)}\X_{(1)}^T\X_{(1)}\bmalpha_{(1)}\bmalpha_{(1)}^T\X_{(1)}^T\X_{(2)}\X_{(2)}^T\X_{(1)}\bmalpha_{(1)}\Big)&\\
&+\tE\Big(\bmalpha_{(1)}^T\X_{(1)}^T\X_{(2)}\X_{(2)}^T\X_{(1)}\bmalpha_{(1)}\bmalpha_{(1)}^T\X_{(1)}^T\X_{(2)}\X_{(2)}^T\X_{(1)}\bmalpha_{(1)}\Big)&\\
&=n_1^2m_{\alpha}^2(p+n_1)^2\cdot{\sigma^4_{\alpha}}\cdot\{1+o(1)\}.&
\end{flalign*}
Similarly, we have 
\begin{flalign*}
&\tE\Big(\bmbeta_{(1)}^T\Z_{(1)}^T\Z\Z^T\Z_{(1)}\bmbeta_{(1)}\bmbeta_{(1)}^T\Z_{(1)}^T\Z\Z^T\Z_{(1)}\bmbeta_{(1)}\Big)
=n_2^2m_{\beta}^2(p+n_2)^2\cdot{\sigma^4_{\beta}}\cdot\{1+o(1)\}.&
\end{flalign*}
Thus, we have  
\begin{flalign*}
&\var\Big(\bmalpha_{(1)}^T\X_{(1)}^T\X\X^T\X_{(1)}\bmalpha_{(1)}\Big)=
o\big\{n_1^2m_{\alpha}^2(p+n_1)^2\cdot{\sigma^4_{\alpha}}\big\}, &\\
&\var\Big(\bmbeta_{(1)}^T\Z_{(1)}^T\Z\Z^T\Z_{(1)}\bmbeta_{(1)}\Big)=o\big\{n_2^2m_{\beta}^2(p+n_2)^2\cdot{\sigma^4_{\beta}}\big\}. & 
\end{flalign*}

\clearpage
\begin{suppfigure}
\includegraphics[page=1,width=1\linewidth]{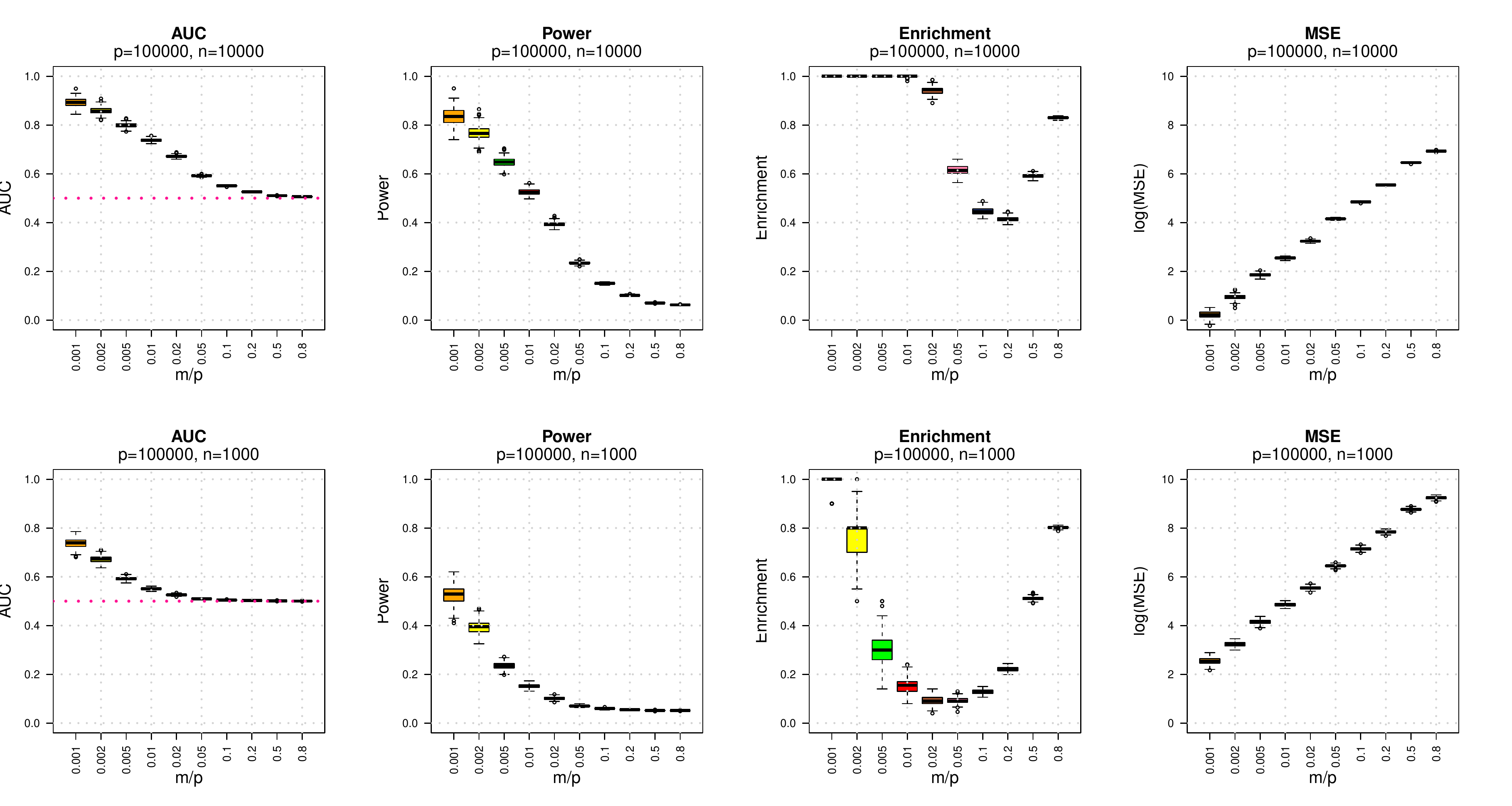}
\caption[]{Trends of GWAS performance when varying sparsity $m/p$ and sample size $n$: AUC of tests, power of tests, enrichment of top-ranked SNP, and MSE of estimated genetic effects.}
\label{fig.s1}
\end{suppfigure}
\clearpage
\begin{suppfigure}
\includegraphics[page=1,width=1\linewidth]{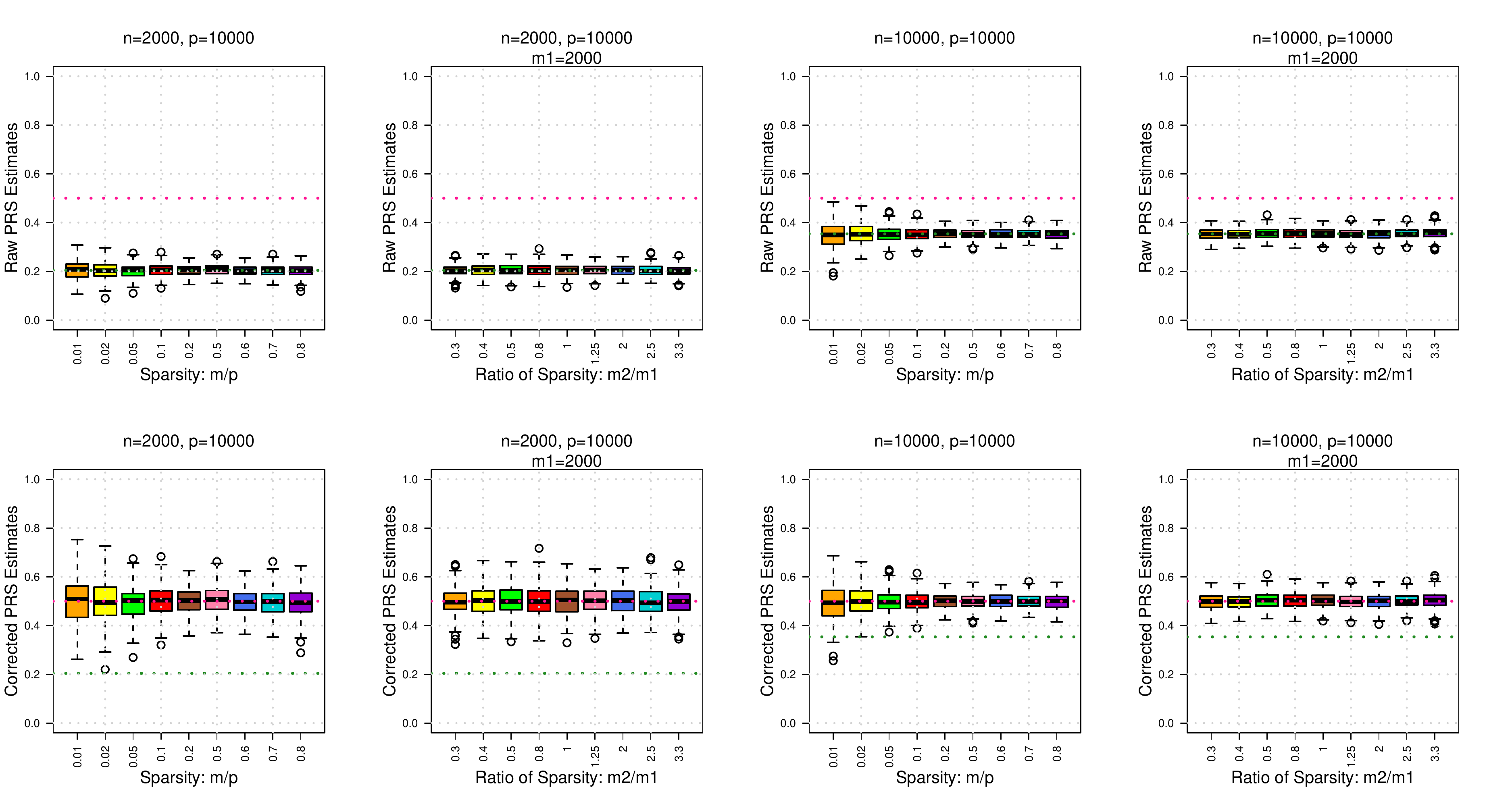}
\caption[]{ Raw genetic correlations estimated by cross-trait PRS with all SNPs ($G_{\alpha\eta}$, upper panels) and corrected ones based on our formulas ($G^A_{\alpha\eta}$, bottom panels). We set $h^2_{\alpha}=h^2_{\eta}=1$, $\varphi_{\alpha\eta}=0.5$, $p=10,000$, and vary $m_{\alpha}$, $m_{\eta}$ and $n$.
}
\label{fig.s2}
\end{suppfigure}
\begin{suppfigure}
\includegraphics[page=1,width=1\linewidth]{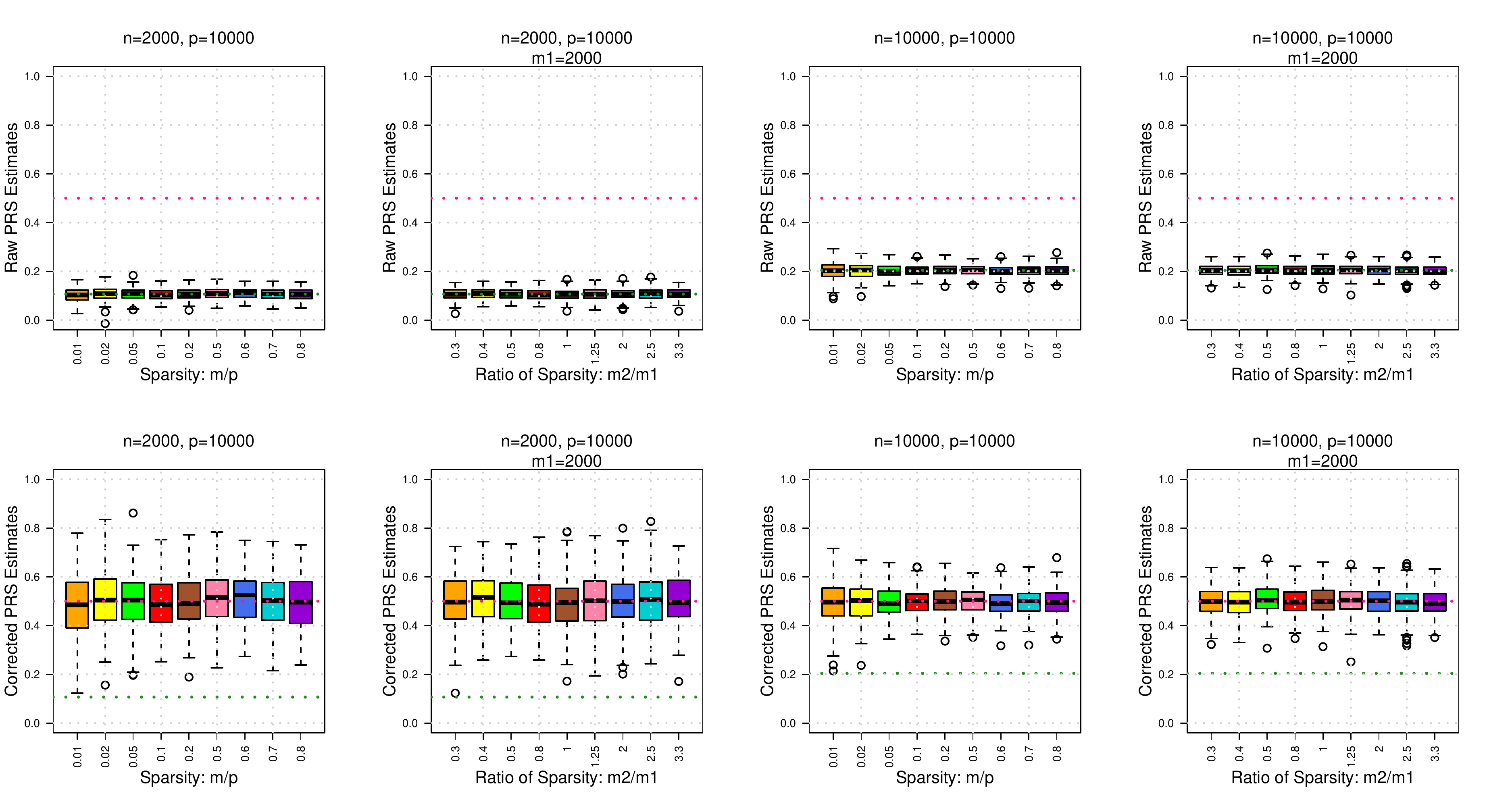}
\caption[]{ Raw genetic correlations estimated by cross-trait PRS with all SNPs ($G_{\alpha\eta}$, upper panels) and corrected ones based on our formulas ($G^A_{\alpha\eta}$, bottom panels). We set $h^2_{\alpha}=h^2_{\eta}=0.5$, $\varphi_{\alpha\eta}=0.5$, $p=10,000$, and vary $m_{\alpha}$, $m_{\eta}$ and $n$.
}
\label{fig.s3}
\end{suppfigure}
\begin{suppfigure}
\includegraphics[page=1,width=1\linewidth]{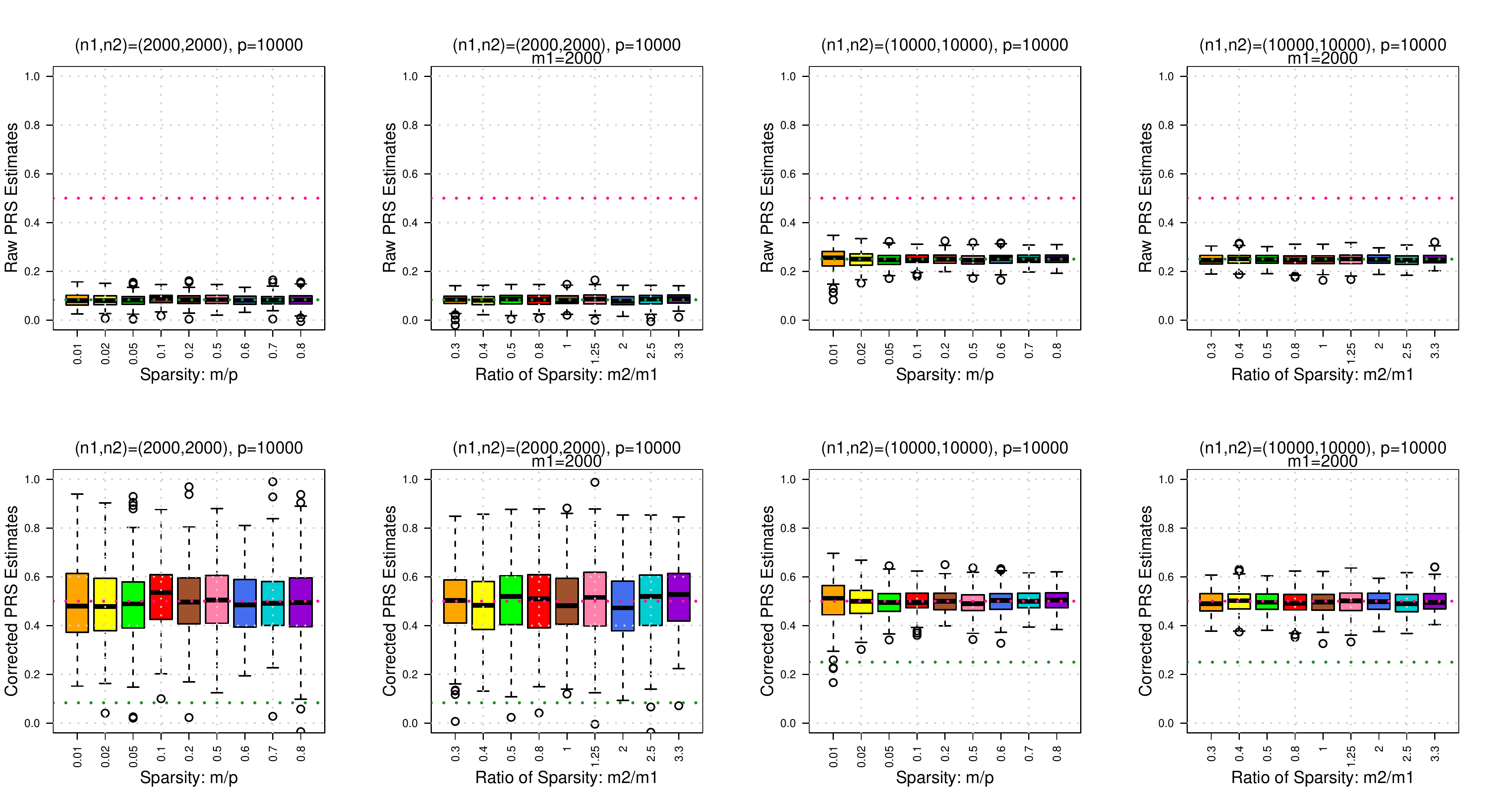}
\caption[]{Raw genetic correlations estimated by cross-trait PRS with all SNPs ($G_{\alpha\beta}$, upper panels) and corrected ones based on our formulas ($G^A_{\alpha\beta}$, bottom panels). We set $h^2_{\alpha}=h^2_{\beta}=1$, $\varphi_{\alpha\beta}=0.5$,
$p=10,000$, and vary $m_{\alpha}$, $m_{\beta}$ and $n$.
}
\label{fig.s4}
\end{suppfigure}
\begin{suppfigure}
\includegraphics[page=1,width=1\linewidth]{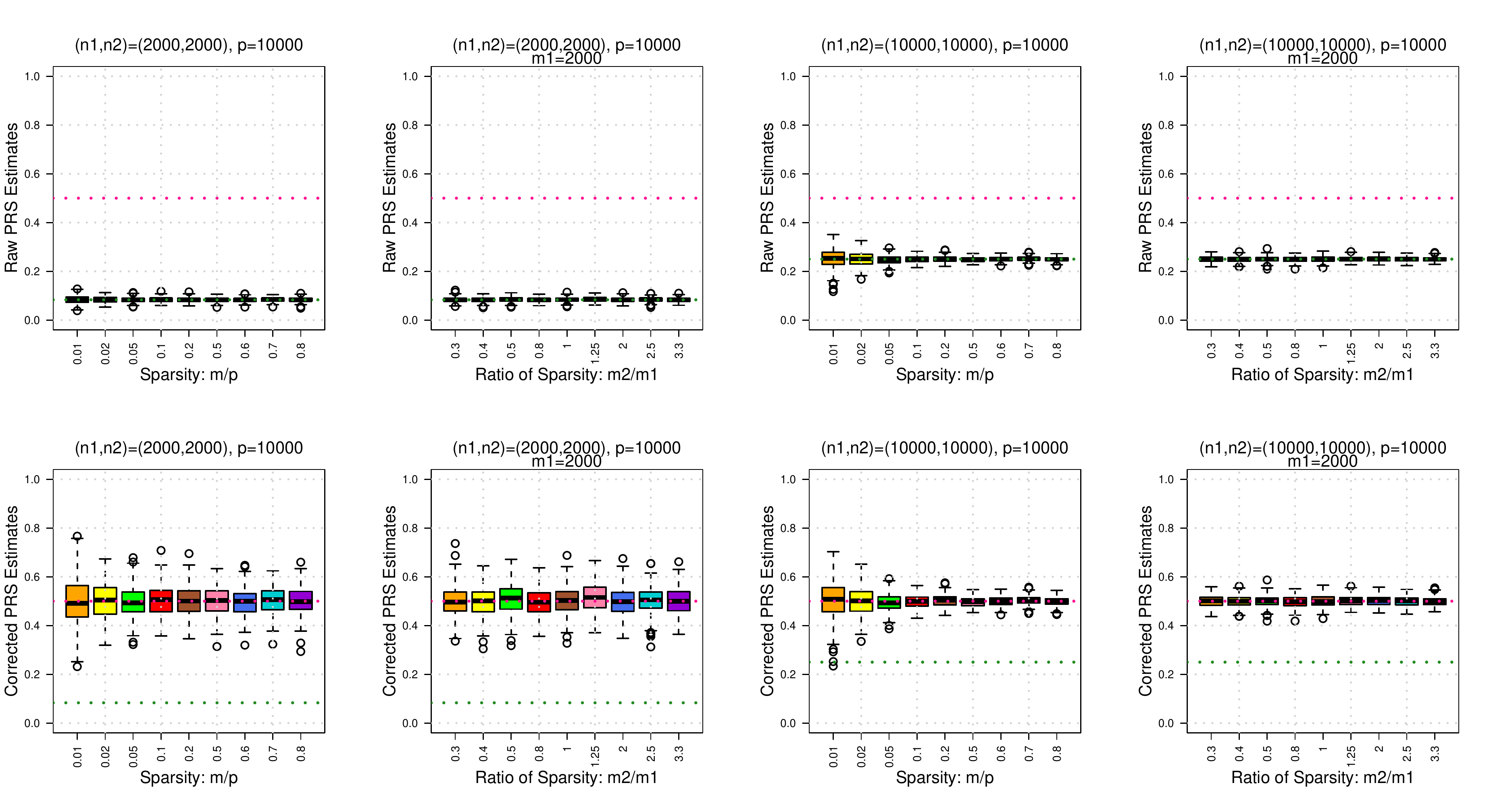}
\caption[]{ Raw genetic correlations estimated by cross-trait PRS directly with all SNPs ($\widehat{\varphi}_{\alpha\beta}$, upper panels) and corrected ones based on our formulas ($\widehat{\varphi}^A_{\alpha\beta}$, bottom panels). We set $h^2_{\alpha}=h^2_{\beta}=1$, $\varphi_{\alpha\beta}=0.5$, $p=10,000$, and vary $m_{\alpha}$, $m_{\beta}$ and $n$.
}
\label{fig.s5}
\end{suppfigure}

 \begin{suppfigure}[ht]
 \includegraphics[page=1,width=1\linewidth]{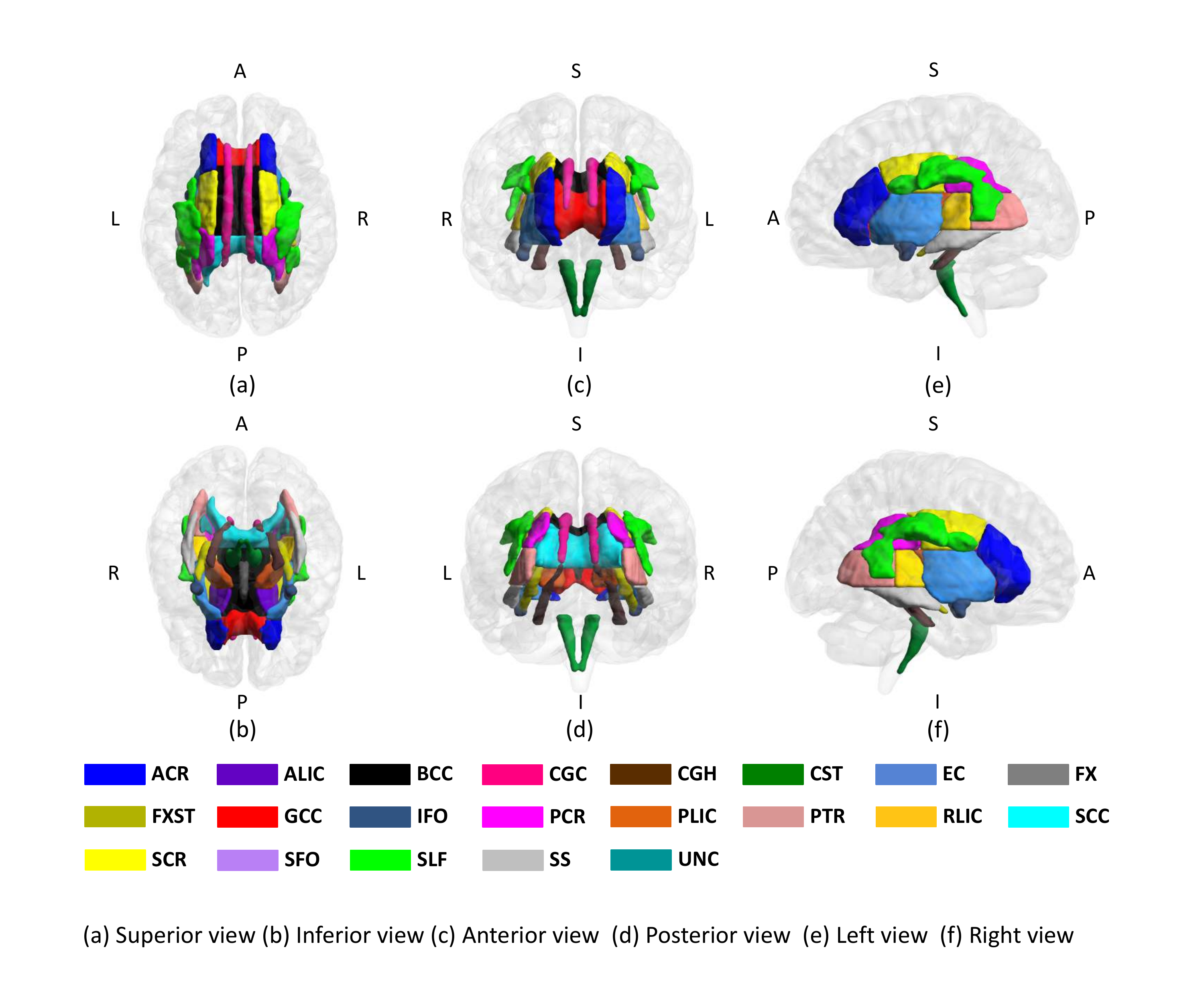}
     \caption{WM main tracts annotation. Originally published in \citet{zhao2018large}. We examine $18$ WM tracts in our real data analysis, whose full names are listed in Supplementary Table~\ref{tab.s2}.}
         \label{fig.s6}
 \end{suppfigure}

\begin{suppfigure}
\includegraphics[page=1,width=1\linewidth]{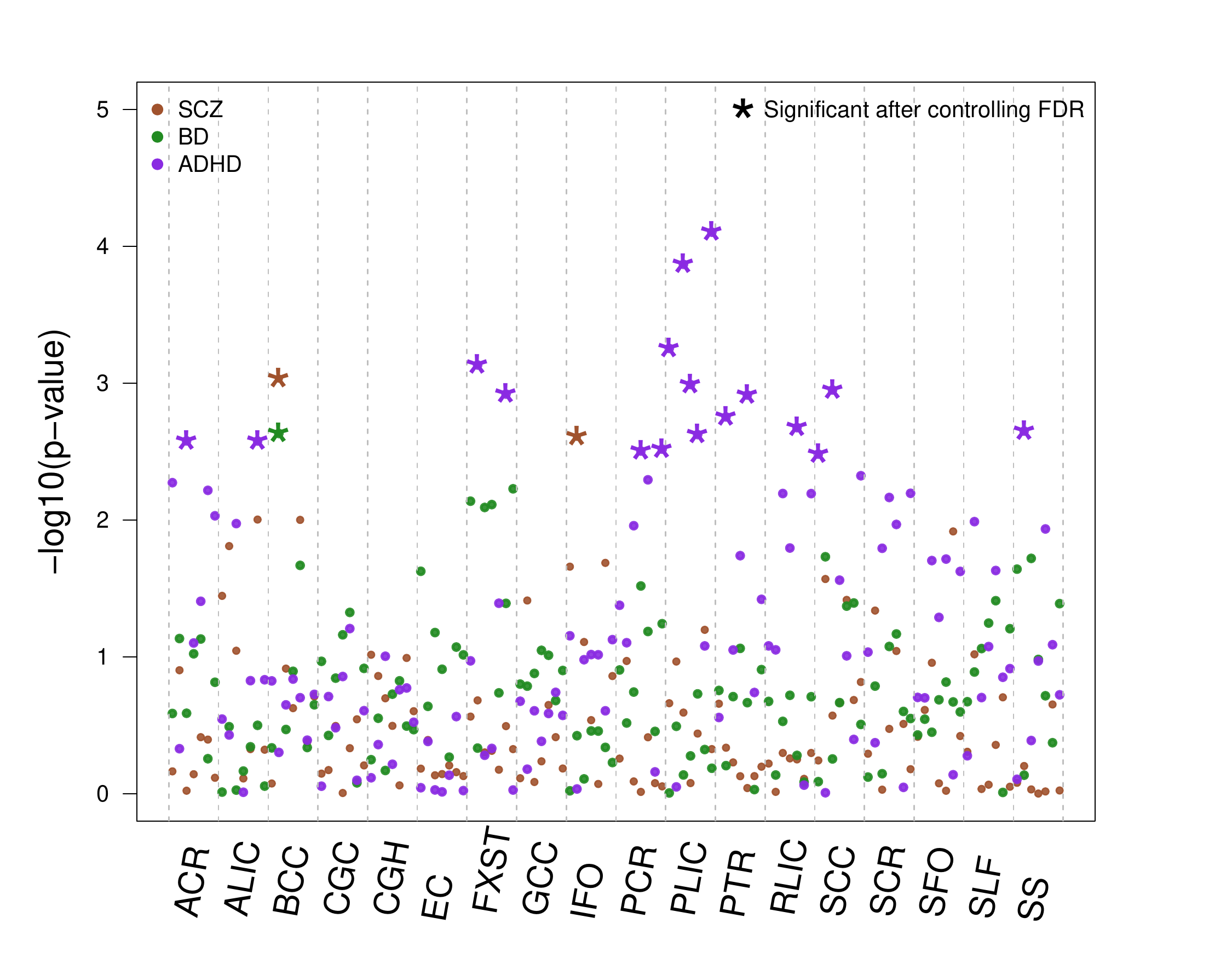}
\caption[]{ Associations between the PRS of four psychiatric disorders created from the published GWAS summary statistics, and $18$ brain WM tracts in the UK Biobank dataset. We control for age, sex, and $10$ genetic
principal components for population structure. FDR is controlled at $0.05$ level.
SCZ,
Schizophrenia; BD, Bipolar
disorder;
ADHD: Attention-decit/hyperactivity disorder;\\
ACR, Anterior corona radiata; ALIC, Anterior limb of internal capsule; BCC, Body of corpus callosum; CGC, Cingulum (cingulate gyrus); CGH, Cingulum (hippocampus); EC, External capsule; FXST, Fornix (cres)/Stria terminalis; GCC, Genu of corpus callosum; IFO, Inferior fronto-occipital fasciculus; PCR, Posterior corona radiata; PLIC, Posterior limb of internal capsule; PTR, Posterior thalamic radiation (include optic radiation); RLIC, Retrolenticular part of internal capsule; SCC, Splenium of corpus callosum; SCR, Superior corona radiata; SFO, Superior fronto-occipital fasciculus; SLF, Superior longitudinal fasciculus; SS, Sagittal stratum.    
}
\label{fig.s7}
\end{suppfigure}

\clearpage
\begin{supptable}[ht]
\centering
\caption[]{Full name and description of DTI parameters.}
\scalebox{0.8}{
\begin{tabular}{rrr}
  \hline
    \hline
DTI parameter & Full name & Description\\ 
  \hline
FA & fractional anisotropy & a summary measure of WM integrity \\ 
MD & mean diffusivities & magnitude of absolute directionality\\ 
AD(L1) & axial diffusivities & eigenvalue of the principal diffusion direction\\ 
RD & radial diffusivities & average of the eigenvalues of the two secondary directions\\ 
MO & mode of anisotropy  & third moment of the tensor\\ 
L2, L3 & &two secondary diffusion direction eigenvalues \\ 
  \hline
   \hline
\end{tabular}
}
\label{tab.s1}
\end{supptable}

\begin{supptable}[ht]
\centering
\caption[]{Full name of $18$ WM tracts.}
\scalebox{0.8}{
\begin{tabular}{rr}
  \hline
    \hline
WM tract & Full name \\ 
  \hline
ACR	& Anterior corona radiata \\ 
ALIC&	Anterior limb of internal capsule\\
BCC	& Body of corpus callosum\\
CGC	& Cingulum (cingulate gyrus) \\
CGH	& Cingulum (hippocampus) \\
EC	& External capsule \\
FXST &	Fornix (cres)/Stria terminalis \\
GCC	& Genu of corpus callosum \\
IFO	& Inferior fronto-occipital fasciculus\\
PCR	& Posterior corona radiata \\
PLIC &	Posterior limb of internal capsule\\
PTR	& Posterior thalamic radiation (include optic radiation) \\
RLIC&	Retrolenticular part of internal capsule\\
SCC	& Splenium of corpus callosum \\
SCR	& Superior corona radiata \\
SFO	& Superior fronto-occipital fasciculus \\
SLF	& Superior longitudinal fasciculus\\
SS &	Sagittal stratum\\
  \hline
   \hline
\end{tabular}
}
\label{tab.s2}
\end{supptable}

\begin{supptable}[ht]
\centering
\caption[]{The $20$ significant associations between the PRS of four psychiatric disorders and brain WM tracts in the UK Biobank dataset (FDR controlled at $0.05$ level). 
Raw $R^2$ stands for the proportion of variance in the DTI parameter that can be explained by psychiatric disorder PRS. Corrected $R^2$s are the ones after correction according our formulas. Seven DTI parameters are considered on each WM tract: FA, fractional anisotropy; MD, mean diffusivities; L1(AD): axial diffusivities, also the eigenvalue of the primary diffusion direction; RD, radial diffusivities; MO, 
mode of anisotropy; L2 and L3, two secondary diffusion direction eigenvalues. 
}
\scalebox{0.8}{
\begin{tabular}{rrrrrr}
  \hline
    \hline
Disorder-Tract-DTI & Estimate & Std. Error & $P$-value & Raw $R^2$ ($\times 100\%$) & Corrected $R^2$ ($\times 100\%$) \\ 
  \hline
ADHD-ACR-L2 & -0.0317 & 0.0105 & 0.0026 & 0.1006 & 4.3005 \\ 
  ADHD-ALIC-MO & 0.0311 & 0.0104 & 0.0026 & 0.0970 & 4.7888 \\ 
  ADHD-FXST-L1 & -0.0373 & 0.0110 & 0.0007 & 0.1392 & 5.9635 \\ 
  ADHD-FXST-MO & -0.0337 & 0.0104 & 0.0012 & 0.1133 & 4.8436 \\ 
  ADHD-PCR-L3 & -0.0319 & 0.0108 & 0.0031 & 0.1017 & 4.9767 \\ 
  ADHD-PCR-RD & -0.0325 & 0.0110 & 0.0030 & 0.1055 & 4.8518 \\ 
  ADHD-PLIC-FA & 0.0388 & 0.0113 & 0.0006 & 0.1509 & 5.4437 \\ 
  ADHD-PLIC-L2 & -0.0431 & 0.0113 & 0.0001 & 0.1862 & 6.6136 \\ 
  ADHD-PLIC-L3 & -0.0361 & 0.0110 & 0.0010 & 0.1306 & 5.4399 \\ 
  ADHD-PLIC-MD & -0.0330 & 0.0109 & 0.0024 & 0.1089 & 5.6350 \\ 
  ADHD-PLIC-RD & -0.0444 & 0.0112 & 0.0001 & 0.1974 & 7.2696 \\ 
  ADHD-PTR-L1 & -0.0351 & 0.0112 & 0.0018 & 0.1233 & 5.6933 \\ 
  ADHD-PTR-MD & -0.0351 & 0.0109 & 0.0012 & 0.1233 & 6.1215 \\ 
  ADHD-RLIC-MD & -0.0336 & 0.0109 & 0.0021 & 0.1127 & 5.6610 \\ 
  ADHD-SCC-FA & 0.0331 & 0.0113 & 0.0033 & 0.1098 & 4.8173 \\ 
  ADHD-SCC-L2 & -0.0367 & 0.0113 & 0.0011 & 0.1347 & 6.4249 \\ 
  ADHD-SS-L1 & -0.0338 & 0.0111 & 0.0022 & 0.1142 & 5.2611 \\ 
  BD-BCC-L1 & 0.0331 & 0.0109 & 0.0023 & 0.1092 & 4.7963 \\ 
  SCZ-BCC-L1 & 0.0363 & 0.0109 & 0.0009 & 0.1315 & 2.8821 \\ 
  SCZ-IFO-L1 & 0.0341 & 0.0113 & 0.0025 & 0.1163 & 3.4237 \\   \hline
   \hline
\end{tabular}
}
\label{tab.s3}
\end{supptable}

\end{document}